\documentclass[longauth]{aa} 

\usepackage[breaklinks, colorlinks, citecolor=blue, linkcolor=blue]{hyperref}
\usepackage{amsmath}
\usepackage{mathtools}
\usepackage{graphicx}
\usepackage{txfonts}
\usepackage{float}

\makeatletter
\DeclareRobustCommand{\ion}[2]{%
  \text{#1\,\check@mathfonts\fontsize\sf@size\z@\selectfont #2}%
}
\makeatother

\newcommand{\hii}{\ion{H}{II}}
\newcommand{\hi}{\ion{H}{I}}

\newcommand{\sii}{[\ion{S}{II}]}
\newcommand{\oiii}{[\ion{O}{III}]}

\begin{document}

   \title{Variations in the $\Sigma_{\rm SFR} {-} \Sigma_{\rm mol} {-} \Sigma_{\rm \star}$ plane across galactic environments in PHANGS galaxies}


   \author{
   I.~Pessa\inst{\ref{mpia}}\thanks{Fellow of the International Max Planck Research School for Astronomy and Cosmic Physics at the University of Heidelberg (IMPRS-HD).} \and 
   E.~Schinnerer\inst{\ref{mpia}} \and 
   A.~K.~Leroy\inst{\ref{ohio}} \and 
   E.~W.~Koch\inst{\ref{cfa}} \and
   E.~Rosolowsky\inst{\ref{alb}} \and
   T.~G.~Williams\inst{\ref{mpia}}
   H.-A.~Pan\inst{\ref{mpia}, \ref{tamkang}} \and 
   A.~Schruba\inst{\ref{mpe}} \and
   A.~Usero\inst{\ref{oan}} \and
   F.~Belfiore\inst{\ref{inaf}} \and 
   F.~Bigiel\inst{\ref{bonn}} \and
   G.A.~Blanc\inst{\ref{carn}, \ref{uch}} \and
   M.~Chevance\inst{\ref{rechen}} \and
   D. Dale\inst{\ref{wyo}} \and 
   E.~Emsellem\inst{\ref{eso},\ref{lyon}}  \and 
   J.~Gensior\inst{\ref{uzh}} \and
   S.~C.~O.~Glover\inst{\ref{zah}} \and 
   K.~Grasha\inst{\ref{Canb}} \and 
   B.~Groves\inst{\ref{ICRA},\ref{Canb}} \and
   R.~S.~Klessen\inst{\ref{zah},\ref{zw}} \and 
   K.~Kreckel\inst{\ref{rechen}} \and
   J.~M.~D.~Kruijssen\inst{\ref{rechen}} \and
   D.~Liu\inst{\ref{mpe}} \and
   S.~E.~Meidt\inst{\ref{ugent}} \and
   J.~Pety\inst{\ref{iram},\ref{lerma}} \and
   M.~Querejeta\inst{\ref{oan}} \and 
   T.~Saito\inst{\ref{mpia}} \and 
   P.~Sanchez-Blazquez\inst{\ref{ucm}} \and 
   E. J. Watkins\inst{\ref{rechen}}
   }
   \institute{Max-Planck-Institute for Astronomy, K\"onigstuhl 17, D-69117 Heidelberg, Germany\label{mpia}\\
        \email{pessa@mpia.de}
        \and Department of Astronomy, The Ohio State University, 140 West 18th Avenue, Columbus, OH 43210, USA\label{ohio}
        \and Center for Astrophysics $\mid$ Harvard \& Smithsonian, 60 Garden St., Cambridge, MA 02138, USA\label{cfa}
        \and Department of Physics, University of Alberta, Edmonton, AB T6G 2E1, Canada\label{alb}
        \and Department of Physics, Tamkang University, No.151, Yingzhuan Rd., Tamsui Dist., New Taipei City 251301, Taiwan\label{tamkang}
        \and Max-Planck-Institute for extraterrestrial Physics, Giessenbachstra{\ss}e 1, D-85748 Garching, Germany\label{mpe}
        \and Observatorio Astronómico Nacional (IGN), C/Alfonso XII, 3, E-28014 Madrid, Spain\label{oan}
        \and INAF — Osservatorio Astrofisico di Arcetri, Largo E. Fermi 5, I-50125, Florence, Italy\label{inaf}
        \and Argelander-Institut f\"ur Astronomie, Universit\"at Bonn, Auf dem H\"ugel 71, D-53121 Bonn, Germany\label{bonn}
        \and Observatories of the Carnegie Institution for Science, Pasadena, CA, USA\label{carn}
        \and Departamento de Astronom\'ia, Universidad de Chile, Santiago,Chile\label{uch}
        \and Astronomisches Rechen-Institut, Zentrum f\"ur Astronomie der Universit\"at Heidelberg, M\"onchhofstra{\ss}e 12-14, D-69120 Heidelberg, Germany\label{rechen}
        \and Department of Physics and Astronomy, University of Wyoming, Laramie, WY 82071, USA\label{wyo}
        \and European Southern Observatory, Karl-Schwarzschild-Stra{\ss}e 2, 85748 Garching, Germany\label{eso}
        \and Univ Lyon, Univ Lyon1, ENS de Lyon, CNRS, Centre de Recherche Astrophysique de Lyon UMR5574, F-69230 Saint-Genis-Laval France\label{lyon}
        \and Institute for Computational Science, Universität Z\"urich, Winterthurerstrasse 190, 8057 Z\"urich, Switzerland\label{uzh}
        \and Universit\"at Heidelberg, Zentrum f\"ur Astronomie, Institut f\"ur theoretische Astrophysik, Albert-Ueberle-Stra{\ss}e 2, D-69120, Heidelberg, Germany\label{zah}
        \and Research School of Astronomy and Astrophysics, Australian National University, Canberra, ACT 2611, Australia\label{Canb}
        \and International Centre for Radio Astronomy Research University of Western Australia 7 Fairway, Crawley WA 6009 Australia\label{ICRA}
        \and Universit\"at Heidelberg, Interdisziplin\"ares Zentrum f\"ur Wissenschaftliches Rechnen, Im Neuenheimer Feld 205, D-69120 Heidelberg, Germany\label{zw}
        \and Sterrenkundig Observatorium, Universiteit Gent, Krijgslaan 281 S9, B-9000 Gent, Belgium\label{ugent}
        \and Institut de Radioastronomie Millim\'etrique (IRAM), 300 Rue de la Piscine, F-38406 Saint Martin d’H`eres, France\label{iram}
        \and LERMA, Observatoire de Paris, PSL Research University, CNRS, Sorbonne Universités, 75014 Paris, France\label{lerma}
        \and Departamento de F\'isica de la Tierra y Astrof\'isica, Universidad Complutense de Madrid, E-28040 Madrid, Spain \label{ucm}
        %
        }

   \date{Received December 3, 2021; Accepted 21 March, 2022}

 
  \abstract
   {}
{There exists some consensus that the stellar mass surface density ($\Sigma_{\star}$) and molecular gas mass surface density ($\Sigma_{\rm mol}$) are the main quantities responsible for locally setting the star formation rate. This regulation is inferred from locally resolved scaling relations between these two quantities and the star formation rate surface density ($\Sigma_{\rm SFR}$), which have been extensively studied in a wide variety of works. However, the universality of these relations is debated. Here, we probe the interplay between these three quantities across different galactic environments at a spatial resolution of $150$~pc.}
   {We performed a hierarchical Bayesian linear regression to find the best set of parameters $C_{\star}$, $C_{\rm mol}$, and $C_{\rm norm}$ that describe the star-forming plane conformed by $\Sigma_{\star}$, $\Sigma_{\rm mol}$, and $\Sigma_{\rm SFR}$, such that $\log \Sigma_{\rm SFR} = C_{\star} \log \Sigma_{\star} + C_{\rm mol} \log \Sigma_{\rm mol} + C_{\rm norm}$. We also explored variations in the determined parameters across galactic environments, focusing our analysis on the $C_{\star}$ and $C_{\rm mol}$ slopes.}
{We find signs of variations in the posterior distributions of $C_{\star}$ and $C_{\rm mol}$ across different galactic environments. The dependence of $\Sigma_{\rm SFR}$ on $\Sigma_{\star}$ spans a wide range of slopes, with negative and positive values, while the dependence of $\Sigma_{\rm SFR}$ on $\Sigma_{\rm mol}$ is always positive. Bars show the most negative value of $C_{\star}$ ($-0.41$), which is a sign of longer depletion times, while spiral arms show the highest $C_{\star}$ among all environments ($0.45$). Variations in $C_{\rm mol}$ also exist, although they are more subtle than those found for $C_{\star}$.}
{We conclude that systematic variations in the interplay of $\Sigma_{\star}$, $\Sigma_{\rm mol}$, and $\Sigma_{\rm SFR}$ across different galactic environments exist at a spatial resolution of $150$~pc, and we interpret these variations to be produced by an additional mechanism regulating the formation of stars that is not captured by either $\Sigma_{\star}$ or $\Sigma_{\rm mol}$. Studying environmental variations in single galaxies, we find that these variations correlate with changes in the star formation efficiency across environments, which could be linked to the dynamical state of the gas that prevents it from collapsing and forming stars, or to changes in the molecular gas fraction.}
   \keywords{galaxies: ISM --
                galaxies: evolution --
                galaxies: star formation --
                galaxies: general
               }
\maketitle

\section{Introduction}
\label{sec:intro}

The conversion of cold dense molecular clouds into stars ultimately occurs when the supporting pressures are insufficient to prevent gravitational collapse. While simple calculations only include gas pressure, both magnetic fields \citep[e.g.,][]{Shu1987} and turbulence \citep[][]{McLow2004} have been proposed as mechanisms able to prevent gravitational collapse. 
In the case of magnetic fields, the collapse is prevented because the neutral hydrogen is coupled with ionized hydrogen, which is tied to the interstellar magnetic field and therefore resists collapse. For the latter proposed mechanism, supersonic turbulent motions act as an additional source of pressure. Once this pressure is no longer sufficient to support the self-gravity of the cloud, it collapses. The collapse of the molecular cloud, which might be triggered by an external source of pressure, leads to its fragmentation \citep{Hoyle1953}, where individual fragments will form stars.

Despite the complexity of this process, several studies have reported correlations between the locally (approximately kpc and sub-kpc scales) measured star formation rate surface density ($\Sigma_{\rm SFR}$), and spatially resolved quantities, such as the local stellar mass surface density ($\Sigma_{\star}$), also known as the ``resolved'' star formation main sequence \citep[rSFMS; ][]{CanoDiaz2016, Abdurro2017, Hsieh2017, Lin2019, Morselli2020, Ellison2021},  or the local molecular gas surface density ($\Sigma_{\rm mol}$), also known as the resolved Kennicutt--Schmidt relation \citep[rKS;][]{Bigiel2008, Leroy2008, Blanc2009, Onodera2010, Bigiel2011, Schruba2011, Ford2013, Leroy2013, Kreckel2018, Williams2018, Dey2019}. The scatter of these correlations is expected to dramatically increase below a critical spatial scale due to statistical undersampling of the time evolution of the star formation process \citep{Schruba2010, Feldmann2011, Kruijssen2014, Kruijssen2018}. \citet{Pessa2021} recently confirmed such an increase in scatter with decreasing physical resolution down to ${\sim}100$~pc in a sample of $18$ nearby galaxies.
These local relationships also have well-studied global manifestations, the star-forming main sequence of galaxies \citep[SFMS;][]{Brinchmann2004, Daddi2007, Noeske2007, Salim2007,Lin2012, Whitaker2012, Speagle2014, Saintonge2016, Popesso2019}, and the galaxy-integrated Kennicutt--Schmidt relation \citep{Schmidt1959, Kennicutt1998, Wyder2009, Genzel2010, Tacconi2010} express relationships among the same quantities for entire galaxies. These galactic-scale relations are key to our current understanding of galaxy evolution and star formation across cosmic time. Studying their spatially resolved versions provides critical insights into their physical origin.

These resolved correlations  arise as a result of some mechanism controlling the formation of stars. Many studies have investigated the dependence on $\Sigma_{\star}$ (through the hydrostatic pressure exerted by the galactic potential) and by $\Sigma_{\rm mol}$ (being the fuel to form stars), as well as a combination of these parameters \citep[e.g.,][]{Matteucci1989, Shi2011, Shi2018, Dey2019, Barrera-Ballesteros2021}. When approaching this topic empirically, some authors have studied a 2D plane in the 3D space spanned by $\Sigma_{\rm SFR}$, $\Sigma_{\star}$ and $\Sigma_{\rm mol}$ \citep{Dib2017,Lin2019,Sanchez2021}, which would imply a scenario where both $\Sigma_{\star}$ and $\Sigma_{\rm mol}$ are responsible for modulating the formation of stars. While there is consensus that these are the main quantities that correlate with SFR to first order, there is debate about the universality of these correlations. Whereas \citet{Sanchez2021}, using data from the CALIFA survey \citep{Sanchez2012}, concluded that the scatter in these relations is fully dominated by individual errors, \citet{Ellison2021} used data from ALMaQUEST \citep{Lin2020} and found that the scatter in these correlations is dominated by galaxy-to-galaxy variations. Similarly, in \citet{Pessa2021}, using high physical resolution data provided by MUSE, the authors found not only galaxy-to-galaxy variations, but also significant variations in these scaling relations across galactic environments. That is, they found different relations for spiral arms, disk, bars, centers and rings. 

In this paper, we aim to test the universality of a two-dimensional (2D) planar star formation relation, defined as $\Sigma_{\rm SFR} \propto \Sigma_{\star}^{a} \Sigma_{\rm mol}^{b}$. If the SFR in a given region is primarily driven by these two quantities, then, this `star-forming plane' should not change significantly between individual galaxies or among different galactic environments. On the other hand, significant variations in the best-fit star-forming plane among different locations would indicate that such a relationship offers an incomplete description of the data and hint at additional quantities or functional forms that are key to regulating star formation in galaxies. 

The Physics at High Angular resolution in Nearby GalaxieS (PHANGS\footnote{\url{http://phangs.org/}}) surveys \citep{Leroy2021b, Emsellem2021, Lee2021} provide the opportunity to explore the relation between these three quantities at a high physical resolution (${\sim}150$~pc), making it possible to isolate galactic environments and study them separately. PHANGS targets probe a diversity of environments (centers and bulges, disks, bars, spiral arms) spanning a range of physical conditions \citep[gas surface density, stellar surface density, dynamical pressure, orbital period, shear rate, streaming motions, gas phase metallicity;][]{Meidt2018, Meidt2020, Jeffreson2020, Kreckel2020, Emsellem2021,Querejeta2021}. Recent works have shown that ${\sim}100{-}150$~pc scale surface density, velocity dispersion, and dynamical state of the molecular gas in these environments are sensitive to the local conditions \citep{Colombo2014, Sun2020b, Rosolowsky2021} in a manner that appears to influence how molecular clouds form stars \citep{Meidt2016, Querejeta2021}. Thus, investigating differences in the interplay of $\Sigma_{\rm SFR}$,  $\Sigma_{\star}$ and $\Sigma_{\rm mol}$ across these galactic environments could be key to understand the galaxy-to-galaxy variations reported by \citet{Ellison2021} and \citet{Pessa2021}, as relative contributions from different environments will vary from one galaxy to another.

The paper is structured as follows: in Sec.~\ref{sec:data} we describe the data and data products used in this work. In Sec.~\ref{sec:methods} we describe in detail the methodology adopted to perform our analyses. In Sec.~\ref{sec:results} and ~\ref{sec:discussion} we present our findings and discussions. Finally, our main conclusions are presented in Sec.~\ref{sec:summary}.

\section{Data}
\label{sec:data}

In this section, we cover only the main aspects of our data set. We refer the reader to \citet{Leroy2021a}, \citet{Emsellem2021} and \citet{Pessa2021} for a more detailed description of our data.
We use a sample of $18$ star-forming galaxies from the overlap of the PHANGS--ALMA and PHANGS--MUSE samples. In order to resolve the typical separation between star-forming regions (${\sim}100$~pc), and limit the effect of extinction, the galaxies studied in this paper have been selected to have distances less than $20$~Mpc and low inclinations ($i<60^{\circ}$). Our sample is summarized in Table~\ref{tab:sample} where we use the global parameters from \citet{Leroy2021b}, which leveraged the distance compilation of \citet{Anand2021} and the galaxy orientations determined by \citet{Lang2020}.

\subsection{VLT MUSE}
We make use of the PHANGS--MUSE survey \citep[PI: E.~Schinnerer;][]{Emsellem2021}. This survey employs the Multi-Unit Spectroscopic Explorer \citep[MUSE;][]{Bacon2014} optical integral field unit (IFU) mounted on the Very Large Telescope (VLT) Unit Telescope~4 to mosaic the star-forming disk of $19$ galaxies from the PHANGS sample. This sample corresponds to a subset of the $90$ galaxies from the PHANGS--ALMA survey \citep[PI: E.~Schinnerer;][]{Leroy2021b}. For the sake of homogeneity of the data set, one galaxy from the PHANGS--MUSE sample (NGC~0628) has been excluded because its MUSE mosaic was obtained using a different observing strategy.

The mosaics consist of $3$ to $15$ individual MUSE pointings, each with a total on-source exposure time of $43$~min. Nine out of the $18$ galaxies were observed with adaptive optics (AO) assistance. These galaxies are marked with a black dot in the first column of Table~\ref{tab:sample}. Each pointing provides a $1\arcmin \times 1\arcmin$ field of view sampled at $0\farcs2$ per pixel, with a typical spectral resolution of ${\sim}2.5$~\AA\ (${\sim}70$~km~s$^{-1}$) covering the wavelength range of $4800{-}9300$~\AA. Observations were reduced using  recipes from the MUSE data reduction pipeline provided by the MUSE consortium \citep{Weilbacher2020}, executed with ESOREX using the python wrapper developed by the PHANGS team\footnote{\url{https://github.com/emsellem/pymusepipe}} \citep{Emsellem2021}. Once the data have been reduced, we have used the PHANGS data analysis pipeline (DAP) to derive various physical quantities, as described in detail in \citet{Emsellem2021}. DAP is based on the GIST pipeline \citep{Bittner2019}, and consists of a series of modules that perform single stellar population (SSP) fitting and emission line measurements to the full MUSE mosaic. Some outputs from the pipeline relevant for this study are described in Secs.~\ref{sec:SSP_fit} and~\ref{sec:SFR}.

\subsection{ALMA CO mapping}
\label{sec:alma}

Our sample of $18$ galaxies have \mbox{CO({\it J}=2--1)} [hereafter \mbox{CO(2--1)}] data from the PHANGS--ALMA survey \citep[PI: E.~Schinnerer;][]{Leroy2021b}. We used the ALMA $12$m and $7$m arrays combined with the total power antennas to map \mbox{CO(2--1)} emission at a spatial resolution of $1\farcs0 {-} 1\farcs5$ (version 4.0 of internal distribution, which is the first public data release). The molecular gas surface density maps have a typical uncertainty of ${\sim}1.2$~$M_{\odot}$~pc$^{-2}$ at a spatial resolution of $150$~pc. 
We use integrated intensity maps and associated statistical uncertainty maps constructed using the PHANGS--ALMA ``broad masking'' scheme. These broad masks are designed to incorporate all emission detected at any scale in the PHANGS--ALMA data. As a result they have high completeness, that is to say, they include most of the flux in the galaxy, at the expense of also having more low-confidence pixels than a more stringent masking technique \cite[see][for details and completeness estimates at 150~pc]{Leroy2021b, Leroy2021a}. A signal-to-noise (S/N) cut of~1 is then applied to drop the most uncertain emission. The strategy for observation, data reduction and product generation are described in \citet{Leroy2021b, Leroy2021a}.

We adopt the local gas-phase metallicity (in solar units) prescription for our fiducial $\alpha_\mathrm{CO}$ conversion factor following \citet{Sun2020} and partially based on \citet{Accurso2017}, following $\alpha_\mathrm{CO} = 4.35 (Z / Z_{\odot})^{-1.6}$ $M_{\odot}$~pc$^{-2}$ (K~km~s$^{-1}$)$^{-1}$, adopting a ratio \mbox{CO(2--1)}-to-\mbox{CO(1--0)} = 0.65 \citep[][T.~Saito et al.\ in prep.]{Leroy2013, denBrok2021, Leroy2021a}. Metallicity gradients are measured from the gas-phase abundances in \hii\ regions, as explained in \citet{Kreckel2020} and \citet{Santoro2021}.
Azimuthal variations in the gas-phase metallicity have been previously reported \citep{Ho2017, Kreckel2020, Williams2022}, however, these variations are small ($0.04{-}0.05$~dex), implying variations of ${\sim}0.06$~dex in $\alpha_{\mathrm{CO}}$ and therefore do not impact our results. We test the robustness of our results against a constant $\alpha_\mathrm{CO(1-0)} = 4.35~M_{\odot}$~pc$^{-2}$ (K~km~s$^{-1}$)$^{-1}$, as the canonical value for our Galaxy \citep{Bolatto2013} in Sec.~\ref{sec:ConvFact}.

\begin{table*}
\centering
\begin{center}
\renewcommand{\arraystretch}{1.2}
\begin{tabular}{lcccccrccc}
\hline
\hline
Target & R.A. & Dec. & $\log_{10} M_{\star}$ & $\log_{10} M_{\mathrm{H}_{2}}$ & $\log_{10} {\rm SFR}$ & $\Delta$MS & Distance & Inclination & Mapped area \\
 & (degrees) & (degrees) & $(M_{\odot})$ & $(M_{\odot})$ & $(M_{\odot}~\mathrm{yr}^{-1})$ & (dex) & (Mpc) & (degrees) & (kpc$^{2})$ \\
 (1) & (2) & (3) & (4) & (5) & (6) & (7) & (8) & (9) & (10)\\
\hline 
NGC~1087 & $41.60492$ & $-0.498717$ & 9.9 & $9.2$ & $0.12$ & $0.33$ & 15.85$\pm$2.08 & 42.9 & 128 \\
NGC~1300$^{\bullet}$ & $49.920815$ & $-19.411114$ & 10.6 & $9.4$ & $0.07$ & $-0.18$ & 18.99$\pm$2.67 & 31.8 & 366 \\
NGC~1365 & $53.40152$ & $-36.140404$ & 11.0 & $10.3$ & $1.23$ & $0.72$ & 19.57$\pm$0.77 & 55.4 & 421 \\
NGC~1385$^{\bullet}$ & $54.369015$ & $-24.501162$ & 10.0 & $9.2$ & $0.32$ & $0.5$ & 17.22$\pm$2.42 & 44.0 & 100 \\
NGC~1433$^{\bullet}$ & $55.506195$ & $-47.221943$ & 10.9 & $9.3$ & $0.05$ & $-0.36$ & 18.63$\pm$1.76 & 28.6 & 441 \\
NGC~1512 & $60.975574$ & $-43.348724$ & 10.7 & $9.1$ & $0.11$ & $-0.21$ & 18.83$\pm$1.78 & 42.5 & 270 \\
NGC~1566$^{\bullet}$ & $65.00159$ & $-54.93801$ & 10.8 & $9.7$ & $0.66$ & $0.29$ & 17.69$\pm$1.91 & 29.5 & 212 \\
NGC~1672 & $71.42704$ & $-59.247257$ & 10.7 & $9.9$ & $0.88$ & $0.56$ & 19.4$\pm$2.72 & 42.6 & 255 \\
NGC~2835 & $139.47044$ & $-22.35468$ & 10.0 & $8.8$ & $0.09$ & $0.26$ & 12.22$\pm$0.9 & 41.3 & 88 \\
NGC~3351 & $160.99065$ & $11.70367$ & 10.4 & $9.1$ & $0.12$ & $0.05$ & 9.96$\pm$0.32 & 45.1 & 76 \\
NGC~3627 & $170.06252$ & $12.9915$ & 10.8 & $9.8$ & $0.58$ & $0.19$ & 11.32$\pm$0.47 & 57.3 & 87 \\
NGC~4254$^{\bullet}$ & $184.7068$ & $14.416412$ & 10.4 & $9.9$ & $0.49$ & $0.37$ & 13.1$\pm$1.87 & 34.4 & 174 \\
NGC~4303$^{\bullet}$ & $185.47888$ & $4.473744$ & 10.5 & $9.9$ & $0.73$ & $0.54$ & 16.99$\pm$2.78 & 23.5 & 220 \\
NGC~4321$^{\bullet}$ & $185.72887$ & $15.822304$ & 10.7 & $9.9$ & $0.55$ & $0.21$ & 15.21$\pm$0.49 & 38.5 & 196 \\
NGC~4535$^{\bullet}$ & $188.5846$ & $8.197973$ & 10.5 & $9.6$ & $0.33$ & $0.14$ & 15.77$\pm$0.36 & 44.7 & 126 \\
NGC~5068 & $199.72807$ & $-21.038744$ & 9.4 & $8.4$ & $-0.56$ & $0.02$ & 5.2$\pm$0.22 & 35.7 & 23 \\
NGC~7496$^{\bullet}$ & $347.44702$ & $-43.42785$ & 10.0 & $9.3$ & $0.35$ & $0.53$ & 18.72$\pm$2.63 & 35.9 & 89 \\
IC5332 & $353.61453$ & $-36.10108$ & 9.7 & $-$ & $-0.39$ & $0.01$ & 9.01$\pm$0.39 & 26.9 & 34 \\
\hline
\end{tabular}
\end{center}
\caption{Summary of the galactic parameters of our sample adopted through this work. $^{\bullet}$:~Galaxies observed with MUSE-AO mode. Values in columns (4), (5) and~(6) correspond to those presented in \citet{Leroy2021b}. Column~(7) shows the vertical offset of the galaxy from the integrated main sequence of galaxies, as defined in \citet{Leroy2019}. Distance measurements are presented in \citet{Anand2021} and inclinations in \citet{Lang2020}. Uncertainties in columns (4), (5), (6) and~(7) are on the order of $0.1$~dex. Column~(10) shows the area mapped by MUSE.}
\label{tab:sample}
\end{table*}

\subsection{Environmental masks}
\label{sec:enviromental_mask}

We used the environmental masks described in \citet{Querejeta2021} to morphologically classify the different environments of each galaxy and label them as disks, spiral arms, rings, bars and centers. This classification was done using photometric data mostly from the Spitzer Survey of Stellar Structure in Galaxies \citep[S$^4$G;][]{Sheth2010}.
In brief, disks and centers are identified via 2D photometric decompositions of $3.6~\mu$m images \citep[see, e.g.,][]{Salo2015}. A central excess of light is labeled as center, independently of its surface brightness profile. The size and orientation of bars and rings are defined visually on the NIR images; for S$^4$G galaxies, the classification follows \citet{Herrera-Endoqui2015}. Finally, spiral arms are only defined when they are clearly dominant features across the galaxy disk (excluding  flocculent spirals). A log-spiral function is fitted to bright regions along arms on the NIR images, and assigned a width determined empirically based on CO emission.

\subsection{Stellar mass surface density maps} 
\label{sec:SSP_fit}

The PHANGS--MUSE DAP \citep{Emsellem2021} includes a stellar population fitting module, where a linear combination of single stellar population templates of specific ages and metallicities are used to reproduce the observed spectrum. We assume a \citet{Calzetti2000} extinction law to account for extragalactic attenuation in the fitting. This permits us to infer stellar population properties from an integrated spectrum, such as mass- or light-weighted ages, metallicities and total stellar masses, together with the underlying star formation history. Before doing the SSP fitting, we correct the full mosaic for Milky Way extinction assuming a \citet[][]{Cardelli89} extinction law and the $E(B{-}V)$ values obtained from the NASA/\linebreak[0]{}IPAC Infrared Science Archive\footnote{\url{https://irsa.ipac.caltech.edu/applications/DUST/}} \citep{Schlafly2011}.

In detail, our spectral fitting pipeline performs the following steps:
First, we use a Voronoi tessellation \citep{Capellari2003} to bin our MUSE data to a minimum S/N of ${\sim}35$, computed at the wavelength range of $5300{-}5500$~\AA. 
We use then the Penalized Pixel-Fitting (pPXF) code \citep{Capellari2004, Capellari2017} to fit the spectrum of each Voronoi bin. To fit our data, we used a grid of templates consisting of $13$ ages, logarithmically spaced and six metallicity bins. We fit the wavelength range $4850{-}7000$~\AA, in order to avoid spectral regions strongly affected by sky residuals. We used templates from the eMILES \citep{Vazdekis2010, Vazdekis2012} database, assuming a \citet{Chabrier2003} IMF and BaSTI isochrone \citep{Pietrinferni2004} with a Galactic abundance pattern. 

The stellar mass map is then reconstructed using the weights assigned by pPXF to each SSP spectrum, given that the current stellar mass of each template is known.
Finally, we have identified foreground stars as velocity outliers in the SSP fitting and we have masked those pixels for the analysis carried out in this paper.

\subsection{Star formation rate measurements}
\label{sec:SFR}

As part of the PHANGS--MUSE DAP \citep{Emsellem2021}, we fit single Gaussian profiles to a number of emission lines for each pixel of the final combined MUSE mosaic of each galaxy in our sample. By integrating the flux of the fitted profile in each pixel, we construct emission line flux maps for every galaxy. We calculate SFR from extinction-corrected H$\alpha$. In detail, we dereddened the H$\alpha$ fluxes, assuming that $\mathrm{H}\alpha_\mathrm{corr} / \mathrm{H}\beta_\mathrm{corr} = 2.86$, as appropriate for a case~B recombination \citep{Osterbrock1989}, temperature $T = 10^{4}$~K and density $n_\mathrm{e} = 100$~cm$^{-3}$, following:
\begin{equation}
    \mathrm{H}\alpha_\mathrm{corr} = \mathrm{H}\alpha_\mathrm{obs}  
    \bigg(\frac{(\mathrm{H}\alpha / \mathrm{H}\beta)_\mathrm{obs}}{2.86}\bigg)^{\frac{k_{\alpha}}{k_{\beta} - k_{\alpha}}}~,
\end{equation}
where H$\alpha_\mathrm{corr}$ and H$\alpha_\mathrm{obs}$ correspond to the extinction-corrected and observed H$\alpha$ fluxes, respectively, and $k_{\alpha}$ and $k_{\beta}$ are the values of a given extinction curve at the wavelengths of H$\alpha$ and H$\beta$.  Opting for an \citet{ODonnell1994} extinction law, we use $k_{\alpha} = 2.52$, $k_{\beta} = 3.66$ and $R_{V} = 3.1$.

Next, we remove pixels that are dominated by active galactic nuclei (AGN) or low-ionization nuclear emission-line regions (LINER) ionization from our sample, performing a cut in the Baldwin--Phillips--Terlevich \citep[BPT;][]{BPT} diagram using the \oiii/H$\beta$ and \sii/H$\alpha$ line ratios, as described in \citet{Kewley2006}. 

For the remaining pixels, we determined the fraction of the H$\alpha$ emission tracing local star formation ($C_\hii$) and the fraction deemed to correspond to the diffuse ionized gas (DIG). DIG is a warm ($10^{4}$~K), low density ($10^{-1}$~cm$^{-3}$) phase of the interstellar medium \citep{Haffner2009, Belfiore2015} produced primarily by photoionization of gas across the galactic disk by photons that escaped from \hii\ regions \citep{Flores-Fajardo2011, Zhang2017, Belfiore2021}. 

To this end, we use the \sii/H$\alpha$ ratio to estimate $C_\hii$ in each pixel, following \citet{Blanc2009} and \cite{Kaplan2016}. After performing this correction and removing the DIG contribution of the H$\alpha$ flux, we rescale the DIG-corrected H$\alpha$ map by~1 plus the fraction of flux removed, to keep the total H$\alpha$ flux constant. We perform this rescaling because photons that ionize the DIG are believe to originally have leaked out from \hii\ regions\footnote{While other sources of DIG ionzing photons exist (e.g., evolved post-AGB stars), and can be relatively important in more passive systems, all galaxies in PHANGS--MUSE sample are actively star froming galaxies, which justifes the assumption that DIG ionization is dominated by UV leakage from \hii\, regions.}. This correction represents then a spatial redistribution of the H$\alpha$ flux. We refer the reader to \citet{Pessa2021} for a detailed description of this procedure. This approach permits us to estimate a star formation rate in pixels contaminated by non-star-forming emission. 
A S/N cut of~3 for H$\alpha$ and~1 for H$\beta$ was then applied before computing the star formation rate surface density map using Eq.~\eqref{eq:sfr}, effectively removing ${\sim}13\%$ of pixels in our sample. While $13\%$ is not insignificant, we note that the majority have $C_\hii \approx 0$, and therefore our S/N cut does not largely impact our results. Pixels below this S/N cut, pixels with $C_\hii \leq 0$ or pixels where $\mathrm{H}\alpha_\mathrm{obs} / \mathrm{H}\beta_\mathrm{obs} < 2.86$, are considered nondetections (see Sec.~\ref{sec:fit}). 
To calculate the corresponding star formation rate from the H$\alpha$ flux, after correcting it for internal extinction and DIG contamination, we adopted the prescription of
\citet{Calzetti-book}:
\begin{equation}
    \label{eq:sfr}
    \frac{\mathrm{SFR}} {M_{\odot}~\mathrm{yr}^{-1}} = 5.5\times10^{-42}\frac{\mathrm{H}\alpha_\mathrm{corr}}{\mathrm{erg~s}^{-1}}~.
\end{equation}
This equation is scaled to a Kroupa universal IMF \citep{Kroupa2001}. Differences with the Chabrier IMF assumed for the SSP fitting are expected to be small \citep[${\sim}1.05$;][]{Kennicutt2012}. With these steps we obtain SFR surface density maps for each galaxy in our sample. We acknowledge that Eq.~\eqref{eq:sfr} assumes a fully sampled IMF, and that the lowest SFR pixels (especially at the ${\sim}150$~pc resolution of our data) may not form enough stars to fully sample the IMF. Hence, the measured SFR is more uncertain in this regime. We decided to exclude from our analysis the regions with low coverage of SFR or molecular gas surface density (see Sec.~\ref{sec:fit}), hence minimizing the impact of the low SFR regime on our results.

\subsection{Probing larger spatial scales}
\label{sec:degrade}

In Sec.~\ref{sec:spatial_res_effect}, we investigate the effect of spatial resolution on our measurements. To do this,  we first convolve our MUSE maps to a common fixed reolution of $150$ pc, and then we rebin our $150$~pc resolution MUSE and ALMA maps to have pixel sizes of $150$~pc, $200$~pc, $300$~pc, $500$~pc and $1$~kpc. Then we replicate our measurements using these rebinned maps. We conduct an identical rebinning process for all three relevant quantities: stellar mass surface density, star formation rate surface density and molecular gas mass surface density \citep[see][for a detailed description]{Pessa2021}. After beginning with matched resolution 150~pc MUSE and ALMA data, we favor this rebinning approach rather than, for example, convolution to a coarser Gaussian beam, because it minimized pixel-to-pixel covariance and yields approximately statistically independent measurements. Our core results use the $150$~pc pixels, which are larger than the native spatial resolution of the maps for all galaxies, except for NGC~1672, which has an ALMA native resolution of ${\sim}180$~pc. 

The rebinning step is followed by an inclination correction, simply using a $\cos(i)$ multiplicative term, where $i$ is the inclination of each galaxy as listed in Table~\ref{tab:sample} \citep[adopted from][]{Lang2020}. All following results and conclusions in the next Sections pertain to a fixed spatial resolution of $150$~pc, unless specifically stated (see Sec.~\ref{sec:spatial_res_effect}).

\section{Methods}
\label{sec:methods}
In this section, we present our methodology to  fit a 2D plane (i.e.\ power law) predicting $\log_{10} \Sigma_{\rm SFR}$ as function of $\log_{10} \Sigma_{\star}$ and $\log_{10} \Sigma_{\rm mol}$. We fit planes separately to the data for each distinct galactic environment, as well as to the full sample. Because the fitting can be potentially biased by the influence of nondetections in either $\Sigma_{\rm SFR}$ or $\Sigma_{\rm mol}$, we explain our approach to nondetections first and then describe the full fitting method.

\subsection{Nondetections in our data}
\label{sec:NDs}

In \citet{Pessa2021}, we found that the adopted treatment of nondetections can have a considerable impact on the derived slope of fitted scaling relations. Here, nondetections (N/Ds) are pixels with a value of $\Sigma_{\rm SFR}$ or $\Sigma_{\rm mol}$ lower than our defined detection threshold. As stellar mass is detected essentially everywhere in our maps at high significance, nondetections in $\Sigma_\star$ are essentially a nonissue.

This impact of N/Ds becomes particularly strong when performing measurements at high spatial resolution, where a larger fraction of pixels are deemed to be N/Ds. For more discussion of how such sparse maps emerge from timescale effects or other stochasticity, see \citet{Pessa2021} and references therein. 

In \citet{Pessa2021}, we overcame this issue by binning the data before fitting a power law. Here we are trying to fit a 2D plane, and discretizing the data in higher dimensions becomes problematic, so we do not proceed in the same way. Instead, we opt here to focus our analysis only on those $\Sigma_{\star}$ ranges that have high detection fractions in both $\Sigma_{\rm SFR}$ and $\Sigma_{\rm mol}$. Here the detection fraction is defined as the fraction of pixels with a measurement of $\Sigma_{\rm SFR}$ (or $\Sigma_{\rm mol}$) above our detection threshold in a given $\Sigma_{\star}$ bin. As a reminder, we required $S/N > 1$ for the ALMA CO intensity, $S/N > 3$ for the H$\alpha$ and $S/N > 1$ for the H$\beta$ at the native resolution, and N/Ds are a nonissue for stellar mass.

\begin{figure*}[h!]
    \includegraphics[width = \textwidth]{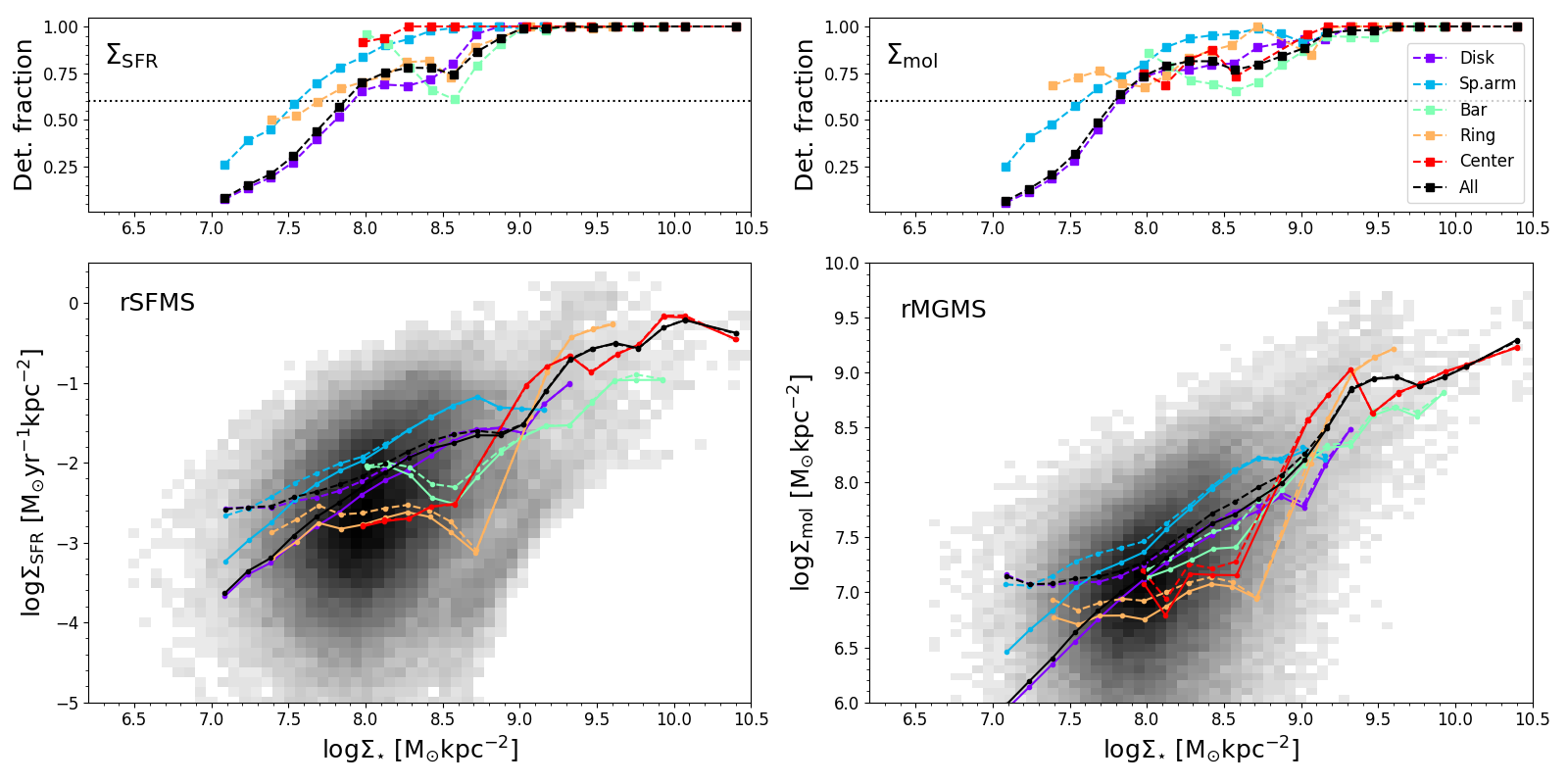}
    
    \caption{rSFMS (bottom left) and rMGMS (bottom right) measured at $150$~pc resolution for all galaxies in our sample.  For reference and to guide the eye, the various solid colored lines represent the binned trends (bins of $0.15$~dex) obtained for each environment, accounting for nondetections as detailed in \citet{Pessa2021}. The black solid line shows the same measurement for all environments simultaneously. The dashed lines show the binned trends obtained when nondetections are neglected from the analysis. 
    The panels on the top row show the detection fraction of $\Sigma_{\rm SFR}$ (left) and $\Sigma_{\rm mol}$ (right), defined as the fraction of pixels with a measurement above our detection threshold in each bin of $\Sigma_{\star}$. The horizontal dotted line marks the detection fraction level of  $60\%$.}
    \label{fig:Det_frac}
\end{figure*}

The top panels of Fig~\ref{fig:Det_frac} show the detection fraction of $\Sigma_{\rm SFR}$ and $\Sigma_{\rm mol}$ in each bin of $\Sigma_{\star}$, and the bottom panels show rSFMS and rMGMS binned for each individual environment (colors) and for the full sample (black).  Two types of lines are shown, solid lines represent the binned trend accounting for N/Ds, and dashed lines represent the binned trend neglecting N/Ds. It is easy to see in the bottom panels that for all environments the two lines deviate, that is, neglecting N/Ds from the analysis systematically leads to flatter slopes. This bias in the slope is due to varying fractions of N/Ds across the $\Sigma_{\star}$ range, usually being higher at the low $\Sigma_{\star}$ end. To minimize the impact of N/Ds on our measurements, we confine our analysis only to those $\Sigma_{\star}$ ranges, where the detection fraction (of both $\Sigma_{\rm SFR}$ and $\Sigma_{\rm mol}$) is higher than 60\%. In Appendix~\ref{sec:threshold_test}, we discuss how using a different detection fraction threshold impacts our results, and we show that a threshold of $60\%$ provides a good compromise between minimizing the impact of N/Ds in our analysis, while maintaining the statistical significance of our data.

We impose the detection fraction threshold independently for each environment, that is, the range of $\Sigma_{\star}$ values used for `spiral arms' is different to that used for `disks'. This approach allows us to maximize the number of data points used in the analysis.

We do caution that our approach to the different galactic environments might impose some bias on the results. The spiral arms defined by \citet{Querejeta2021} tend to have more gas and more star formation at fixed $\Sigma_\star$. Because they were identified partially based on multiwavelength data that trace gas and SFR, this is somewhat by construction. It is not \textit{a priori} certain that spiral arms must behave distinctly from other environments in the $\Sigma_\star{-}\Sigma_{\rm SFR}{-}\Sigma_{\rm mol}$ space, but some bias might be expected from how we set up the analysis. This also manifests in the detection threshold cuts: because spiral arms have higher $\Sigma_{\rm mol}$ and $\Sigma_{\rm SFR}$ at fixed $\Sigma_\star$, detections extend to lower $\Sigma_\star$ for these environments and we might also expect a different best-fitting plane. This is already somewhat evident in Fig.~\ref{fig:Det_frac}, where the ``Sp. arm'' environment shows higher detection fractions at fixed $\Sigma_\star$ than other cases.

It is worth noting that we are not dropping a large fraction of our data by imposing these thresholds. Indeed, only ${\sim}20$\% of our $150$~pc pixels are removed from the entire sample across all environments. Given that different galactic environments sample different parts of the galactic disk and therefore different (typical) surface densities, the adopted N/D thresholds are most relevant for the disk environment, which extends to the largest galactic radii and therefore (also) samples low surface densities. As a result, ${\sim}38$\% of its original pixels do not satisfy the threshold. For all other environments, the fraction of pixels dropped is only ${<}5$\%. After performing this cut, the disk still remains the galactic environment with the largest number of pixels. Thus, we are only mildly increasing the statistical uncertainty of our results with this approach, and it permits us to reduce any bias in the reported quantities by considering only those $\Sigma_{\star}$ regimes not dominated by nondetections. 

\subsection{Fitting technique}
\label{sec:fit}

In this work, we aim at finding the best set of parameters $C_{\star}$, $C_{\rm mol}$ and $C_{\rm norm}$ that describe the star-forming plane conformed by the quantities $\log \Sigma_{\rm SFR}$, $\log \Sigma_{\star}$ and $\log \Sigma_{\rm mol}$, such that:
\begin{equation}
    \log \Sigma_{\mathrm{SFR},i} = C_{\star,i} \log \Sigma_{\star} + C_{\mathrm{mol},i} \log \Sigma_{\rm mol} + C_{\mathrm{norm},i}~,
    \label{eq:plane_eq}
\end{equation}
where the subindex~$i$ stands for each galactic environment. If $\Sigma_{\star}$ and $\Sigma_{\rm mol}$ are the only quantities that determine $\Sigma_{\rm SFR}$ in a given region of a galaxy, then this plane should not change between different galactic environments. However, given that galactic environments represent a diversity of other physical conditions including gas phase metallicity, stellar age, stellar geometry (flattened vs. triaxial), shear rate, radial flows and the gas  structure and organization, then we might expect there to be additional factors that influence gas stability and star formation at fixed $\Sigma_{\rm mol}$ and $\Sigma_{\star}$ \citep[i.e.,][]{Hunter1998, Martig2009, Krumholz2018, Meidt2018, Meidt2020, Gensior2020, Gensior2021}. Thus, the question we aim to address in this paper is whether there is a single relationship between $\Sigma_{\rm SFR}$, $\Sigma_{\star}$ and $\Sigma_{\rm mol}$ which is valid in all environments, or if this relation differs between (some) environments. In the latter case, can we also aim at identifying the parameter(s) setting the level of $\Sigma_{\rm SFR}$, besides $\Sigma_{\star}$ and $\Sigma_{\rm mol}$.


To minimize the covariance between model parameters, we normalize the distributions of the three involved variables, $\log \Sigma_{\rm SFR}$, $\log \Sigma_{\star}$ and $\log \Sigma_{\rm mol}$, by their (full sample) mean before fitting the data (i.e.\ centering). After the data are normalized, we find the best set of parameters $C_{\star}$, $C_{\rm mol}$ and $C_{\rm norm}$ such that:
\begin{multline}
    \log \Sigma_{\rm SFR} - \langle \log \Sigma_{\rm SFR} \rangle = 
    C_{\star} (\log \Sigma_{\star} - \langle \log \Sigma_{\star} \rangle) \\
    + C_{\rm mol} (\log \Sigma_{\rm mol} - \langle \log \Sigma_{\rm mol} \rangle) + C_{\rm norm}~,
\label{eq:plane_eq_norm}
\end{multline}
where the quantities inside  the `$\langle \, \rangle$' brackets represent the mean of each distribution, across the full sample. Collecting these terms, we can define a new recentered normalization $\hat{C}_{\rm norm}$ so that:
\begin{equation}
    \hat{C}_{\rm norm} = C_{\rm norm} +  \langle \log \Sigma_{\rm SFR} \rangle - C_{\star}  \langle \log \Sigma_{\star} \rangle - C_{\rm mol} \langle \log \Sigma_{\rm mol} \rangle~.
\label{eq:C2real}
\end{equation}
For consistency we apply a fixed normalization to the data ($\langle \log \Sigma_{\star} \rangle = 8.11$, $\langle \log \Sigma_{\rm mol} \rangle = 7.25$ and $\langle \log \Sigma_{\rm SFR} \rangle = -2.48$) throughout this work, that is, we use the same average values at different spatial resolutions (Sec.~\ref{sec:spatial_res_effect}), and when exploring environments in individual galaxies (Sec.~\ref{sec:third_param}).

We choose to fit a single $C_{\rm norm}$ value for all different galactic environments to avoid degeneracies between the normalization ($C_{\rm norm}$) and the slopes  of the planes ($C_{\star}$ and $C_{\rm mol}$). This choice focuses our analysis specifically on variations of the dependency on $\Sigma_{\star}$ and $\Sigma_{\rm mol}$. However, we stress that the main conclusions of this paper are not affected by this approach. 

We designed our methodology (i.e.\ unique normalization and detection threshold) to test for environ\-ment-driven differences in the plane relating $\log \Sigma_\star$, $\log \Sigma_{\rm SFR}$, and $\log \Sigma_{\rm mol}$. The existence of such differences is of considerable interest, because it can reveal the degree to which physics covariant with the environment definitions impact the star formation process. However, we caution that our methodology is not necessarily the optimal approach to determine the `real’ slopes of the star-forming plane. Environment may also affect the normalization ($C_{\rm norm}$). Because we keep $C_{\rm norm}$ fixed, our fitting method may allow some of the signal associated with normalization variation to the derived slopes ($C_{\star}$ and/or $C_{\rm mol}$).

In order to find the best-fitting parameters that describe the star-forming plane in each environment, as defined in Eq.~\eqref{eq:plane_eq}, we perform a Bayesian hierarchical linear regression, based on routines from the {\sc PyMC3} python package \citep{pymc3}. The hierarchical linear regression represents the middle ground between assuming that different environments are completely independent populations, and assuming that they all are identical and described by the same model.
Instead, it assumes that the parameters of the models that describe the data sets of different galactic environments have some underlying similarity. That is, the coefficients that describe the star-forming plane in different galactic environments are assumed to follow the same underlying `hyperprior' distribution, defined by a common set of `hyperparameters'.
We choose weakly informative hyperpriors in order to ensure that our results are not dominated by the priors. 
The hyperprior distributions considered are the following:
\begin{gather}
C_{\star,\mu} \sim \mathcal{N}(1, 10^{2})~, \\
C_{\star,\sigma} \sim \mathcal{H}(5^{2})~, \\
C_{\mathrm{mol},\mu} \sim \mathcal{N}(1, 10^{2})~, \\
C_{\mathrm{mol},\sigma} \sim \mathcal{H}(5^{2})~,
\end{gather}
where $\mathcal{N}(\mu,\sigma^{2})$ and $\mathcal{H}(\sigma^{2})$ stand for Normal and Half-Normal distributions, respectively, with mean $\mu$ and variance $\sigma^{2}$.
The prior distribution of each coefficient is then defined as
\begin{equation}
    C_{j} \sim \mathcal{N}(C_{j,\mu}, C_{j,\sigma})~,
\end{equation}
for $j$ in $\{\star,\mathrm{mol}\}$. In addition to these hyperpriors, we adopt the following priors for the normalization coefficient. 
\begin{gather}
C_{\mathrm{norm},\mu} \sim \mathcal{N}(0, 10^{2})~, \\
C_{\mathrm{norm},\sigma} \sim \mathcal{H}(5^{2})~.
\end{gather}
Finally, we include an additional term to account for intrinsic dispersion, which is common for all environments, and its prior is defined as
\begin{equation}
    \epsilon \sim \mathcal{N}(0, \sigma_{\mathrm{intr}})~,
\end{equation}
where $\sigma_{\mathrm{intr}}$ corresponds to the intrinsic dispersion of the data with respect to the model, and for its prior distribution we use:
\begin{equation}
    \sigma_{\mathrm{intr}} \sim \mathcal{HC}(5^{2})~,
\end{equation}
where $\mathcal{HC}(\sigma^{2})$ stands for Half-Cauchy distribution, a common choice for a prior distribution of a scaling parameter like the intrinsic scatter. We find that $\sigma_{\mathrm{\rm intr}}$ describes well the standard deviation in the fit residuals of our data. 

We use four Markov chain Monte Carlo (MCMC) chains to sample the posterior distributions, each one of them having $2000$ iterations plus $2000$ additional burn-in iterations. The posterior is sampled using the {\sc NUTS} algorithm \citep{homan2014}. The convergence of the MCMC chains is ensured by using the $\hat{R} \approx 1$ criterion \citep{Gelman1992}. Essentially, $\hat{R}$ corresponds to the ratio of the between-chain variance and the within-chain variance. In Appendix~\ref{sec:ToyModel}, we use a toy model to test the accuracy of the hierarchical modeling, and find that this approach correctly recovers the test values in mock data sets.

\begin{table}
\centering
\begin{center}
\renewcommand{\arraystretch}{1.2}
\begin{tabular}{lcccc}
\hline
\hline
 & $C_{\star}$ & $C_{\rm mol}$ & $N_{\rm pix}$\\
\hline 
Disk & $0.17\pm 0.02$ & $0.87\pm 0.01$ &  18977\\
Sp. arm & $0.45\pm 0.02$ & $0.89\pm 0.01$&  15743\\
Bar & $-0.41\pm 0.01$ & $0.83\pm 0.01$ &  7286\\
Ring & $-0.08\pm 0.02$ & $1.01\pm 0.01$&  7399\\
Center & $0.20\pm 0.05$ & $0.88\pm 0.04$ &  587\\
All & $-0.04\pm 0.01$ & $0.92\pm 0.01$ &  49992\\
\hline
\end{tabular}
\end{center}
\caption{Mean and standard deviation of the posteriors distributions in Fig.~\ref{fig:posteiors_fidu}, for each parameter and each galactic environment. Additionally to the above parameters, we obtain $C_{\rm norm}= -0.01$ and an overall intrinsic dispersion $\sigma_{\rm intr} \approx 0.63$ dex. The number of pixels included in each environment is indicated in the last column.}
\label{tab:coefs_fidu}
\end{table}

\section{Results}
\label{sec:results}

\subsection{Star-forming plane across galactic environments at 150~pc resolution}

\begin{figure*}[h!]
    \includegraphics[width = \textwidth]{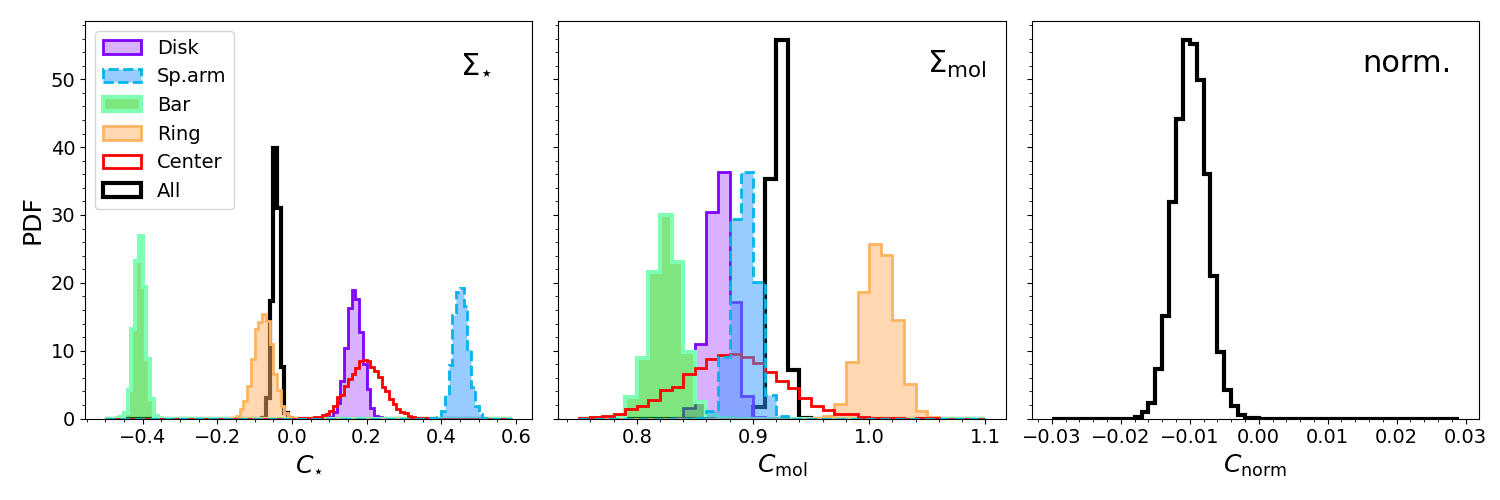}
    
    \caption{Posterior distributions for the coefficients $C_{\star}$, $C_{\rm mol}$ and $C_{\rm norm}$ that define the star-forming plane in each galactic environment, measured at a spatial resolution of $150$ pc. As the posterior distribution of $C_{\rm norm}$ is considerably narrower than that of $C_{\star}$ and $C_{\rm mol}$, the $x$-axis has been binned in smaller bins and the distribution renormalized for an easier visualization.}
    \label{fig:posteiors_fidu}
\end{figure*}

Figure~\ref{fig:posteiors_fidu} shows the posterior distributions of each one of the parameters that define the star-forming plane in each galactic environment, following the same color code as in Fig.~\ref{fig:Det_frac}. The mean and standard deviation of each one of these distributions are reported in Table~\ref{tab:coefs_fidu}, as well as the number of pixels used in the fit for each environment. 
We firstly note that the marginalized posteriors of $C_{\star}$ and $C_{\rm mol}$ of different environments are significantly different (${>}1\sigma$ in most of cases). The posterior distributions of $C_{\star}$ and $C_{\rm mol}$ are significantly broader for galaxy centers than for the rest of environments. This could either be statistical (i.e.,\ due to the smaller number of pixels probing this environments) or intrinsic (i.e.\ different galaxy centers have different scaling relations). The posterior of $C_{\rm norm}$ is only shown for the full sample, since we model a single normalization for all environments as explained in Sec.~\ref{sec:fit}. As a single value of $C_{\rm norm}$ is fitted for the full sample, the posterior distribution of $C_{\rm norm}$ is considerably narrower than that of $C_{\star}$ and $C_{\rm mol}$. Thus, the $x$-axis in the right-most panel has smaller bins, and the distribution of $C_{\rm norm}$ has been renormalized for an easier visualization. After correcting for the centering of the data (Eq.~\ref{eq:C2real}), this value of $C_{\rm norm}$ implies an absolute normalization $\hat{C}_{\rm norm}$ of $-8.83$.  This value represents a characteristic depletion time of ${\sim}1.5$~Gyr, well within the range of values reported in \citet{Querejeta2021}.

We note that $C_{\star}$ parametrizes the rate of change of the rKS normalization, associated with $\Sigma_{\star}$, and the index of the power law spans a wide range of values, changing sign significantly across the sample, from about $-0.41$ to~$0.45$. Interestingly, bars have a very negative value of $C_{\star}$ ($\sim -0.4$), meaning that they show longer depletion times at higher $\Sigma_{\star}$ values. While this low value of $C_{\star}$ is certainly an indication of lower star formation efficiencies, compared to other environments,  we remind the reader that due to our adopted approach of using a single normalization across different galactic environments, a negative $C_{\star}$ could be interpreted also as the slope trying to capture a vertical offset between environments.

In Sec.~\ref{sec:third_param}, we discuss the possibility of this low value of $C_{\star}$ being linked to lower star formation efficiencies (SFE) driven by radial and/or turbulent motion of gas in bars. In contrast, spiral arms show the highest values of $C_{\star}$ (${\sim}0.45$), leading to shorter depletion times at higher $\Sigma_{\star}$ values. As $\Sigma_{\star}$ varies strongly as a function of radius, $C_{\star}$ expresses a radial trend in the normalization of the molecular gas scaling relation. Analogous variations were reported in \citet{Muraoka2019}, where the authors report radial variations of a factor $2{-}3$ in the SFE of individual  nearby galaxies from the CO Multi-line Imaging of Nearby
Galaxies (COMING) project \citep{Sorai2019}.
 
On the other hand, $C_{\rm mol}$ shows a more homogeneous behavior with values in the range of $0.83$ to $1.01$, which implies that higher $\Sigma_{\rm mol}$ leads to higher $\Sigma_{\rm SFR}$ everywhere in the galaxy. Subtle variations are found for bars having the lowest values ($C_{\rm mol}{}\sim0.83$) or rings having the largest value ($C_{\rm mol} \sim 1.01$). It is interesting to note that while rings show nearly a linear slope in molecular gas, other galactic environments exhibit a sublinear behavior. While a linear relation implies a constant depletion time (defined as $\tau = \Sigma_{\rm mol} / \Sigma_{\rm SFR}$), a sublinear slope leads to a depletion time that increases with gas surface density (neglecting changes in $C_{\star}$). \citet{Shetty2014} examined different possibilities to physically interpret this sublinearity. On one hand, differences in the molecular gas properties such as star formation efficiency or volume density \citep[see][]{Bacchini2019a, Bacchini2020} could lead to variations of the depletion time.  Alternatively, variations in the depletion time could be produced by a diffuse component of the molecular gas (i.e., molecular gas not actively forming stars). \citet{Schinnerer2019} and \citet{Pan2022} found that a significant fraction of molecular gas is indeed decoupled from the H$\alpha$ emission in PHANGS galaxies. Although for this analysis, we choose only pixels that show both molecular and ionized gas emission, quiescent gas could still be present in our data due to projection effects. In this scenario, the sublinearity is an indication that the fraction of diffuse molecular gas grows with $\Sigma_{\rm mol}$. We highlight that this sublinearity persists for bars, disks and spiral arms, when we let $C_{\rm norm}$ free for each individual environment.

A third possibility is that variations in the depletion times are driven by variations in the real underlying  $\alpha_{\rm CO}$ conversion factor or \mbox{CO(2--1)}-to-\mbox{CO(1--0)} line ratios. \citet{Leroy2021c} studied variations in the $R_{21}\equiv$ \mbox{CO(2--1)}-to-\mbox{CO(1--0)} line ratio within PHANGS galaxies, and reported a central enhancement of $R_{21}$ of a factor of ${\sim}0.18$~dex. Such anticorrelation of $R_{21}$ with galactocentric radius (and thus, correlation with $\Sigma_{\rm SFR}$) implies our fiducial choice of $R_{21} = 0.65$ could potentially lead to an overestimation of the total molecular gas content in the inner and denser regions of the galaxy, and to an underestimation of the gas content in the outer regions. This can effectively depress the measured $C_{\rm mol}$ slope by up to ${\sim}0.15$. This effect likely plays a more relevant role in the global measurement, considering all environments, due to the larger dynamic range of $\Sigma_{\star}$. However, it can not be ruled out that it is also, partially, driving the sublinearity found in the  single-environment measurements.

\begin{figure}[h!]
    \includegraphics[width = \columnwidth]{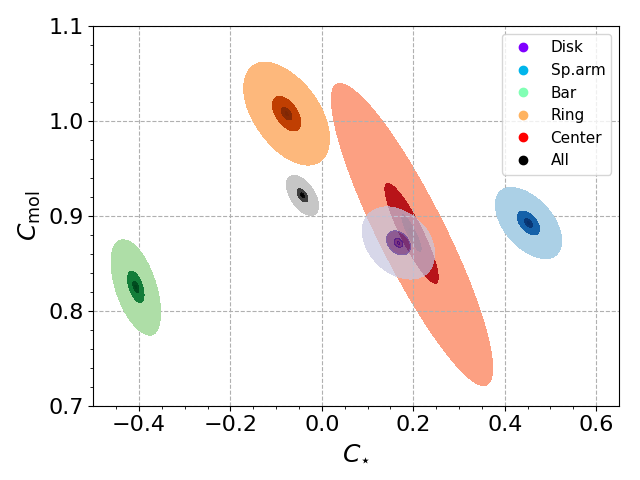}
    
    \caption{Posteriors distributions of the coefficients $C_{\star}$ and $C_{\rm mol}$ for each separate environment, , measured at an spatial resolution of $150$ pc, to show the covariance between them. The posteriors have been smoothed using a Gaussian kernel. The color scale for each environment indicates the 1-, 2- and 3-sigma confidence intervals.}
    \label{fig:cov}
\end{figure}

Figure~\ref{fig:cov} shows the covariance between the posterior distributions of the parameters $C_{\star}$ and $C_{\rm mol}$ measured for each galactic environment. As $C_{\star}$ and $C_{\rm mol}$ are the coefficients of the linear combination to predict $\log \Sigma_{\rm SFR}$ of a given pixel from its $\log \Sigma_{\star}$ and $\log \Sigma_{\rm mol}$ measurements, they show a negative covariance, that is, a higher $C_{\star}$ implies a lower value of $C_{\rm mol}$. The covariance is much larger for centers likely due to the lower number of pixels in these environments (see Table~\ref{tab:coefs_fidu}), but could also be connected to a physically more heterogeneous behavior of centers. On the other hand, the covariance in disks is significantly smaller. In Appendix~\ref{sec:ToyModel}, we show that this covariance arises, in part, as an artifact of the fitting procedure. However, the covariance could be also partially due to the fact that $\Sigma_{\star}$ and $\Sigma_{\rm mol}$ are physically highly correlated quantities, through the rMGMS, thus the covariance reflects the slope between them.

\begin{figure}[h!]
    \includegraphics[width = \columnwidth]{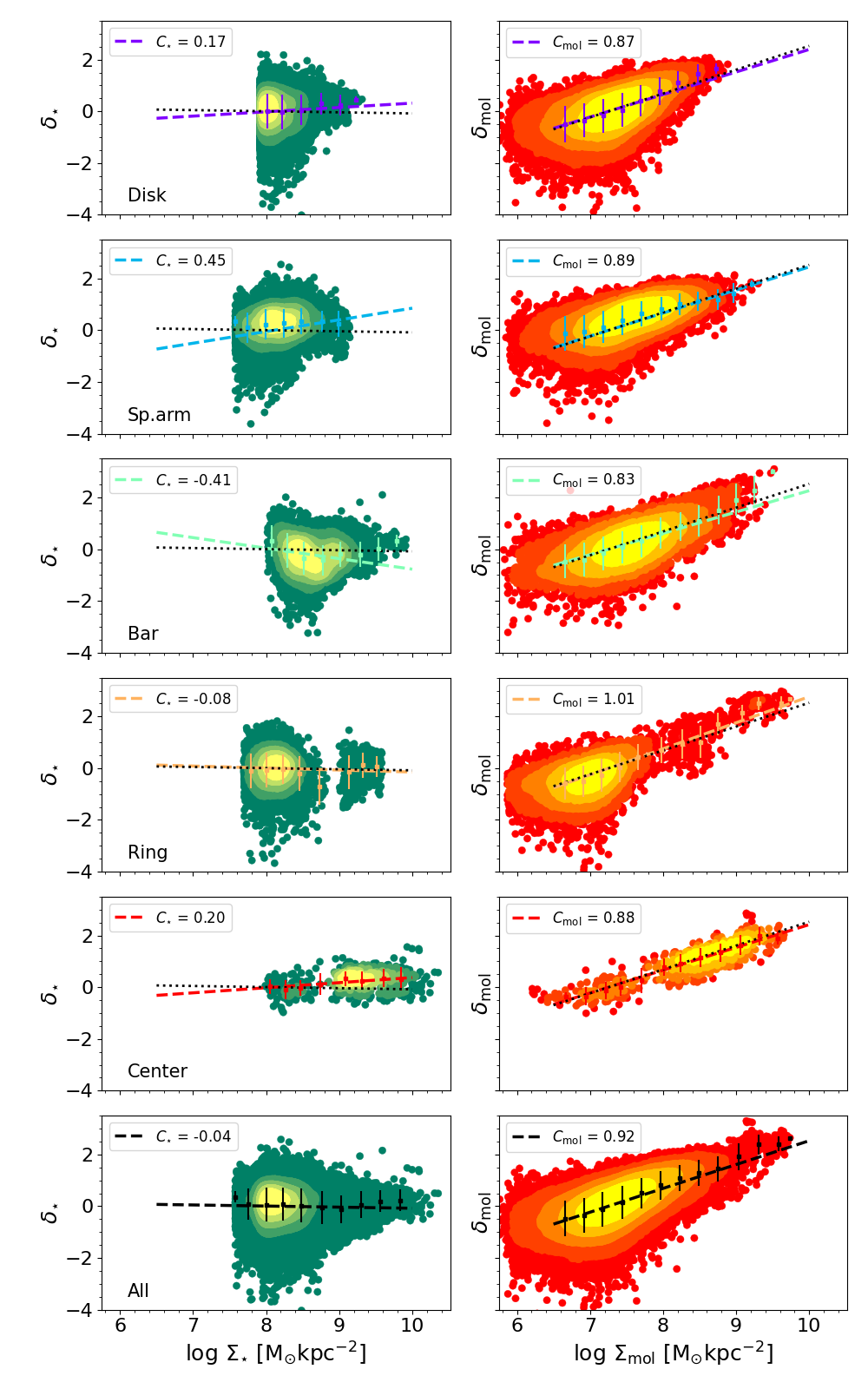}
    
    \caption{Partial residual  for each independent variable, dropping the dependency from the second independent variable, found for each individual environment and for all environments together as a function of $\Sigma_{\star}$ (left) and $\Sigma_{\rm mol}$ (right), measured at an spatial resolution of $150$ pc. The binned data are shown for each corresponding environment. The trend for the full sample is overplotted as a black dotted line.}
    \label{fig:projection}
\end{figure}

Figure~\ref{fig:projection} provides an alternative visualization of the 2D planes obtained with the fitting procedure. It shows the partial residual  for each independent variable after removing the dependency on the second independent variable. Specifically, we define $\delta_{\star}$ and $\delta_{\rm mol}$ as
\begin{align}
    \delta_{\star} &= \log \Sigma_{\rm SFR} - C_{\rm norm} + C_{\rm mol} \log \Sigma_{\rm mol}~, \\
    \delta_{\rm mol} &= \log \Sigma_{\rm SFR} - C_{\rm norm} + C_{\star} \log \Sigma_{\star}~,
\end{align}
to visualize the partial residuals as a function of $\Sigma_{\star}$ and $\Sigma_{\rm mol}$, respectively. The figure shows in each row the residuals for a given environment as a 2D histogram. The bottom row shows the residuals for the full sample. The slope in each panel corresponds to the best-fitting value of $C_{\star}$ (left column) and $C_{\rm mol}$ (right column) for each environment. 

The sharp cut at low $\Sigma_{\star}$ in the $x$-axis of the left panels is a consequence of the adopted detection threshold for each galactic environment as described in Sec.~\ref{sec:fit}. The figure shows the range of slopes derived for the different environments probed. Spiral arms are the environment that show the most positive trend for $\Sigma_{\star}$, while centers and disks show more subtle positive trends. In contrast, bars show the most negative $C_{\star}$ values of all environments. Similarly, bars also exhibit the lowest values of $C_{\rm mol}$, which can be understood as a lower efficiency at converting molecular gas into stars toward higher values of $\Sigma_{\rm mol}$. On the other hand, star-forming rings show the largest $C_{\rm mol}$ value ($\approx 1.01$), which indicates a nearly constant depletion time across the $\Sigma_{\rm mol}$ range. We inspect these trends per environment and find that they are not driven by individual galaxies and that, indeed, the trends of the multiple galaxies are consistent with each other. These trends reveal that the galaxy-to-galaxy variations in the scaling relations reported in \citet{Ellison2021} and \citet{Pessa2021} could plausibly be explained by a different relative contribution of the different galactic environments across the sample galaxies.

\subsection{Effect of spatial resolution}
\label{sec:spatial_res_effect}

\begin{figure}[h!]
    \includegraphics[width = \columnwidth]{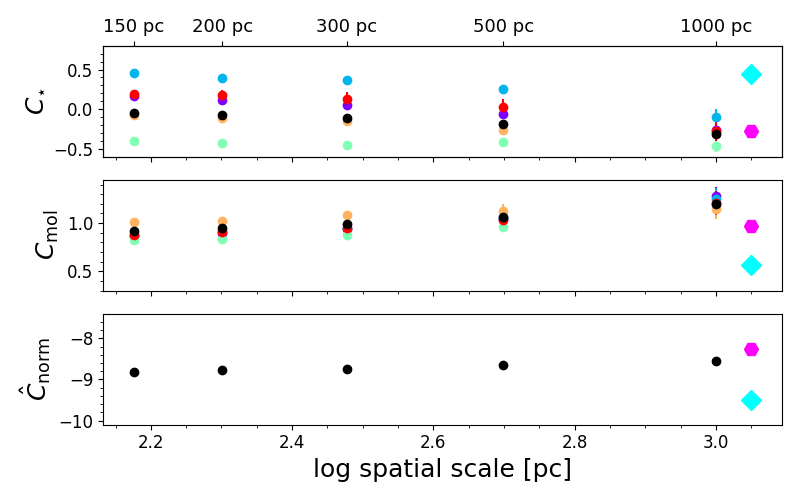}
    
    \caption{Change of the coefficients that define the star-forming planes in each galactic environment with spatial scale. The color code for each environment is the same as used in Fig.~\ref{fig:Det_frac}. The cyan diamond and a magenta hexagon indicate the results reported by \citet{Sanchez2021} (EDGE--CALIFA sample) and \citet{Lin2019}, respectively.}
    \label{fig:scale_evo}
\end{figure}

In this section, we explore how the spatial scale of the data impacts our measurements. We degraded our data, as explained in Sec.~\ref{sec:degrade}, and repeated the measurement of $C_{\star}$, $C_{\rm mol}$ and $C_{\rm norm}$, following the same procedure as detailed in Sec.~\ref{sec:fit}.
Figure~\ref{fig:scale_evo} displays the mean of the distribution of each parameter, for each galactic environment, as a function of the spatial scale of the data. The error-bars show the standard deviation of the corresponding posterior distribution. The bottom panel shows the `real' normalization value $\hat{C}_{\rm norm}$, as defined in Eq.~\eqref{eq:C2real}. It is clear that different environments have different coefficients in their scaling relations at ${\sim}150$~pc resolution, but the differences are reduced when looking at larger spatial scales. This is likely due to a combination of two effects: (i) At larger spatial scales, the light from different environments is blended, and (ii) the number of available pixels decreases drastically at larger spatial scales, from a total sample size of ${\sim}50{,}000$ to ${\sim}1200$ making the measurement statistically less certain. Whereas (ii) would have an effect across the whole range of spatial scales probed, (i) will be relevant only at spatial scales larger than that of the typical structural size of each environment.

\begin{figure}[h!]
    \includegraphics[width = \columnwidth]{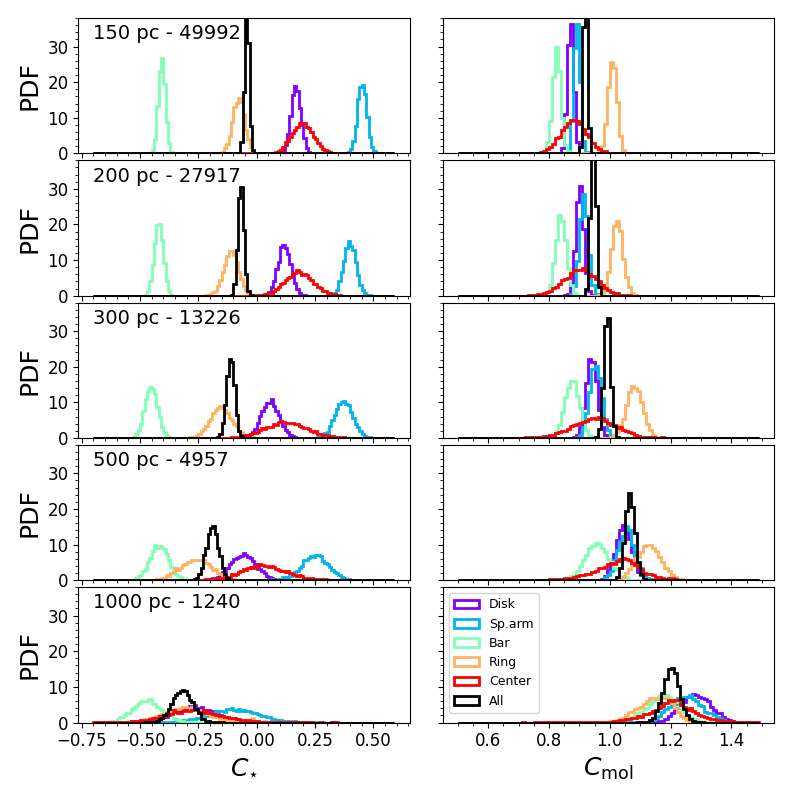}
    
    \caption{Posterior distribution of the coefficients that define the star-forming planes in each galactic environment, at each one of the spatial scales probed, from $150$~pc to $1$~kpc. The spatial scale and the number of pixels used for each measurement are indicated in the top left corner for each row.}    
    \label{fig:scale_evo2}
\end{figure}

Figure~\ref{fig:scale_evo2} explicitly shows the posterior distributions of $C_{\star}$ and $C_{\rm mol}$, for each environment and spatial scale probed. It shows that at spatial scales larger than $300$~pc, the differences between the posteriors of different environments become smaller than $1\sigma$ (with respect to the full-sample measurement), especially for $C_{\rm mol}$. This spatial scale roughly matches the width of the spiral arms in our Galaxy \citep{Reid2014}. In our sample, rings, spiral arms and centers are found to have sizes on the order of a few hundreds parsecs \citep{Querejeta2021}, and therefore, their emission is expected to blend  at spatial scales larger than this. On the other hand, bars have larger typical sizes, on the order of several kpc, which is consistent with finding that differences in the posterior distributions of their scaling relation parameters persist up to scales of ${\sim}500$~pc in both, $C_{\star}$ and $C_{\rm mol}$. This homogeneization toward lower spatial resolutions agrees well with the absence of systematic galaxy-to-galaxy variations reported in \citet{Sanchez2021}, using data from the CALIFA \citep{Sanchez2012} and EDGE \citep{Bolatto2017} surveys.

It is also worth noting the general trends of $C_{\star}$ and $C_{\rm mol}$ toward larger spatial scales. While the posteriors of $C_{\star}$ are shifted toward more negative values, the posteriors of $C_{\rm mol}$ move toward steeper slopes. These changes are likely a combination of the effect of N/Ds in our data (even though we remove the $\Sigma_{\star}$ ranges more critically affected by N/Ds to minimize their impact), and the  $C_{\star} {-} C_{\rm mol}$ covariance. The former could induce a steepening in these slopes, while the latter would drive their (a)symmetry. 

\begin{figure}[h!]
    \includegraphics[width = \columnwidth]{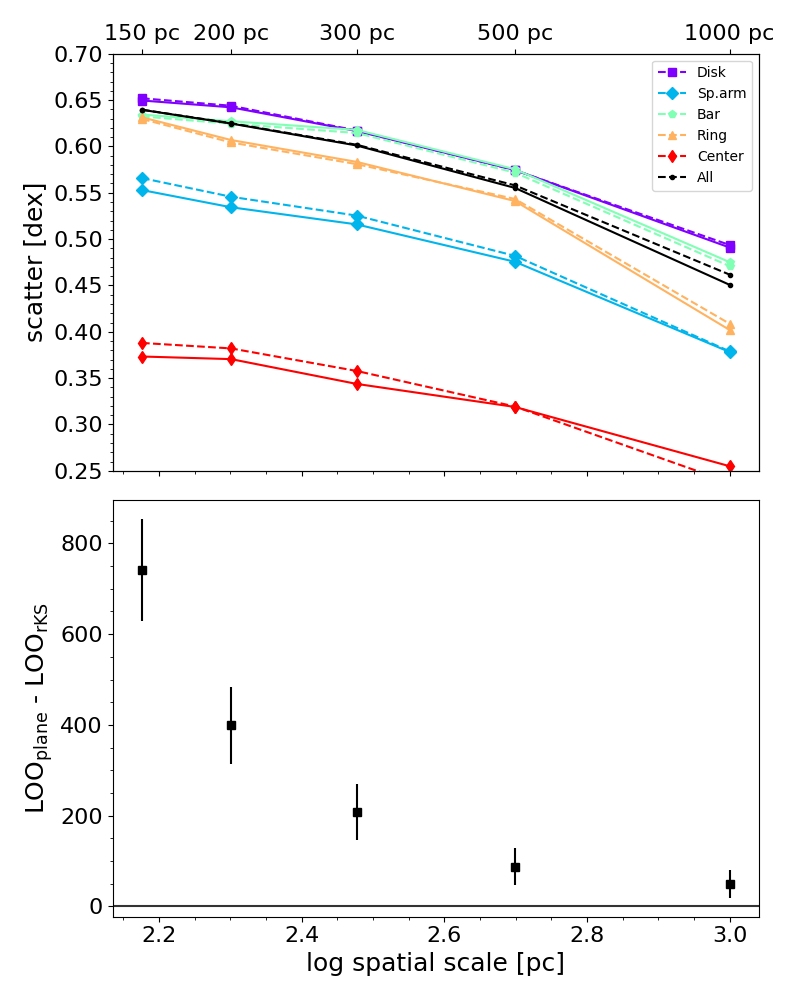}
    
    \caption{Comparison between plane and single power law models. \textit{Top}: Scatter with respect to the star-forming 2D plane (solid) and rKS (dashed) determined for each galactic environment, at the different spatial scales probed.
    \textit{Bottom}: Difference of the LOO score measured for the plane and the rKS models, where a $>0$ indicates a preference for the plane model. The error bars are the $3\sigma$ standard errors. The black horizontal line at $0$ marks where there would be no preference between the models. The plane model is preferred over the rKS at all spatial scales, increasingly so at finer spatial scales.}
    \label{fig:scale_scatter}
\end{figure}

Finally, we find that galactic environments are not only different in terms of the coefficients describing their star-forming plane, but also in terms of the scatter around it. The top panel of Fig.~\ref{fig:scale_scatter} shows the scatter (defined as the standard deviation of the residuals)  of each environment with respect to its modeled star-forming plane, at the different spatial scales probed (solid lines). 

The first noticeable feature is how the scatter decreases toward larger spatial scales. The same trend is reported in \citet{Pessa2021} for the 1D scaling relations. This effect has been associated with the decoupling of different stages of the star-forming cycle at high spatial resolution, in other words, a given aperture can be either dominated by a peak in the CO emission (i.e.\ early in the star-forming cycle) or a peak in the H$\alpha$ emission (i.e.\ late in the star-forming cycle) \citep[e.g.,][]{Schruba2010,Feldmann2011, Leroy2013, Chevance2020b, Chevance2020}. At larger spatial scales, these peaks are averaged, which diminishes the sampling effect. \citep{Schruba2010, Kruijssen2014, Semenov2017, Kruijssen2018}. The fraction of pixels containing only molecular gas or only ionized gas has been empirically characterized in \citet{Schinnerer2019} and \citet{Pan2022}. Both works found that ${\sim}500$~pc is a critical spatial scale, below which distinguishing particular stages of the star formation process is possible, in nearby star-forming galaxies. In this line, \citet{kruijssen2019} and \citet{Chevance2020b} found that the typical distance between independent star-forming regions is $100{-}300$~pc. J.~Kim et al. (in prep.) and  Machado et al. (in prep.) report an average distance of $250-300$~pc in the full PHANGS sample.

The figure also shows that at ${\sim}150$~pc, the scatter in certain environments is significantly lower than in others. Centers and spiral arms show a particularly low scatter, whereas disk is the only environment that shows a scatter larger than the full sample. It is also worth noticing that the rate at which the scatter decreases with increasing spatial scale also varies across environments, with bars showing the flattest trend between $150$ and $300$~pc, while rings and centers showing the steepest trend. Together with the scatter, this indicates variations in the properties of the molecular clouds, such as differences in their lifetimes, or a different typical separation between clouds \citep[as suggested by][]{Chevance2020b}. In this line, \citet{Henshaw2020} found evidence that fragmentation indeed occurs on smaller size scales in centers and potentially also (but to a lesser extent) in spiral arms. This idea would be consistent with the higher scatter, and its flatter decrease toward larger spatial scale that we see in bars and disks, resulting from a lower spatial density of star-forming regions, as compared to rings, spiral arms or centers. Additional variations in the intrinsic properties of molecular clouds have been reported in previous works \citep{Sun2018,Sun2020, Sun2020b, Rosolowsky2021}. \citet{Sun2020}, using data from PHANGS--ALMA, found that cloud-scale molecular gas surface density, velocity dispersion and turbulent pressure of molecular clouds depend on local environmental conditions. Similarly, \citet{Colombo2014} used data from the PdBI Arcsecond Whirlpool Survey \citep[PAWS;][]{Schinnerer2013} to characterize variations of molecular gas properties across galactic environments, and concluded that this environmental variations are a consequence of the combined action of large-scale dynamical processes and feedback from high-mass star formation. A similar conclusion was obtained by \cite{Renaud2015}, where the authors used hydrodynamical simulations to study the regulation of the star formation in bars.

\subsection{Plane versus single power law }

Here we explore whether a 2D plane offers a more accurate prediction of $\Sigma_{\rm SFR}$ than the rKS. For this comparison, we find the best-fitting power law in each environment following the same prescription described in Sec.~\ref{sec:fit} (i.e., centering the data and using a single $C_{\rm norm}$ value), dropping the dependency on $\Sigma_{\star}$ (i.e., removing $C_{\star}$ from the model). The top panel of Fig.~\ref{fig:scale_scatter} shows as dashed lines the scatter with respect to the rKS measured for each galactic environment, at each one of the spatial scales probed. The figure shows that when we separate the data by environment, the additional dependency of $\Sigma_{\star}$ does not reduce the scatter significantly. Moreover, rKS shows a slightly smaller scatter than the 2D plane in some environments. This is a consequence of the low predictive power found for $\Sigma_{\star}$, with respect to $\Sigma_{\rm mol}$ (i.e.\ $C_{\rm mol} > C_{\star}$), to predict $\Sigma_{\rm SFR}$. However, it should be kept in mind that the lower predictive power found for $\Sigma_{\star}$ could be, at least partially, related to its limited dynamic range, as shown by Fig.~\ref{fig:projection}. Nevertheless, this finding is consistent with \citet{Lin2019}, where the authors find that the `extended' version of the rKS (i.e.\ with an additional dependency of $\Sigma_{\star}$) has its scatter only slightly reduced with respect to the conventional rKS.

Despite offering only a modest improvement in the scatter, we find that the plane model is formally favored over the rKS at all spatial scales. We compare the models using leave-one-out \citep[LOO;][]{LOO} cross-validation, a standard model comparison method \citep{Vehtari2017}. The LOO cross validation is computationally more expensive than other forms of cross validations (e.g., K-fold, random subsampling), but it offers a more robust estimate of the predictive accuracy of a model, and it is suitable for relatively small datasets. 

The bottom panel of Fig.~\ref{fig:scale_scatter} shows the difference of the LOO statistics for both models (the environmental separation is included in each model), where a higher LOO statistic indicates the preferred model. The figure shows that the LOO statistic for the plane model is higher than for the rKS model at all spatial scales at the ${>}3\sigma$ level, and that the preference for the plane model becomes stronger at high spatial resolution. For this comparison, we checked that the measured $\sigma_{\rm intr}$ at each spatial scales captures the different levels of scatter of the data across the range of spatial scales probed, making the decrease in the LOO statistic not primarily driven by the intrinsic scatter.

This result is consistent with Fig.~\ref{fig:scale_evo2}, which shows that the width of the posterior distributions of $C_{\star}$ increases toward larger spatial scales in all environments, and thus nonzero $C_{\star}$ are more constrained on finer spatial scales. This means that including the $C_{\star}$ in the model is more important at smaller scales, but less critical at larger spatial scales. An explanation for this is that in the high spatial resolution measurements, the high $\Sigma_{\rm SFR}$ (and $\Sigma_{\rm mol}$) regions are often located in the inner regions of the galaxy (and vice-versa). However, at lower spatial resolutions, $\Sigma_{\rm SFR}$ (and $\Sigma_{\rm mol}$) is diluted in a larger regions, due to its intrinsic patchy configuration. As this does not occur with $\Sigma_{\rm \star}$, this results in $\Sigma_{\rm \star}$ providing less information to predict $\Sigma_{\rm SFR}$, compared to $\Sigma_{\rm mol}$, toward lower resolutions.

\section{Discussion}
\label{sec:discussion}
\subsection{Comparison to previous findings}
\label{sec:PrevFind}

Here we compare our results with those in the literature.
Our results agree relatively well with those reported in \citet{Lin2019}, where the authors used ALMaQUEST \citep{Lin2020} data to explore the extended version of the rKS relation \citep{Shi2011, Shi2018}. In terms of the coefficients we use in this paper, they found $C_{\star} = -0.29$, $C_{\rm mol} = 0.97$ and $C_{\rm norm} = -8.27$. These numbers agree with those we report here (see Table~\ref{tab:coefs_fidu}), not only in their absolute values but also in terms of the relative predicting power of $\log \Sigma_{\star}$ and $\log \Sigma_{\rm mol}$ to predict $\log \Sigma_{\rm SFR}$. \citet{Shi2018} performed similar measurements using data from The \hi\ Nearby Galaxy Sample \citep[THINGS;][]{Walter2008} and from the \textit{GALEX} data archive (using total gas surface density rather than molecular gas). Interestingly, in terms of the same coefficients, they found  $C_{\star} = 0.55$, $C_{\rm mol} = 1.09$ and $C_{\rm norm} = -10.47$. This is roughly consistent with the idea that mid-plane pressure helps to regulate current star-formation, as pressure scales as $\Sigma_{\star}^{0.5}\Sigma_{\rm gas}$ \citep{Shi2018}. Furthermore, they found that outer disks of dwarfs galaxies and local luminous infrared galaxies show the largest offset in this relation, whereas local spirals show almost no offset. This is an intriguing result, as we also find a relatively similar behavior for spiral arms. However, we stress that this comparison has the caveats of (i) local spiral galaxies host different environments (not only spiral arms) and (ii) the study carried out in \citet{Shi2018} uses data with different spatial resolutions. Nevertheless, this is an interesting comparison that could indicate that mid-plane hydrostatic pressure plays a more relevant role in spiral arms than in other environments.


On the other hand, in \citet{Sanchez2021}, the authors performed a similar measurement using data from the CALIFA and EDGE surveys. For the EDGE--CALIFA sample, which represents a better reference for comparison in terms of molecular gas tracer and spatial scale, they found $C_{\star} = 0.44\pm0.01$, $C_{\rm mol} = 0.57\pm0.01$ and $C_{\rm norm} = -9.5\pm0.01$. These coefficients partially agree with our results, in terms of the lower relative importance of $\log \Sigma_{\star}$ to predict the $\log \Sigma_{\rm SFR}$ value with respect to $\log \Sigma_{\rm mol}$. However, significant quantitative differences between the measured coefficient exist. The differences persist even if we exclude the detection-fraction-threshold step from the fitting (this does not change our results significantly at a spatial resolution of ${\sim}1$~kpc).

Part of the differences we see here are in the statistical questions we are exploring with respect to the variation of these scaling relationships with environment. \citet{Sanchez2021} demonstrate the importance of the statistical framework for interpreting the results of these regressions. Our hierarchical Bayesian approach for this analysis provides a framework for analyzing the data within well-resolved environments that could not be addressed at coarser resolution by the complementary EDGE--CALIFA sample, despite its much larger size, compared to our PHANGS--MUSE sample. A direct comparison is not possible since PHANGS--MUSE galaxies all have distances lower than 20 Mpc, and an overlap between both samples does not exist. Nevertheless, it is worth mentioning that \citet{Sanchez2021} also report measurements using their CALIFA--only sample, where the molecular gas estimates are based on the ISM dust attenuation prescription reported in \citet{Barrera-Ballesteros2021}. For this sample, they obtained $C_{\star} = 0.66\pm0.02$, $C_{\rm mol} = 0.38\pm0.01$ and $C_{\rm norm} = -9.9\pm0.05$. These numbers are in stronger disagreement with our results, not only due to the differences in the values of the measured slopes, but also in the relative importance of $\log \Sigma_{\star}$ and  $\log \Sigma_{\rm mol}$ to predict the $\log \Sigma_{\rm SFR}$ Hence, any comparison of results should be made carefully, as differences in the sample, such as the molecular gas tracer, could lead to different conclusions. The results reported in \citet{Sanchez2021} (EDGE--CALIFA sample) and in \citet{Lin2019} are indicated in Fig.~\ref{fig:scale_evo} as a cyan diamond and a magenta hexagon, respectively.


Other studies have also explored systematic difference across galactic environments. In \citet{Querejeta2021}, the authors used 74 galaxies of the PHANGS sample to measure the distributions of $\Sigma_{\rm mol}$, $\Sigma_{\rm SFR}$ and depletion times across galactic environments at a fixed spatial scale of ${\sim}1.5$~kpc. They found a strong correlation between molecular gas and SFR surface densities, with a global slope of $N = 0.97$, but little variation across galactic environments. However, they reported a slight offset toward shorter depletion times for centers ($\tau_{\rm dep} = 1.2$~Gyr), and longer depletion times for bars ($\tau_{\rm dep} = 2.1$~Gyr). This result is consistent with our measurement of the lowest $C_{\star}$ in this environment at a spatial resolution of ${\sim}1$~kpc. On the other hand, we do not find that $C_{\star}$ in centers is higher than in the other environments. However, this specific measurement is highly uncertain (due to the low number of `center' pixels at this resolution in our study) and thus, a proper comparison is not possible. Overall, they did not find evidence of strong variations in how efficiently different environment form stars. Not finding strong differences across environments is also consistent with the homogenization of galactic structure at large spatial scales which we report in this work (see Sec.~\ref{sec:spatial_res_effect}).

\subsection{What drives the environmental variations we see?}
\label{sec:third_param}

In previous sections, we reported the coefficients that describe the star-forming plane for each individual galactic environment. Furthermore, we find significant differences in the star-forming planes associated with these different environments, particularly for bars, spiral arms and rings. In this section, we explore which parameter(s) could be driving the differences we report in Sec.~\ref{sec:results}. 

To perform this exploration, we measure $C_{\star}$, $C_{\rm mol}$ and $C_{\rm norm}$ in each environment of each individual galaxy in our sample (following the same procedure detailed in Sec.~\ref{sec:fit}), and search for correlations between the coefficients obtained in each environment, and a set of additional parameters:
\begin{itemize}
    \item Molecular gas fraction; $\log f_{\rm mol}$, calculated as $\log (\Sigma_{\rm mol} / \Sigma_{\star})$
    \item Star formation efficiency; $\log \mathrm{SFE}$, calculated as $\log (\Sigma_{\rm SFR} / \Sigma_{\rm mol})$
    \item Free fall time ($t_{\rm ff}$), calculated following \cite{Utomo2018}, i.e.\ $t_{\rm ff} = \sqrt{\frac{3\pi}{32G} \Big(\frac{H}{\Sigma_{\rm mol}}\Big)}$, where $G$ is the gravitational constant and $H$ is the vertical scale height of the molecular gas layer which can be estimated as $H = \sqrt{\frac{\sigma_{\rm mol}^{2}h_{\star}}{G\Sigma_{\star}}}$, with $\sigma_{\rm mol}$ corresponding to the velocity dispersion of the molecular gas component, and $h_{\star}$ is the typical stellar scale height, for which we adopt a value of $300$~pc \citep{Utomo2018}.
    \item Depletion time in units of free fall time; $\tau / t_{\rm ff}$, where $\tau =  \Sigma_{\rm mol} / \Sigma_{\rm SFR}$ and $t_{\rm ff}$ is calculated as described above.
    \item Mid-plane hydrostatic pressure; $\log P_{\rm h}$, calculated following \citet{Elmergreen1989} as $P_{\rm h} = \frac{\pi}{2}G\Sigma_{\rm mol} \left( \Sigma_{\rm mol} + \frac{\sigma_{\rm mol}}{\sigma_{\star}} \Sigma_{\star} \right)$. This approach neglects the contribution from atomic gas, however, the relative contribution of atomic gas with respect to molecular gas in these galaxies and this regime of $\Sigma_{\rm mol}$ values is expected to be subdominant \citep[][Leroy et al (in prep.)]{Bigiel2008, Schruba2019}. \citet{Barrera-Ballesteros2021b} explore the role of $P_{\rm h}$ in regulating $\Sigma_{\rm SFR}$, and they found a tight (scatter $\sim 0.2$~dex) correlation between these two quantities. \citet{Sun2020b} also found that $\Sigma_{\rm SFR}$ correlates with the dynamical equilibrium pressure of the ISM. This is encouraging to explore if differences between environments could be driven by differences in the mid-plane hydrostatic pressure, which has the additional dependencies on $\sigma_{\rm mol}$ and $\sigma_{\star}$.
    \item Gas-phase metallicity; $[Z/H]_{\rm gas}$, modeled from emission-line measurements in the MUSE data, allowing for azimuthal variations, as detailed in \citet{Williams2021}.
    \item Stellar velocity dispersion; $\sigma_{\star}$, resulting of SSP fitting of the MUSE data (see Sec.~\ref{sec:SSP_fit}).
    \item H$\alpha$ velocity dispersion; $\sigma_{\mathrm{H}\alpha}$, measured from the MUSE data (see Sec.~\ref{sec:SFR}).
    \item Molecular gas velocity dispersion; $\sigma_{\rm mol}$, measured from the ALMA data \citep{Leroy2021a}.
\end{itemize}
We have additionally considered stellar population parameters measured by SSP fitting of the MUSE data \citep[][I.~Pessa et al.\ in prep.]{Emsellem2021}
\begin{itemize}
    \item light-weighted stellar age; $\log \mathrm{AGE}_{\rm LW}$
    \item mass-weighted stellar age; $\log \mathrm{AGE}_{\rm MW}$
    \item light-weighted stellar metallicity; ${[Z/H]}_{\rm LW}$
    \item mass-weighted stellar metallicity; ${[Z/H]}_{\rm MW}$
\end{itemize}

A similar exploration of parameters was carried out by \citet{Dey2019}, using data from the EDGE--CALIFA sample, performing a data-driven approach to investigate what shapes the SFR, finding that $\Sigma_{\rm SFR}$ scales primarily with  $\Sigma_{\star}$ and $\Sigma_{\rm mol}$. Conversely, we perform this exploration in order to understand what is driving the environmental differences we see in the plane spanned by these three quantities.

We compute the mean of the distribution of these quantities in each environment (considering only those pixels that were used for the fitting of the plane), and measure the weighted Pearson's correlation coefficient ($\rho$) between these environ\-mental-averaged quantities and the corresponding coefficients. For this latter step, we consider only the subset of galaxies that satisfy the following criteria:
\begin{enumerate}
    \item Probe at least three different environments.
    \item Host galaxy has a bar and spiral arms.
    \item Fraction of pixels removed due to the imposed detection threshold is $<50\%$.
\end{enumerate}
These selection criteria ensure that we select galaxies in which we can simultaneously probe several environments, including bars and spiral arms, which are the environments that exhibit the largest differences, especially in terms of $C_{\star}$ (see Fig.~\ref{fig:posteiors_fidu}), and that most of the pixels of these galaxies are actually used in the fitting. Out of our sample, 8~galaxies satisfy the listed conditions (NGC~1300, NGC~1365, NGC~1512, NGC~1566, NGC~1672, NGC~3627, NGC~4303, NGC~4321). As a result of these conditions, we drop from our sample galaxies with total stellar mass of $\log M_{\star}\,[M_{\odot}] < 10.6$. We perform this exploration on an individual-galaxy basis because we acknowledge that, while different environments usually exhibit different properties in a given galaxy, the same environments are not necessarily identical across different galaxies.

For this subset of galaxies, we quantify the level of correlation between each of the parameters listed, and the coefficients $C_{\star}$ and $C_{\rm mol}$ obtained for each environment, using the `overall' Pearson's correlation coefficient, defined as:
\begin{equation}
    \langle \rho \rangle = \frac{\lvert\sum_{i=1}^{\mathrm{n_{gal}}}  \rho_{i}\rvert}{\mathrm{n_{gal}}}~,
\end{equation}
which corresponds to the average $\rho$ across the subset of galaxies. For each galaxy, the uncertainty in its correlation coefficient $\rho_{i}$ (i.e., between the plane coefficients and each one of the parameters listed) is estimated by performing $50$ Monte-Carlo simulations, perturbing the measured $C_{\star}$ and $C_{\rm mol}$  for each environment in each iteration. The uncertainty in  $\langle \rho \rangle$ is then calculated using standard error propagation. 

\begin{figure}[h!]
    \includegraphics[width=\columnwidth]{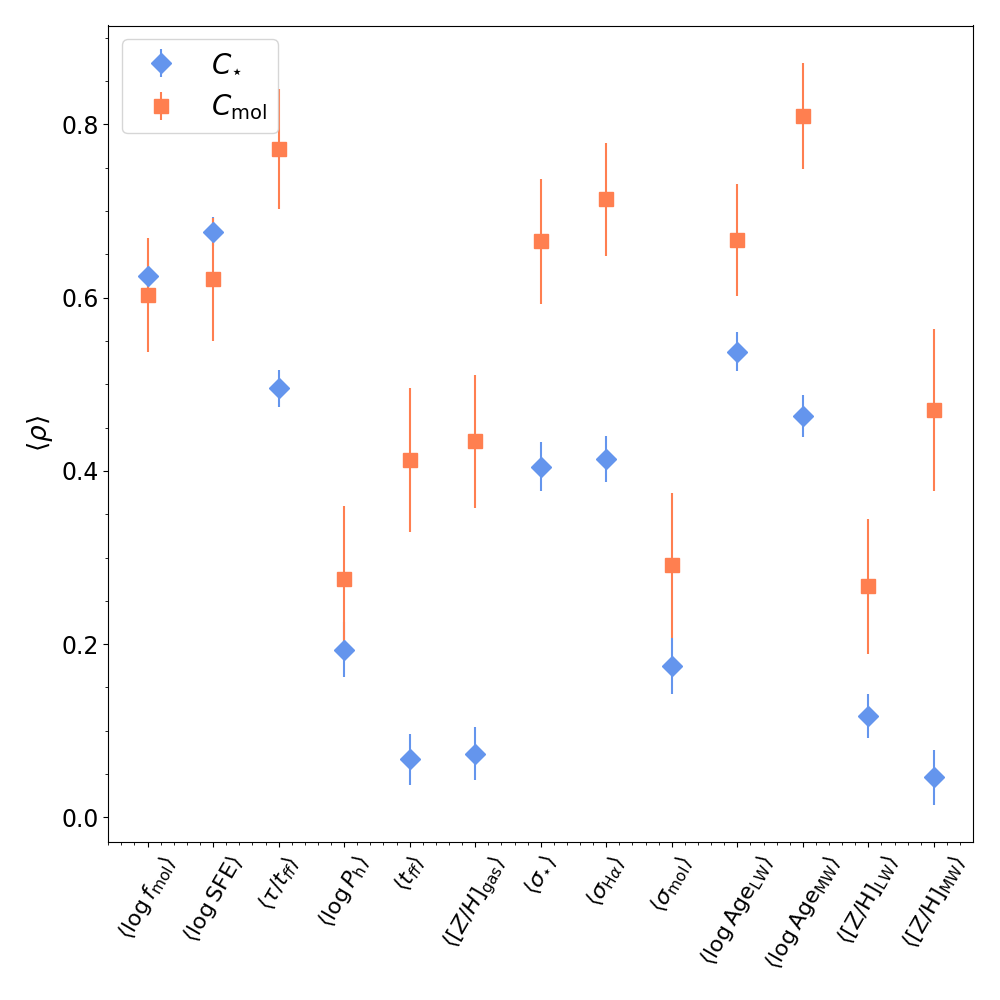}
    \caption{Overall Pearson coefficient to quantify the level of correlation between the $C_{\star}$ and $C_{\rm mol}$ parameters that define the star-forming plane in each environment, and the set of different parameters explored.}
    \label{fig:OverallP}
\end{figure}

Figure~\ref{fig:OverallP} shows the overall Pearson's correlation coefficient, that quantifies the average level of correlation between the median of each one of the quantities listed on the $x$-axis for a given environment and its corresponding $C_{\star}$ (blue) and $C_{\rm mol}$ (orange) value.
It is clear that some parameters show a high level of correlation with $C_{\rm mol}$, the highest being the stellar mass-weighted age of the underlying stellar populations. However, we acknowledge that a possible correlation with the average age of the underlying stellar population would likely be a consequence of differences in the star-forming process, rather than driving it. For this reason, we focus on the following high-$\rho$ parameter, \ $\langle \tau / t_{\rm ff} \rangle$ ($\langle \rho \rangle \sim 0.77$). 

\begin{figure*}[h!]
    \includegraphics[width=\textwidth]{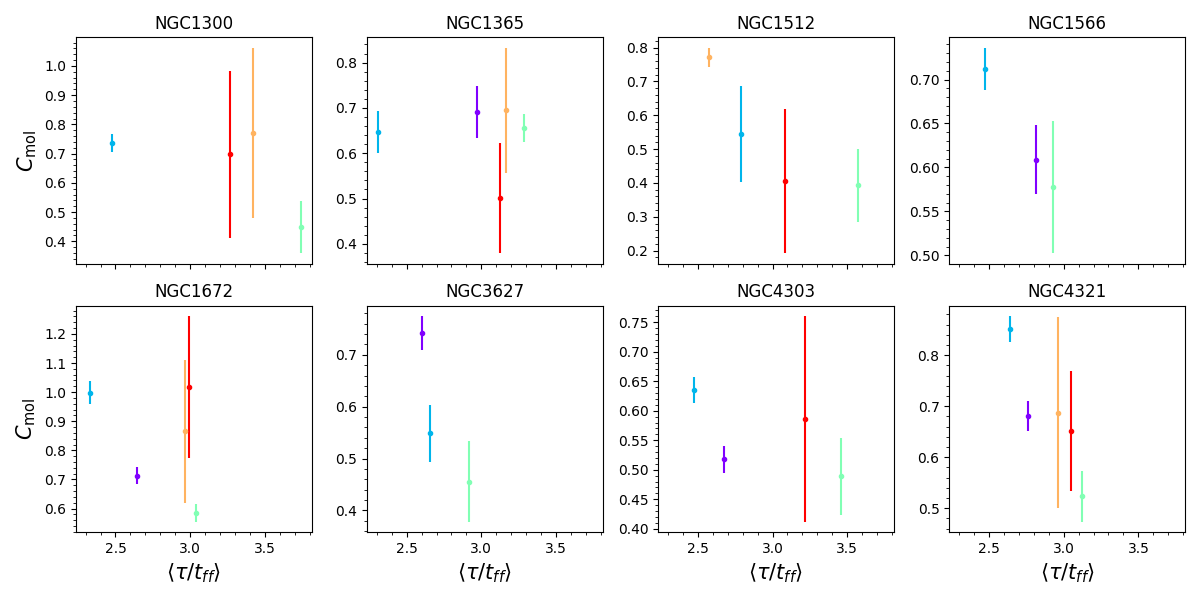}
    \caption{$C_{\rm mol} {-} \langle \tau / t_{\rm ff} \rangle$ correlation across different galactic environments, for the galaxies that satisfy our single-galaxy selection criteria. The color code for each environment is the same as used in Fig.~\ref{fig:Det_frac}.}
    \label{fig:C1_bestpar}
\end{figure*}

As $\langle \tau / t_{\rm ff} \rangle$ corresponds to the depletion time normalized by the free fall time, that is, the characteristic time that it would take a gas cloud to collapse under its own gravitational attraction, it is a metric that describes how efficiently the gas is collapsing and forming stars. A high value can be interpreted as a relative excess of molecular gas with respect to SFR, in other words, a larger fraction of the gas is not forming stars (quiescent), compared to an environment with a shorter average depletion time.
Figure~\ref{fig:C1_bestpar} shows explicitly the correlation between $\langle \tau / t_{\rm ff}  \rangle$ measured in each environment, and the corresponding $C_{\rm mol}$ coefficient for each individual galaxy in our subset. For most of the subset of galaxies, we see a negative trend (except for NGC~1365). Bars tend to have longer depletion times and lower values of $C_{\rm mol}$, as opposed to spiral arms, that show shorter depletion times and higher values of $C_{\rm mol}$. However, this correlation is not surprising, as $C_{\rm mol}$ quantifies how efficiently molecular gas forms stars, and $\langle \tau \rangle$ corresponds to the inverse of $\langle \log \mathrm{SFE} \rangle$ in a given environment. Thus, environments with on average longer depletion times (or lower SFE) will be naturally described by a lower $C_{\rm mol}$. Nevertheless, this correlation shows that quantifiable differences in depletion time across environments exists, and that these differences lead to variations in the star-forming plane.

Therefore, it is interesting that other parameters that show a high level of correlation with $C_{\rm mol}$ are $\sigma_{\star}$ and $\sigma_{\mathrm{H}\alpha}$. Velocity dispersion encodes information about noncircular motion of gas and stars in the galaxy, such as radial motions along the bar, or turbulence induced by star formation feedback or by AGN activity, for galaxies hosting an AGN (NGC~1365, NGC~1566, NGC~1672, NGC~3627, NGC~7496). Thus, the correlations between $C_{\rm mol}$ and $\sigma_{\star, \mathrm{H}\alpha}$ could be an imprint of how the noncircular motion of the gas prevents collapse and efficient star formation (leading to longer depletion times). The correlation with $\sigma_{\mathrm{H}\alpha}$ is presented in Appendix~\ref{sec:corr_Ha_GasFrac} for completeness.

Of the remaining parameters, only two show a potential correlation with $C_{\rm mol}$, $\langle \log \mathrm{SFE} \rangle$ and $\langle f_{\mathrm{mol}} \rangle$. The first one is not surprising, as it corresponds to the inverse of $\langle \tau \rangle$. For $\langle f_{\mathrm{mol}} \rangle$, we find hints of a positive correlation, that is, a higher gas fraction leads to a higher $C_{\rm mol}$. This implies that stars are formed more efficiently in galactic environments which are more gas-rich.

For $t_{\rm ff}$, $\log P_{\rm h}$, $[Z/H]_{\rm gas}$, $\sigma_{\rm mol}$,
$\log [Z/H]_{\rm LW}$ and $\log [Z/H]_{\rm MW}$, we do not find a correlation with $C_{\rm mol}$.

\begin{figure*}[h!]
    \includegraphics[width=\textwidth]{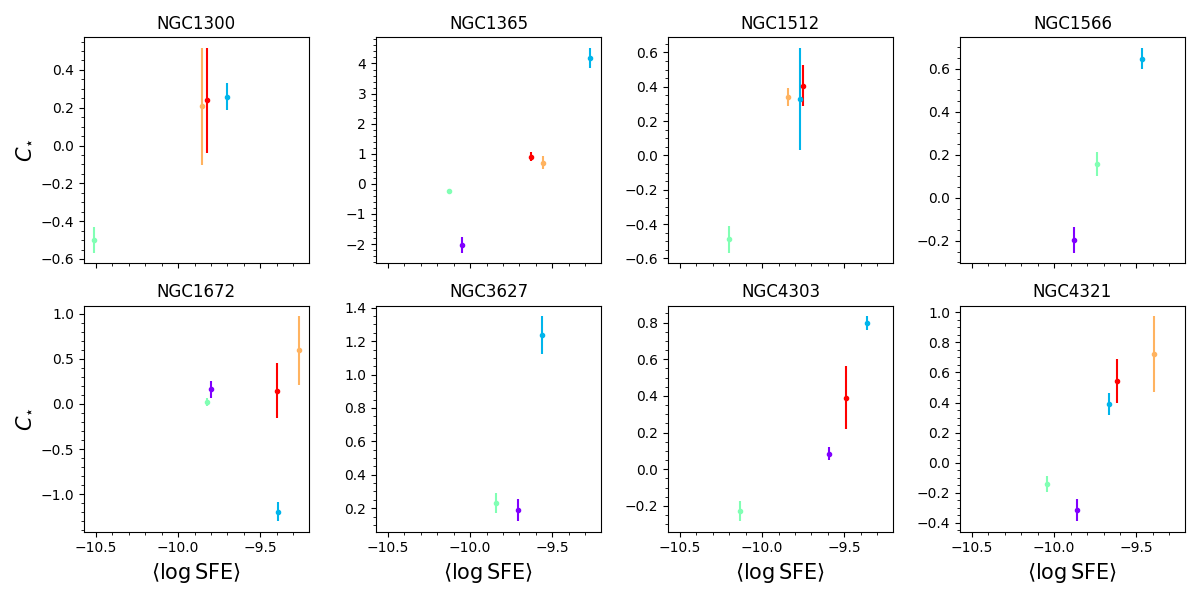}
    \caption{$C_{\star} {-} \langle \log$ SFE$\rangle$ correlation across different galactic environments, for the galaxies that satisfy our single-galaxy selection criteria. The color code for each environment is the same as used in Fig.~\ref{fig:Det_frac}.}
    \label{fig:C0_bestpar}
\end{figure*}

On the other hand, $C_{\star}$ shows in general lower levels of correlation for most of the parameters probed. This is not unexpected, due to the lower predictive power of $\Sigma_{\star}$ to predict $\Sigma_{\rm SFR}$ as compared to $\Sigma_{\rm mol}$. Nevertheless, we still see a correlation of $C_{\star}$ with $\log \mathrm{SFE}$ ($\langle \rho \rangle \sim 0.68$). Figure~\ref{fig:C0_bestpar} shows how environments with higher average star formation efficiency exhibit higher measured $C_{\star}$ values. A similar correlation is found for $\langle f_{\mathrm{mol}} \rangle$, which is shown in Appendix~\ref{sec:corr_Ha_GasFrac} for completeness.

$\langle \tau / t_{\rm ff}  \rangle$ and $\log \mathrm{AGE}_{\rm LW}$ show hints of a mild correlation with $C_{\star}$ ($\langle \rho \rangle \sim 0.5$), but as discussed earlier, the first is expected as depletion time correspond to the inverse of SFE, and the second, if exists, is likely a consequence (rather than a cause) of differences in the star-forming process. None of the remaining parameters show hints of correlation with $C_{\star}$.

The poor level of correlation of $C_{\star}$ and $C_{\rm mol}$ with $\langle \log P_{\rm h} \rangle$ can be explained by the fact that $P_{\rm h}$ scales primarily with $\Sigma_{\rm mol}$ and $\Sigma_{\star}$. These dependencies are already captured by $C_{\star}$ and $C_{\rm mol}$, thus, $P_{\rm h}$ contributes little additional information.

Finding these correlations with physical conditions in a given galactic environment suggests that an additional parameter is indeed causing the variations we measure across galactic environments (and also can explain galaxy-to-galaxy variations), and thus, playing a relevant role in regulating the SFR. However, determining what is (or are) the key parameter(s) is beyond the scope of this paper, as we would need a larger sample of galaxies that show a variety of galactic environments, and have, at the same time, high coverage of H$\alpha$ and CO emission. 
 
Finally, we stress here that due to the relatively low number of galaxies where we can reliably probe several galactic environments (8), and since each one of the correlations has only as many points as different environments the galaxy has, we can only speculate about the influence of an additional parameter in the coefficients of the star-forming plane, rather than robustly establishing a causal relation. Thus, we present this as a possible line of exploration, with intriguing dependencies that need confirmation with larger samples. 


\subsection{Choice of \texorpdfstring{$\alpha_{\rm CO}$}{alphaCO} conversion factor}
\label{sec:ConvFact}

\begin{figure*}[h!]
    \includegraphics[width=\textwidth]{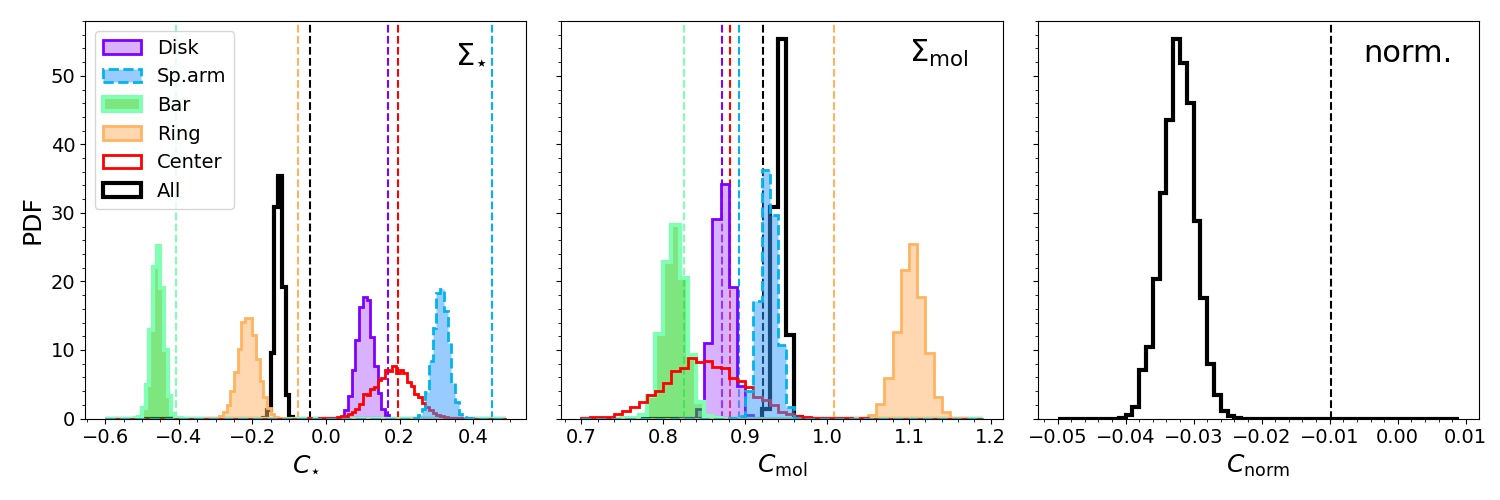}
    \caption{Posterior distributions for the coefficients $C_{\star}$, $C_{\rm mol}$ and $C_{\rm norm}$ that define the star-forming plane in each separate environment, using a constant $\alpha_\mathrm{CO}$. The vertical dashed lines show the centers of the posteriors distribution obtained under our fiducial choice of $\alpha_\mathrm{CO}$. As the posterior distribution of $C_{\rm norm}$ is considerably narrower than that of $C_{\star}$ and $C_{\rm mol}$, the $x$-axis has been binned in smaller bins and renormalized for an easier visualization.}
    \label{fig:alphaCO}
\end{figure*}

We have tested our main conclusions against a constant $\alpha_\mathrm{CO}$ conversion factor of $4.35~M_{\odot}$~pc$^{-2}$ (K~km~s$^{-1}$)$^{-1}$ \citep{Bolatto2013}, instead of the metal\-licity-depen\-dent prescription (described in Sec.~\ref{sec:data}). We found that qualitatively, none of our conclusions are affected by our choice of $\alpha_\mathrm{CO}$. Figure~\ref{fig:alphaCO} shows the posterior distributions of the coefficients determined at a spatial scale of $150$~pc, under the assumption of a constant $\alpha_\mathrm{CO}$ conversion factor.
It is easy to see that even though some specific coefficients may change, the relative difference between environments, and the relative predictive power of $C_{\star}$ and $C_{\rm mol}$ remains qualitatively the same.

\section{Summary}
\label{sec:summary}

We have investigated the star-forming plane, conformed by $\Sigma_{\rm SFR}$, $\Sigma_{\star}$, and $\Sigma_{\rm mol}$ in 18 galaxies from the PHANGS sample at a physical resolution of $150$~pc, and explored potential variations driven by galactic environment. Our main conclusions are as follows:

\begin{enumerate}
    \item We found significant differences (${>}1\sigma$) in the coefficients that describe the star-forming plane across galactic environments. These differences are particularly significant for bars, spiral arms and rings. We interpret these variations as evidence for an additional regulator mechanism that is not captured by neither $\Sigma_{\star}$ nor $\Sigma_{\rm mol}$.
    \item These variations between environments homogenize toward lower spatial resolutions. Differences are no longer significant at spatial scales larger than ${\sim}500$~pc. This is because the combined effect of blending of environments at lower spatial resolutions, and the lower number of pixels per environment at large spatial scales.
    \item We find a good agreement with similar measurements done using data from the ALMaQUEST survey. On the other hand, we find a moderate agreement with measurements reported using data from the EDGE--CALIFA sample. Differences with coarser resolution studies are not surprising, as our statistical  framework is specifically designed to quantify differences that would be measurable at high spatial resolution only.
    \item We used a subset of galaxies from our sample, where we can probe (${\geq}3$) several galactic environments, and search for correlations between variations in the coefficients that define the star-forming plane in each environment, and the median of the distribution of a number of variables measured in those environments. We find a strong correlation of $C_{\rm mol}$ with depletion time and H$\alpha$ velocity dispersion ($\langle \rho \rangle \sim 0.77$ and $0.72$, respectively), which could be an imprint of longer depletion times driven by increased turbulent or radial motion in some galactic environments, leading to the environmental differences we observe for the star-forming plane.
\end{enumerate}

Our results are consistent with the existence of additional physics being at play in the regulation of the star formation. However, a larger sample of galaxies where we can simulataneously probe different galactic environments is required to confirm the correlations explored in this work and, thus, to provide a definitive answer as to what is (or are) the additonal parameter(s) modulating the formation of stars.

\begin{acknowledgements}
This work was carried out as part of the PHANGS collaboration. Based on observations collected at the European Organisation for Astronomical Research in the Southern Hemisphere under ESO programmes IDs 094.C-0623, 098.C-0484 and 1100.B-0651. This paper also makes use of the following ALMA data: ADS/JAO.ALMA\#2013.1.01161.S, ADS/JAO.ALMA\#2015.1.00925.S, ADS/JAO.ALMA\#2015.1.00956.S and ADS/JAO.ALMA\#2017.1.00886.L. ALMA is a partnership of ESO (representing its member states), NSF (USA) and NINS (Japan), together with NRC (Canada), MOST and ASIAA (Taiwan), and  KASI (Republic of Korea), in cooperation with the Republic of Chile. The Joint ALMA Observatory is operated by ESO, AUI/NRAO and NAOJ. The National Radio Astronomy Observatory is a facility of the National Science Foundation operated under cooperative agreement by Associated Universities, Inc.

MC and JMDK gratefully acknowledge funding from the Deutsche Forschungsgemeinschaft (DFG, German Research Foundation) through an Emmy Noether Research Group (grant number KR4801/1-1), as well as from the European Research Council (ERC) under the European Union's Horizon 2020 research and innovation programme via the ERC Starting Grant MUSTANG (grant agreement number 714907).

JG gratefully acknowledges financial support from the Swiss National Science Foundation (grant no CRSII5\_193826).

KK gratefully acknowledges funding from the German Research Foundation (DFG) in the form of an Emmy Noether Research Group (grant number KR4598/2-1, PI Kreckel). 

EWK acknowledges support from the Smithsonian Institution as a Submillimeter Array (SMA) Fellow.

ER acknowledges the support of the Natural Sciences and Engineering Research Council of Canada (NSERC), funding reference number RGPIN-2017-03987.

MQ acknowledges support from the Spanish grant PID2019-106027GA-C44, funded by MCIN/AEI/10.13039/501100011033.

GAB gratefully acknowledges support by the ANID BASAL project FB210003.

JPe acknowledges support by the Programme National ``Physique et Chimie du Milieu Interstellaire'' (PCMI) of CNRS/INSU with INC/INP, co-funded by CEA and CNES.

JPe acknowledges support from the ANR grant ANR-21-CE31-0010.

RSK and SCOG acknowledge support from DFG via the Collaborative Research Center (SFB 881, Project-ID 138713538) ``The Milky Way System'' (subprojects A1, B1, B2 and B8) and from the Heidelberg Cluster of Excellence (EXC 2181 - 390900948) ``STRUCTURES: A unifying approach to emergent phenomena in the physical world, mathematics, and complex dat'', funded by the German Excellence Strategy. They also thank for funding form the European Research Council in the ERC Synergy Grant ``ECOGAL -- Understanding our Galactic ecosystem: From the disk of the Milky Way to the formation sites of stars and planets'' (project ID 855130). 

FB acknowledges funding from the European Research Council (ERC) under the European Union’s Horizon 2020 research and innovation programme (grant agreement No.726384/Empire)

PSB acknowledge support from grant PID2019-107427GB-C31 from the Spanish Ministry of Science and Innovation

AU acknowledges support from the Spanish grants PGC2018-094671-B-I00, funded by MCIN/AEI/10.13039/501100011033 and by "ERDF A way of making Europe", and PID2019-108765GB-I00, funded by MCIN/AEI/10.13039/501100011033. 

This research made use of \texttt{Astropy}, a community-developed core Python package for Astronomy \citep{Astropy2013, Astropy2018}.

\end{acknowledgements}

\bibliographystyle{aa}
\bibliography{biblio} 

\begin{thebibliography}{135}
\expandafter\ifx\csname natexlab\endcsname\relax\def\natexlab#1{#1}\fi

\bibitem[{{Abdurro'uf} \& {Akiyama}(2017)}]{Abdurro2017}
{Abdurro'uf} \& {Akiyama}, M. 2017, \mnras, 469, 2806

\bibitem[{{Accurso} {et~al.}(2017){Accurso}, {Saintonge}, {Catinella},
  {Cortese}, {Dav{\'e}}, {Dunsheath}, {Genzel}, {Gracia-Carpio}, {Heckman},
  {Jimmy}, {Kramer}, {Li}, {Lutz}, {Schiminovich}, {Schuster}, {Sternberg},
  {Sturm}, {Tacconi}, {Tran}, \& {Wang}}]{Accurso2017}
{Accurso}, G., {Saintonge}, A., {Catinella}, B., {et~al.} 2017, \mnras, 470,
  4750

\bibitem[{{Anand} {et~al.}(2021){Anand}, {Lee}, {Van Dyk}, {Leroy},
  {Rosolowsky}, {Schinnerer}, {Larson}, {Kourkchi}, {Kreckel}, {Scheuermann},
  {Rizzi}, {Thilker}, {Tully}, {Bigiel}, {Blanc}, {Boquien}, {Chandar}, {Dale},
  {Emsellem}, {Deger}, {Glover}, {Grasha}, {Groves}, {Klessen}, {Kruijssen},
  {Querejeta}, {S{\'a}nchez-Bl{\'a}zquez}, {Schruba}, {Turner}, {Ubeda},
  {Williams}, \& {Whitmore}}]{Anand2021}
{Anand}, G.~S., {Lee}, J.~C., {Van Dyk}, S.~D., {et~al.} 2021, \mnras, 501,
  3621

\bibitem[{{Astropy Collaboration} {et~al.}(2018){Astropy Collaboration},
  {Price-Whelan}, {Sip{\H{o}}cz}, {G{\"u}nther}, {Lim}, {Crawford}, {Conseil},
  {Shupe}, {Craig}, {Dencheva}, {Ginsburg}, {VanderPlas}, {Bradley},
  {P{\'e}rez-Su{\'a}rez}, {de Val-Borro}, {Aldcroft}, {Cruz}, {Robitaille},
  {Tollerud}, {Ardelean}, {Babej}, {Bach}, {Bachetti}, {Bakanov}, {Bamford},
  {Barentsen}, {Barmby}, {Baumbach}, {Berry}, {Biscani}, {Boquien}, {Bostroem},
  {Bouma}, {Brammer}, {Bray}, {Breytenbach}, {Buddelmeijer}, {Burke},
  {Calderone}, {Cano Rodr{\'\i}guez}, {Cara}, {Cardoso}, {Cheedella}, {Copin},
  {Corrales}, {Crichton}, {D'Avella}, {Deil}, {Depagne}, {Dietrich}, {Donath},
  {Droettboom}, {Earl}, {Erben}, {Fabbro}, {Ferreira}, {Finethy}, {Fox},
  {Garrison}, {Gibbons}, {Goldstein}, {Gommers}, {Greco}, {Greenfield},
  {Groener}, {Grollier}, {Hagen}, {Hirst}, {Homeier}, {Horton}, {Hosseinzadeh},
  {Hu}, {Hunkeler}, {Ivezi{\'c}}, {Jain}, {Jenness}, {Kanarek}, {Kendrew},
  {Kern}, {Kerzendorf}, {Khvalko}, {King}, {Kirkby}, {Kulkarni}, {Kumar},
  {Lee}, {Lenz}, {Littlefair}, {Ma}, {Macleod}, {Mastropietro}, {McCully},
  {Montagnac}, {Morris}, {Mueller}, {Mumford}, {Muna}, {Murphy}, {Nelson},
  {Nguyen}, {Ninan}, {N{\"o}the}, {Ogaz}, {Oh}, {Parejko}, {Parley}, {Pascual},
  {Patil}, {Patil}, {Plunkett}, {Prochaska}, {Rastogi}, {Reddy Janga},
  {Sabater}, {Sakurikar}, {Seifert}, {Sherbert}, {Sherwood-Taylor}, {Shih},
  {Sick}, {Silbiger}, {Singanamalla}, {Singer}, {Sladen}, {Sooley},
  {Sornarajah}, {Streicher}, {Teuben}, {Thomas}, {Tremblay}, {Turner},
  {Terr{\'o}n}, {van Kerkwijk}, {de la Vega}, {Watkins}, {Weaver}, {Whitmore},
  {Woillez}, {Zabalza}, \& {Astropy Contributors}}]{Astropy2018}
{Astropy Collaboration}, {Price-Whelan}, A.~M., {Sip{\H{o}}cz}, B.~M., {et~al.}
  2018, \aj, 156, 123

\bibitem[{{Astropy Collaboration} {et~al.}(2013){Astropy Collaboration},
  {Robitaille}, {Tollerud}, {Greenfield}, {Droettboom}, {Bray}, {Aldcroft},
  {Davis}, {Ginsburg}, {Price-Whelan}, {Kerzendorf}, {Conley}, {Crighton},
  {Barbary}, {Muna}, {Ferguson}, {Grollier}, {Parikh}, {Nair}, {Unther},
  {Deil}, {Woillez}, {Conseil}, {Kramer}, {Turner}, {Singer}, {Fox}, {Weaver},
  {Zabalza}, {Edwards}, {Azalee Bostroem}, {Burke}, {Casey}, {Crawford},
  {Dencheva}, {Ely}, {Jenness}, {Labrie}, {Lim}, {Pierfederici}, {Pontzen},
  {Ptak}, {Refsdal}, {Servillat}, \& {Streicher}}]{Astropy2013}
{Astropy Collaboration}, {Robitaille}, T.~P., {Tollerud}, E.~J., {et~al.} 2013,
  \aap, 558, A33

\bibitem[{{Bacchini} {et~al.}(2019){Bacchini}, {Fraternali}, {Iorio}, \&
  {Pezzulli}}]{Bacchini2019a}
{Bacchini}, C., {Fraternali}, F., {Iorio}, G., \& {Pezzulli}, G. 2019, \aap,
  622, A64

\bibitem[{{Bacchini} {et~al.}(2020){Bacchini}, {Fraternali}, {Pezzulli}, \&
  {Marasco}}]{Bacchini2020}
{Bacchini}, C., {Fraternali}, F., {Pezzulli}, G., \& {Marasco}, A. 2020, \aap,
  644, A125

\bibitem[{{Bacon} {et~al.}(2014){Bacon}, {Vernet}, {Borisova}, {Bouch{\'e}},
  {Brinchmann}, {Carollo}, {Carton}, {Caruana}, {Cerda}, {Contini}, {Franx},
  {Girard}, {Guerou}, {Haddad}, {Hau}, {Herenz}, {Herrera}, {Husemann},
  {Husser}, {Jarno}, {Kamann}, {Krajnovic}, {Lilly}, {Mainieri}, {Martinsson},
  {Palsa}, {Patricio}, {P{\'e}contal}, {Pello}, {Piqueras}, {Richard},
  {Sandin}, {Schroetter}, {Selman}, {Shirazi}, {Smette}, {Soto}, {Streicher},
  {Urrutia}, {Weilbacher}, {Wisotzki}, \& {Zins}}]{Bacon2014}
{Bacon}, R., {Vernet}, J., {Borisova}, E., {et~al.} 2014, The Messenger, 157,
  13

\bibitem[{{Baldwin} {et~al.}(1981){Baldwin}, {Phillips}, \& {Terlevich}}]{BPT}
{Baldwin}, J.~A., {Phillips}, M.~M., \& {Terlevich}, R. 1981, \pasp, 93, 5

\bibitem[{{Barrera-Ballesteros}
  {et~al.}(2021{\natexlab{a}}){Barrera-Ballesteros}, {Heckman}, {S{\'a}nchez},
  {Drory}, {Cruz-Gonzalez}, {Carigi}, {Riffel}, {Boquien}, {Tissera},
  {Bizyaev}, {Rong}, {Boardman}, {Alvarez Hurtado}, \& {MaNGA
  Team}}]{Barrera-Ballesteros2021}
{Barrera-Ballesteros}, J.~K., {Heckman}, T., {S{\'a}nchez}, S.~F., {et~al.}
  2021{\natexlab{a}}, \apj, 909, 131

\bibitem[{{Barrera-Ballesteros}
  {et~al.}(2021{\natexlab{b}}){Barrera-Ballesteros}, {S{\'a}nchez}, {Heckman},
  {Wong}, {Bolatto}, {Ostriker}, {Rosolowsky}, {Carigi}, {Vogel}, {Levy},
  {Colombo}, {Luo}, \& {Cao}}]{Barrera-Ballesteros2021b}
{Barrera-Ballesteros}, J.~K., {S{\'a}nchez}, S.~F., {Heckman}, T., {et~al.}
  2021{\natexlab{b}}, \mnras, 503, 3643

\bibitem[{{Belfiore} {et~al.}(2015){Belfiore}, {Maiolino}, {Bundy}, {Thomas},
  {Maraston}, {Wilkinson}, {S{\'a}nchez}, {Bershady}, {Blanc}, {Bothwell},
  {Cales}, {Coccato}, {Drory}, {Emsellem}, {Fu}, {Gelfand}, {Law}, {Masters},
  {Parejko}, {Tremonti}, {Wake}, {Weijmans}, {Yan}, {Xiao}, {Zhang}, {Zheng},
  {Bizyaev}, {Kinemuchi}, {Oravetz}, \& {Simmons}}]{Belfiore2015}
{Belfiore}, F., {Maiolino}, R., {Bundy}, K., {et~al.} 2015, \mnras, 449, 867

\bibitem[{{Belfiore} {et~al.}(2022){Belfiore}, {Santoro}, {Groves},
  {Schinnerer}, {Kreckel}, {Glover}, {Klessen}, {Emsellem}, {Blanc}, {Congiu},
  {Barnes}, {Boquien}, {Chevance}, {Dale}, {Diederik Kruijssen}, {Leroy},
  {Pan}, {Pessa}, {Schruba}, \& {Williams}}]{Belfiore2021}
{Belfiore}, F., {Santoro}, F., {Groves}, B., {et~al.} 2022, \aap, 659, A26

\bibitem[{{Bigiel} {et~al.}(2008){Bigiel}, {Leroy}, {Walter}, {Brinks}, {de
  Blok}, {Madore}, \& {Thornley}}]{Bigiel2008}
{Bigiel}, F., {Leroy}, A., {Walter}, F., {et~al.} 2008, \aj, 136, 2846

\bibitem[{{Bigiel} {et~al.}(2011){Bigiel}, {Leroy}, {Walter}, {Brinks}, {de
  Blok}, {Kramer}, {Rix}, {Schruba}, {Schuster}, {Usero}, \&
  {Wiesemeyer}}]{Bigiel2011}
{Bigiel}, F., {Leroy}, A.~K., {Walter}, F., {et~al.} 2011, \apjl, 730, L13

\bibitem[{{Bittner} {et~al.}(2019){Bittner}, {Falc{\'o}n-Barroso}, {Nedelchev},
  {Dorta}, {Gadotti}, {Sarzi}, {Molaeinezhad}, {Iodice}, {Rosado-Belza}, {de
  Lorenzo-C{\'a}ceres}, {Fragkoudi}, {Gal{\'a}n-de Anta}, {Husemann},
  {M{\'e}ndez-Abreu}, {Neumann}, {Pinna}, {Querejeta},
  {S{\'a}nchez-Bl{\'a}zquez}, \& {Seidel}}]{Bittner2019}
{Bittner}, A., {Falc{\'o}n-Barroso}, J., {Nedelchev}, B., {et~al.} 2019, \aap,
  628, A117

\bibitem[{{Blanc} {et~al.}(2009){Blanc}, {Heiderman}, {Gebhardt}, {Evans}, \&
  {Adams}}]{Blanc2009}
{Blanc}, G.~A., {Heiderman}, A., {Gebhardt}, K., {Evans}, Neal~J., I., \&
  {Adams}, J. 2009, \apj, 704, 842

\bibitem[{{Bolatto} {et~al.}(2013){Bolatto}, {Wolfire}, \&
  {Leroy}}]{Bolatto2013}
{Bolatto}, A.~D., {Wolfire}, M., \& {Leroy}, A.~K. 2013, \araa, 51, 207

\bibitem[{{Bolatto} {et~al.}(2017){Bolatto}, {Wong}, {Utomo}, {Blitz}, {Vogel},
  {S{\'a}nchez}, {Barrera-Ballesteros}, {Cao}, {Colombo}, {Dannerbauer},
  {Garc{\'\i}a-Benito}, {Herrera-Camus}, {Husemann}, {Kalinova}, {Leroy},
  {Leung}, {Levy}, {Mast}, {Ostriker}, {Rosolowsky}, {Sandstrom}, {Teuben},
  {van de Ven}, \& {Walter}}]{Bolatto2017}
{Bolatto}, A.~D., {Wong}, T., {Utomo}, D., {et~al.} 2017, \apj, 846, 159

\bibitem[{{Brinchmann} {et~al.}(2004){Brinchmann}, {Charlot}, {White},
  {Tremonti}, {Kauffmann}, {Heckman}, \& {Brinkmann}}]{Brinchmann2004}
{Brinchmann}, J., {Charlot}, S., {White}, S.~D.~M., {et~al.} 2004, \mnras, 351,
  1151

\bibitem[{{Calzetti}(2013)}]{Calzetti-book}
{Calzetti}, D. 2013, {Star Formation Rate Indicators}, ed.
  J.~{Falc{\'o}n-Barroso} \& J.~H. {Knapen}, 419

\bibitem[{{Calzetti} {et~al.}(2000){Calzetti}, {Armus}, {Bohlin}, {Kinney},
  {Koornneef}, \& {Storchi-Bergmann}}]{Calzetti2000}
{Calzetti}, D., {Armus}, L., {Bohlin}, R.~C., {et~al.} 2000, \apj, 533, 682

\bibitem[{{Cano-D{\'\i}az} {et~al.}(2016){Cano-D{\'\i}az}, {S{\'a}nchez},
  {Zibetti}, {Ascasibar}, {Bland-Hawthorn}, {Ziegler}, {Gonz{\'a}lez Delgado},
  {Walcher}, {Garc{\'\i}a-Benito}, {Mast}, {Mendoza-P{\'e}rez},
  {Falc{\'o}n-Barroso}, {Galbany}, {Husemann}, {Kehrig}, {Marino},
  {S{\'a}nchez-Bl{\'a}zquez}, {L{\'o}pez-Cob{\'a}}, {L{\'o}pez-S{\'a}nchez}, \&
  {Vilchez}}]{CanoDiaz2016}
{Cano-D{\'\i}az}, M., {S{\'a}nchez}, S.~F., {Zibetti}, S., {et~al.} 2016,
  \apjl, 821, L26

\bibitem[{{Cappellari}(2017)}]{Capellari2017}
{Cappellari}, M. 2017, \mnras, 466, 798

\bibitem[{{Cappellari} \& {Copin}(2003)}]{Capellari2003}
{Cappellari}, M. \& {Copin}, Y. 2003, \mnras, 342, 345

\bibitem[{{Cappellari} \& {Emsellem}(2004)}]{Capellari2004}
{Cappellari}, M. \& {Emsellem}, E. 2004, \pasp, 116, 138

\bibitem[{{Cardelli} {et~al.}(1989){Cardelli}, {Clayton}, \&
  {Mathis}}]{Cardelli89}
{Cardelli}, J.~A., {Clayton}, G.~C., \& {Mathis}, J.~S. 1989, \apj, 345, 245

\bibitem[{{Chabrier}(2003)}]{Chabrier2003}
{Chabrier}, G. 2003, \pasp, 115, 763

\bibitem[{{Chevance} {et~al.}(2020{\natexlab{a}}){Chevance}, {Kruijssen},
  {Hygate}, {Schruba}, {Longmore}, {Groves}, {Henshaw}, {Herrera}, {Hughes},
  {Jeffreson}, {Lang}, {Leroy}, {Meidt}, {Pety}, {Razza}, {Rosolowsky},
  {Schinnerer}, {Bigiel}, {Blanc}, {Emsellem}, {Faesi}, {Glover}, {Haydon},
  {Ho}, {Kreckel}, {Lee}, {Liu}, {Querejeta}, {Saito}, {Sun}, {Usero}, \&
  {Utomo}}]{Chevance2020b}
{Chevance}, M., {Kruijssen}, J.~M.~D., {Hygate}, A. P.~S., {et~al.}
  2020{\natexlab{a}}, \mnras, 493, 2872

\bibitem[{{Chevance} {et~al.}(2020{\natexlab{b}}){Chevance}, {Kruijssen},
  {Vazquez-Semadeni}, {Nakamura}, {Klessen}, {Ballesteros-Paredes}, {Inutsuka},
  {Adamo}, \& {Hennebelle}}]{Chevance2020}
{Chevance}, M., {Kruijssen}, J.~M.~D., {Vazquez-Semadeni}, E., {et~al.}
  2020{\natexlab{b}}, \ssr, 216, 50

\bibitem[{{Colombo} {et~al.}(2014){Colombo}, {Hughes}, {Schinnerer}, {Meidt},
  {Leroy}, {Pety}, {Dobbs}, {Garc{\'\i}a-Burillo}, {Dumas}, {Thompson},
  {Schuster}, \& {Kramer}}]{Colombo2014}
{Colombo}, D., {Hughes}, A., {Schinnerer}, E., {et~al.} 2014, \apj, 784, 3

\bibitem[{{Daddi} {et~al.}(2007){Daddi}, {Dickinson}, {Morrison}, {Chary},
  {Cimatti}, {Elbaz}, {Frayer}, {Renzini}, {Pope}, {Alexander}, {Bauer},
  {Giavalisco}, {Huynh}, {Kurk}, \& {Mignoli}}]{Daddi2007}
{Daddi}, E., {Dickinson}, M., {Morrison}, G., {et~al.} 2007, \apj, 670, 156

\bibitem[{{den Brok} {et~al.}(2021){den Brok}, {Chatzigiannakis}, {Bigiel},
  {Puschnig}, {Barnes}, {Leroy}, {Jimenez-Donaire}, {Usero}, {Schinnerer},
  {Rosolowsky}, {Faesi}, {Grasha}, {Hughes}, {Kruijssen}, {Liu}, {Neumann},
  {Pety}, {Querejeta}, T., {Schruba}, \& {Stuber}}]{denBrok2021}
{den Brok}, J.~S., {Chatzigiannakis}, D., {Bigiel}, F., {et~al.} 2021, \mnras\
  submitted

\bibitem[{{Dey} {et~al.}(2019){Dey}, {Rosolowsky}, {Cao}, {Bolatto}, {Sanchez},
  {Utomo}, {Colombo}, {Kalinova}, {Wong}, {Blitz}, {Vogel}, {Loeppky}, \&
  {Garc{\'\i}a-Benito}}]{Dey2019}
{Dey}, B., {Rosolowsky}, E., {Cao}, Y., {et~al.} 2019, \mnras, 488, 1926

\bibitem[{{Dib} {et~al.}(2017){Dib}, {Hony}, \& {Blanc}}]{Dib2017}
{Dib}, S., {Hony}, S., \& {Blanc}, G. 2017, \mnras, 469, 1521

\bibitem[{{Ellison} {et~al.}(2021){Ellison}, {Lin}, {Thorp}, {Pan}, {Scudder},
  {S{\'a}nchez}, {Bluck}, \& {Maiolino}}]{Ellison2021}
{Ellison}, S.~L., {Lin}, L., {Thorp}, M.~D., {et~al.} 2021, \mnras, 501, 4777

\bibitem[{{Elmegreen}(1989)}]{Elmergreen1989}
{Elmegreen}, B.~G. 1989, \apj, 338, 178

\bibitem[{{Emsellem} {et~al.}(2021){Emsellem}, {Schinnerer}, {Santoro},
  {Belfiore}, {Pessa}, {McElroy}, {Blanc}, {Congiu}, {Groves}, {Ho}, {Kreckel},
  {Razza}, {Sanchez-Blazquez}, {Egorov}, {Faesi}, {Klessen}, {Leroy}, {Meidt},
  {Querejeta}, {Rosolowsky}, {Scheuermann}, {Anand}, {Barnes},
  {Be{\v{s}}li{\'c}}, {Bigiel}, {Boquien}, {Cao}, {Chevance}, {Dale},
  {Eibensteiner}, {Glover}, {Grasha}, {Henshaw}, {Hughes}, {Koch}, {Kruijssen},
  {Lee}, {Liu}, {Pan}, {Pety}, {Saito}, {Sandstrom}, {Schruba}, {Sun},
  {Thilker}, {Usero}, {Watkins}, \& {Williams}}]{Emsellem2021}
{Emsellem}, E., {Schinnerer}, E., {Santoro}, F., {et~al.} 2021, arXiv e-prints,
  arXiv:2110.03708

\bibitem[{{Feldmann} {et~al.}(2011){Feldmann}, {Gnedin}, \&
  {Kravtsov}}]{Feldmann2011}
{Feldmann}, R., {Gnedin}, N.~Y., \& {Kravtsov}, A.~V. 2011, \apj, 732, 115

\bibitem[{{Flores-Fajardo} {et~al.}(2011){Flores-Fajardo}, {Morisset},
  {Stasi{\'n}ska}, \& {Binette}}]{Flores-Fajardo2011}
{Flores-Fajardo}, N., {Morisset}, C., {Stasi{\'n}ska}, G., \& {Binette}, L.
  2011, \mnras, 415, 2182

\bibitem[{{Ford} {et~al.}(2013){Ford}, {Gear}, {Smith}, {Eales}, {Baes},
  {Bendo}, {Boquien}, {Boselli}, {Cooray}, {De Looze}, {Fritz}, {Gentile},
  {Gomez}, {Gordon}, {Kirk}, {Lebouteiller}, {O'Halloran}, {Spinoglio},
  {Verstappen}, \& {Wilson}}]{Ford2013}
{Ford}, G.~P., {Gear}, W.~K., {Smith}, M. W.~L., {et~al.} 2013, \apj, 769, 55

\bibitem[{{Gelman} \& {Rubin}(1992)}]{Gelman1992}
{Gelman}, A. \& {Rubin}, D.~B. 1992, Statistical Science, 7, 457

\bibitem[{{Gensior} \& {Kruijssen}(2021)}]{Gensior2021}
{Gensior}, J. \& {Kruijssen}, J.~M.~D. 2021, \mnras, 500, 2000

\bibitem[{{Gensior} {et~al.}(2020){Gensior}, {Kruijssen}, \&
  {Keller}}]{Gensior2020}
{Gensior}, J., {Kruijssen}, J.~M.~D., \& {Keller}, B.~W. 2020, \mnras, 495, 199

\bibitem[{{Genzel} {et~al.}(2010){Genzel}, {Tacconi}, {Gracia-Carpio},
  {Sternberg}, {Cooper}, {Shapiro}, {Bolatto}, {Bouch{\'e}}, {Bournaud},
  {Burkert}, {Combes}, {Comerford}, {Cox}, {Davis}, {F{\"o}rster Schreiber},
  {Garcia-Burillo}, {Lutz}, {Naab}, {Neri}, {Omont}, {Shapley}, \&
  {Weiner}}]{Genzel2010}
{Genzel}, R., {Tacconi}, L.~J., {Gracia-Carpio}, J., {et~al.} 2010, \mnras,
  407, 2091

\bibitem[{{Haffner} {et~al.}(2009){Haffner}, {Dettmar}, {Beckman}, {Wood},
  {Slavin}, {Giammanco}, {Madsen}, {Zurita}, \& {Reynolds}}]{Haffner2009}
{Haffner}, L.~M., {Dettmar}, R.~J., {Beckman}, J.~E., {et~al.} 2009, Reviews of
  Modern Physics, 81, 969

\bibitem[{{Henshaw} {et~al.}(2020){Henshaw}, {Kruijssen}, {Longmore}, {Riener},
  {Leroy}, {Rosolowsky}, {Ginsburg}, {Battersby}, {Chevance}, {Meidt},
  {Glover}, {Hughes}, {Kainulainen}, {Klessen}, {Schinnerer}, {Schruba},
  {Beuther}, {Bigiel}, {Blanc}, {Emsellem}, {Henning}, {Herrera}, {Koch},
  {Pety}, {Ragan}, \& {Sun}}]{Henshaw2020}
{Henshaw}, J.~D., {Kruijssen}, J.~M.~D., {Longmore}, S.~N., {et~al.} 2020,
  Nature Astronomy, 4, 1064

\bibitem[{{Herrera-Endoqui} {et~al.}(2015){Herrera-Endoqui},
  {D{\'\i}az-Garc{\'\i}a}, {Laurikainen}, \& {Salo}}]{Herrera-Endoqui2015}
{Herrera-Endoqui}, M., {D{\'\i}az-Garc{\'\i}a}, S., {Laurikainen}, E., \&
  {Salo}, H. 2015, \aap, 582, A86

\bibitem[{{Ho} {et~al.}(2017){Ho}, {Seibert}, {Meidt}, {Kudritzki},
  {Kobayashi}, {Groves}, {Kewley}, {Madore}, {Rich}, {Schinnerer},
  {D'Agostino}, \& {Poetrodjojo}}]{Ho2017}
{Ho}, I.~T., {Seibert}, M., {Meidt}, S.~E., {et~al.} 2017, \apj, 846, 39

\bibitem[{{Homan} \& {Gelman}(2014)}]{homan2014}
{Homan}, M.~D. \& {Gelman}, A. 2014, Journal of Machine Learning Research, 15,
  1593

\bibitem[{{Hoyle}(1953)}]{Hoyle1953}
{Hoyle}, F. 1953, \apj, 118, 513

\bibitem[{{Hsieh} {et~al.}(2017){Hsieh}, {Lin}, {Lin}, {Pan}, {Hsu},
  {S{\'a}nchez}, {Cano-D{\'\i}az}, {Zhang}, {Yan}, {Barrera-Ballesteros},
  {Boquien}, {Riffel}, {Brownstein}, {Cruz-Gonz{\'a}lez}, {Hagen}, {Ibarra},
  {Pan}, {Bizyaev}, {Oravetz}, \& {Simmons}}]{Hsieh2017}
{Hsieh}, B.~C., {Lin}, L., {Lin}, J.~H., {et~al.} 2017, \apjl, 851, L24

\bibitem[{{Hunter} {et~al.}(1998){Hunter}, {Elmegreen}, \&
  {Baker}}]{Hunter1998}
{Hunter}, D.~A., {Elmegreen}, B.~G., \& {Baker}, A.~L. 1998, \apj, 493, 595

\bibitem[{{Jeffreson} {et~al.}(2020){Jeffreson}, {Kruijssen}, {Keller},
  {Chevance}, \& {Glover}}]{Jeffreson2020}
{Jeffreson}, S. M.~R., {Kruijssen}, J.~M.~D., {Keller}, B.~W., {Chevance}, M.,
  \& {Glover}, S. C.~O. 2020, \mnras, 498, 385

\bibitem[{{Kaplan} {et~al.}(2016){Kaplan}, {Jogee}, {Kewley}, {Blanc},
  {Weinzirl}, {Song}, {Drory}, {Luo}, \& {van den Bosch}}]{Kaplan2016}
{Kaplan}, K.~F., {Jogee}, S., {Kewley}, L., {et~al.} 2016, \mnras, 462, 1642

\bibitem[{{Kennicutt}(1998)}]{Kennicutt1998}
{Kennicutt}, Robert~C., J. 1998, \apj, 498, 541

\bibitem[{{Kennicutt} \& {Evans}(2012)}]{Kennicutt2012}
{Kennicutt}, R.~C. \& {Evans}, N.~J. 2012, \araa, 50, 531

\bibitem[{{Kewley} {et~al.}(2006){Kewley}, {Groves}, {Kauffmann}, \&
  {Heckman}}]{Kewley2006}
{Kewley}, L.~J., {Groves}, B., {Kauffmann}, G., \& {Heckman}, T. 2006, \mnras,
  372, 961

\bibitem[{{Kreckel} {et~al.}(2018){Kreckel}, {Faesi}, {Kruijssen}, {Schruba},
  {Groves}, {Leroy}, {Bigiel}, {Blanc}, {Chevance}, {Herrera}, {Hughes},
  {McElroy}, {Pety}, {Querejeta}, {Rosolowsky}, {Schinnerer}, {Sun}, {Usero},
  \& {Utomo}}]{Kreckel2018}
{Kreckel}, K., {Faesi}, C., {Kruijssen}, J.~M.~D., {et~al.} 2018, \apjl, 863,
  L21

\bibitem[{{Kreckel} {et~al.}(2020){Kreckel}, {Ho}, {Blanc}, {Glover}, {Groves},
  {Rosolowsky}, {Bigiel}, {Boqu{\'\i}en}, {Chevance}, {Dale}, {Deger},
  {Emsellem}, {Grasha}, {Kim}, {Klessen}, {Kruijssen}, {Lee}, {Leroy}, {Liu},
  {McElroy}, {Meidt}, {Pessa}, {Sanchez-Blazquez}, {Sandstrom}, {Santoro},
  {Scheuermann}, {Schinnerer}, {Schruba}, {Utomo}, {Watkins}, \&
  {Williams}}]{Kreckel2020}
{Kreckel}, K., {Ho}, I.~T., {Blanc}, G.~A., {et~al.} 2020, \mnras, 499, 193

\bibitem[{{Kroupa}(2001)}]{Kroupa2001}
{Kroupa}, P. 2001, \mnras, 322, 231

\bibitem[{{Kruijssen} \& {Longmore}(2014)}]{Kruijssen2014}
{Kruijssen}, J.~M.~D. \& {Longmore}, S.~N. 2014, \mnras, 439, 3239

\bibitem[{{Kruijssen} {et~al.}(2019){Kruijssen}, {Schruba}, {Chevance},
  {Longmore}, {Hygate}, {Haydon}, {McLeod}, {Dalcanton}, {Tacconi}, \& {van
  Dishoeck}}]{kruijssen2019}
{Kruijssen}, J.~M.~D., {Schruba}, A., {Chevance}, M., {et~al.} 2019, \nat, 569,
  519

\bibitem[{{Kruijssen} {et~al.}(2018){Kruijssen}, {Schruba}, {Hygate}, {Hu},
  {Haydon}, \& {Longmore}}]{Kruijssen2018}
{Kruijssen}, J.~M.~D., {Schruba}, A., {Hygate}, A. P.~S., {et~al.} 2018,
  \mnras, 479, 1866

\bibitem[{{Krumholz} {et~al.}(2018){Krumholz}, {Burkhart}, {Forbes}, \&
  {Crocker}}]{Krumholz2018}
{Krumholz}, M.~R., {Burkhart}, B., {Forbes}, J.~C., \& {Crocker}, R.~M. 2018,
  \mnras, 477, 2716

\bibitem[{{Lang} {et~al.}(2020){Lang}, {Meidt}, {Rosolowsky}, {Nofech},
  {Schinnerer}, {Leroy}, {Emsellem}, {Pessa}, {Glover}, {Groves}, {Hughes},
  {Kruijssen}, {Querejeta}, {Schruba}, {Bigiel}, {Blanc}, {Chevance},
  {Colombo}, {Faesi}, {Henshaw}, {Herrera}, {Liu}, {Pety}, {Puschnig}, {Saito},
  {Sun}, \& {Usero}}]{Lang2020}
{Lang}, P., {Meidt}, S.~E., {Rosolowsky}, E., {et~al.} 2020, \apj, 897, 122

\bibitem[{{Lee} {et~al.}(2022){Lee}, {Whitmore}, {Thilker}, {Deger}, {Larson},
  {Ubeda}, {Anand}, {Boquien}, {Chandar}, {Dale}, {Emsellem}, {Leroy},
  {Rosolowsky}, {Schinnerer}, {Schmidt}, {Lilly}, {Turner}, {Van Dyk}, {White},
  {Barnes}, {Belfiore}, {Bigiel}, {Blanc}, {Cao}, {Chevance}, {Congiu},
  {Egorov}, {Glover}, {Grasha}, {Groves}, {Henshaw}, {Hughes}, {Klessen},
  {Koch}, {Kreckel}, {Kruijssen}, {Liu}, {Lopez}, {Mayker}, {Meidt}, {Murphy},
  {Pan}, {Pety}, {Querejeta}, {Razza}, {Saito}, {S{\'a}nchez-Bl{\'a}zquez},
  {Santoro}, {Sardone}, {Scheuermann}, {Schruba}, {Sun}, {Usero}, {Watkins}, \&
  {Williams}}]{Lee2021}
{Lee}, J.~C., {Whitmore}, B.~C., {Thilker}, D.~A., {et~al.} 2022, \apjs, 258,
  10

\bibitem[{{Leroy} {et~al.}(2021{\natexlab{a}}){Leroy}, {Schinnerer}, {Hughes},
  {Rosolowsky}, {Pety}, {Schruba}, {Usero}, {Blanc}, {Chevance}, {Emsellem},
  {Faesi}, {Herrera}, {Liu}, {Meidt}, {Querejeta}, {Saito}, {Sandstrom}, {Sun},
  {Williams}, {Anand}, {Barnes}, {Behrens}, {Belfiore}, {Benincasa},
  {Be{\v{s}}li{\'{c}}}, {Bigiel}, {Bolatto}, {den_Brok}, {Cao}, {Chandar},
  {Chastenet}, {Chiang}, {Congiu}, {Dale}, {Deger}, {Eibensteiner}, {Egorov},
  {Garc{\'{\i}}a-Rodr{\'{\i}}guez}, {Glover}, {Grasha}, {Henshaw}, {Ho},
  {Kepley}, {Kim}, {Klessen}, {Kreckel}, {Koch}, {Kruijssen}, {Larson}, {Lee},
  {Lopez}, {Machado}, {Mayker}, {McElroy}, {Murphy}, {Ostriker}, {Pan},
  {Pessa}, {Puschnig}, {Razza}, {S{\'{a}}nchez-Bl{\'{a}}zquez}, {Santoro},
  {Sardone}, {Scheuermann}, {Sliwa}, {Sormani}, {Stuber}, {Thilker}, {Turner},
  {Utomo}, {Watkins}, \& {Whitmore}}]{Leroy2021b}
{Leroy}, A., {Schinnerer}, E., {Hughes}, A., {et~al.} 2021{\natexlab{a}},
  \apjs, 257, 43

\bibitem[{{Leroy} {et~al.}(2021{\natexlab{b}}){Leroy}, {Hughes}, {Liu}, {Pety},
  {Rosolowsky}, {Saito}, {Schinnerer}, {Schruba}, {Usero}, {Faesi}, {Herrera},
  {Chevance}, {Hygate}, {Kepley}, {Koch}, {Querejeta}, {Sliwa}, {Will},
  {Wilson}, {Anand}, {Barnes}, {Belfiore}, {Be{\v{s}}li{\'c}}, {Bigiel},
  {Blanc}, {Bolatto}, {Boquien}, {Cao}, {Chandar}, {Chastenet}, {Chiang},
  {Congiu}, {Dale}, {Deger}, {den Brok}, {Eibensteiner}, {Emsellem},
  {Garc{\'\i}a-Rodr{\'\i}guez}, {Glover}, {Grasha}, {Groves}, {Henshaw},
  {Jim{\'e}nez Donaire}, {Kim}, {Klessen}, {Kreckel}, {Kruijssen}, {Larson},
  {Lee}, {Mayker}, {McElroy}, {Meidt}, {Mok}, {Pan}, {Puschnig}, {Razza},
  {S{\'a}nchez-Bl'azquez}, {Sandstrom}, {Santoro}, {Sardone}, {Scheuermann},
  {Sun}, {Thilker}, {Turner}, {Ubeda}, {Utomo}, {Watkins}, \&
  {Williams}}]{Leroy2021a}
{Leroy}, A.~K., {Hughes}, A., {Liu}, D., {et~al.} 2021{\natexlab{b}}, \apjs,
  255, 19

\bibitem[{{Leroy} {et~al.}(2022){Leroy}, {Rosolowsky}, {Usero}, {Sandstrom},
  {Schinnerer}, {Schruba}, {Bolatto}, {Sun}, {Barnes}, {Belfiore}, {Bigiel},
  {den Brok}, {Cao}, {Chiang}, {Chevance}, {Dale}, {Eibensteiner}, {Faesi},
  {Glover}, {Hughes}, {Jim{\'e}nez Donaire}, {Klessen}, {Koch}, {Kruijssen},
  {Liu}, {Meidt}, {Pan}, {Pety}, {Puschnig}, {Querejeta}, {Saito}, {Sardone},
  {Watkins}, {Weiss}, \& {Williams}}]{Leroy2021c}
{Leroy}, A.~K., {Rosolowsky}, E., {Usero}, A., {et~al.} 2022, \apj, 927, 149

\bibitem[{{Leroy} {et~al.}(2019){Leroy}, {Sandstrom}, {Lang}, {Lewis}, {Salim},
  {Behrens}, {Chastenet}, {Chiang}, {Gallagher}, {Kessler}, \&
  {Utomo}}]{Leroy2019}
{Leroy}, A.~K., {Sandstrom}, K.~M., {Lang}, D., {et~al.} 2019, \apjs, 244, 24

\bibitem[{{Leroy} {et~al.}(2008){Leroy}, {Walter}, {Brinks}, {Bigiel}, {de
  Blok}, {Madore}, \& {Thornley}}]{Leroy2008}
{Leroy}, A.~K., {Walter}, F., {Brinks}, E., {et~al.} 2008, \aj, 136, 2782

\bibitem[{{Leroy} {et~al.}(2013){Leroy}, {Walter}, {Sandstrom}, {Schruba},
  {Munoz-Mateos}, {Bigiel}, {Bolatto}, {Brinks}, {de Blok}, {Meidt}, {Rix},
  {Rosolowsky}, {Schinnerer}, {Schuster}, \& {Usero}}]{Leroy2013}
{Leroy}, A.~K., {Walter}, F., {Sandstrom}, K., {et~al.} 2013, \aj, 146, 19

\bibitem[{{Lin} {et~al.}(2012){Lin}, {Dickinson}, {Jian}, {Merson}, {Baugh},
  {Scott}, {Foucaud}, {Wang}, {Yan}, {Yan}, {Cheng}, {Guo}, {Helly}, {Kirsten},
  {Koo}, {Lagos}, {Meger}, {Messias}, {Pope}, {Simard}, {Grogin}, \&
  {Wang}}]{Lin2012}
{Lin}, L., {Dickinson}, M., {Jian}, H.-Y., {et~al.} 2012, \apj, 756, 71

\bibitem[{{Lin} {et~al.}(2020){Lin}, {Ellison}, {Pan}, {Thorp}, {Su},
  {S{\'a}nchez}, {Belfiore}, {Bothwell}, {Bundy}, {Chen}, {Concas}, {Hsieh},
  {Hsieh}, {Li}, {Maiolino}, {Masters}, {Newman}, {Rowlands}, {Shi},
  {Smethurst}, {Stark}, {Xiao}, \& {Yu}}]{Lin2020}
{Lin}, L., {Ellison}, S.~L., {Pan}, H.-A., {et~al.} 2020, \apj, 903, 145

\bibitem[{{Lin} {et~al.}(2019){Lin}, {Pan}, {Ellison}, {Belfiore}, {Shi},
  {S{\'a}nchez}, {Hsieh}, {Rowlands}, {Ramya}, {Thorp}, {Li}, \&
  {Maiolino}}]{Lin2019}
{Lin}, L., {Pan}, H.-A., {Ellison}, S.~L., {et~al.} 2019, \apjl, 884, L33

\bibitem[{{Mac Low} \& {Klessen}(2004)}]{McLow2004}
{Mac Low}, M.-M. \& {Klessen}, R.~S. 2004, Reviews of Modern Physics, 76, 125

\bibitem[{{Martig} {et~al.}(2009){Martig}, {Bournaud}, {Teyssier}, \&
  {Dekel}}]{Martig2009}
{Martig}, M., {Bournaud}, F., {Teyssier}, R., \& {Dekel}, A. 2009, \apj, 707,
  250

\bibitem[{{Matteucci} {et~al.}(1989){Matteucci}, {Franco}, {Francois}, \&
  {Treyer}}]{Matteucci1989}
{Matteucci}, F., {Franco}, J., {Francois}, P., \& {Treyer}, M.-A. 1989, \rmxaa,
  18, 145

\bibitem[{{Meidt}(2016)}]{Meidt2016}
{Meidt}, S.~E. 2016, \apj, 818, 69

\bibitem[{{Meidt} {et~al.}(2020){Meidt}, {Glover}, {Kruijssen}, {Leroy},
  {Rosolowsky}, {Hughes}, {Schinnerer}, {Schruba}, {Usero}, {Bigiel}, {Blanc},
  {Chevance}, {Pety}, {Querejeta}, \& {Utomo}}]{Meidt2020}
{Meidt}, S.~E., {Glover}, S. C.~O., {Kruijssen}, J.~M.~D., {et~al.} 2020, \apj,
  892, 73

\bibitem[{{Meidt} {et~al.}(2018){Meidt}, {Leroy}, {Rosolowsky}, {Kruijssen},
  {Schinnerer}, {Schruba}, {Pety}, {Blanc}, {Bigiel}, {Chevance}, {Hughes},
  {Querejeta}, \& {Usero}}]{Meidt2018}
{Meidt}, S.~E., {Leroy}, A.~K., {Rosolowsky}, E., {et~al.} 2018, \apj, 854, 100

\bibitem[{{Morselli} {et~al.}(2020){Morselli}, {Rodighiero}, {Enia},
  {Corbelli}, {Casasola}, {Rodr{\'\i}guez-Mu{\~n}oz}, {Renzini}, {Tacchella},
  {Baronchelli}, {Bianchi}, {Cassata}, {Franceschini}, {Mancini}, {Negrello},
  {Popesso}, \& {Romano}}]{Morselli2020}
{Morselli}, L., {Rodighiero}, G., {Enia}, A., {et~al.} 2020, \mnras, 496, 4606

\bibitem[{{Muraoka} {et~al.}(2019){Muraoka}, {Sorai}, {Miyamoto}, {Yoda},
  {Morokuma-Matsui}, {Kobayashi}, {Kuroda}, {Kaneko}, {Kuno}, {Takeuchi},
  {Nakanishi}, {Watanabe}, {Tanaka}, {Yasuda}, {Yajima}, {Shibata}, {Salak},
  {Espada}, {Matsumoto}, {Noma}, {Kita}, {Komatsuzaki}, {Kajikawa}, {Yashima},
  {Pan}, {Oi}, {Seta}, \& {Nakai}}]{Muraoka2019}
{Muraoka}, K., {Sorai}, K., {Miyamoto}, Y., {et~al.} 2019, \pasj, 71, S15

\bibitem[{{Noeske} {et~al.}(2007){Noeske}, {Faber}, {Weiner}, {Koo}, {Primack},
  {Dekel}, {Papovich}, {Conselice}, {Le Floc'h}, {Rieke}, {Coil}, {Lotz},
  {Somerville}, \& {Bundy}}]{Noeske2007}
{Noeske}, K.~G., {Faber}, S.~M., {Weiner}, B.~J., {et~al.} 2007, \apjl, 660,
  L47

\bibitem[{{O'Donnell}(1994)}]{ODonnell1994}
{O'Donnell}, J.~E. 1994, \apj, 422, 158

\bibitem[{{Onodera} {et~al.}(2010){Onodera}, {Kuno}, {Tosaki}, {Kohno},
  {Nakanishi}, {Sawada}, {Muraoka}, {Komugi}, {Miura}, {Kaneko}, {Hirota}, \&
  {Kawabe}}]{Onodera2010}
{Onodera}, S., {Kuno}, N., {Tosaki}, T., {et~al.} 2010, \apjl, 722, L127

\bibitem[{{Osterbrock}(1989)}]{Osterbrock1989}
{Osterbrock}, D.~E. 1989, {Astrophysics of gaseous nebulae and active galactic
  nuclei}

\bibitem[{{Pan} {et~al.}(2022){Pan}, {Schinnerer}, {Hughes}, {Leroy}, {Groves},
  {Barnes}, {Belfiore}, {Bigiel}, {Blanc}, {Cao}, {Chevance}, {Congiu}, {Dale},
  {Eibensteiner}, {Emsellem}, {Faesi}, {Glover}, {Grasha}, {Herrera}, {Ho},
  {Klessen}, {Kruijssen}, {Lang}, {Liu}, {McElroy}, {Meidt}, {Murphy}, {Pety},
  {Querejeta}, {Razza}, {Rosolowsky}, {Saito}, {Santoro}, {Schruba}, {Sun},
  {Tomi{\v{c}}i{\'c}}, {Usero}, {Utomo}, \& {Williams}}]{Pan2022}
{Pan}, H.-A., {Schinnerer}, E., {Hughes}, A., {et~al.} 2022, \apj, 927, 9

\bibitem[{{Pessa} {et~al.}(2021){Pessa}, {Schinnerer}, {Belfiore}, {Emsellem},
  {Leroy}, {Schruba}, {Kruijssen}, {Pan}, {Blanc}, {Sanchez-Blazquez},
  {Bigiel}, {Chevance}, {Congiu}, {Dale}, {Faesi}, {Glover}, {Grasha},
  {Groves}, {Ho}, {Jim{\'e}nez-Donaire}, {Klessen}, {Kreckel}, {Koch}, {Liu},
  {Meidt}, {Pety}, {Querejeta}, {Rosolowsky}, {Saito}, {Santoro}, {Sun},
  {Usero}, {Watkins}, \& {Williams}}]{Pessa2021}
{Pessa}, I., {Schinnerer}, E., {Belfiore}, F., {et~al.} 2021, \aap, 650, A134

\bibitem[{{Pietrinferni} {et~al.}(2004){Pietrinferni}, {Cassisi}, {Salaris}, \&
  {Castelli}}]{Pietrinferni2004}
{Pietrinferni}, A., {Cassisi}, S., {Salaris}, M., \& {Castelli}, F. 2004, \apj,
  612, 168

\bibitem[{{Popesso} {et~al.}(2019){Popesso}, {Concas}, {Morselli}, {Schreiber},
  {Rodighiero}, {Cresci}, {Belli}, {Erfanianfar}, {Mancini}, {Inami},
  {Dickinson}, {Ilbert}, {Pannella}, \& {Elbaz}}]{Popesso2019}
{Popesso}, P., {Concas}, A., {Morselli}, L., {et~al.} 2019, \mnras, 483, 3213

\bibitem[{{Querejeta} {et~al.}(2021){Querejeta}, {Schinnerer}, {Meidt}, {Sun},
  {Leroy}, {Emsellem}, {Klessen}, {Mu{\~n}oz-Mateos}, {Salo}, {Laurikainen},
  {Be{\v{s}}li{\'c}}, {Blanc}, {Chevance}, {Dale}, {Eibensteiner}, {Faesi},
  {Garc{\'\i}a-Rodr{\'\i}guez}, {Glover}, {Grasha}, {Henshaw}, {Herrera},
  {Hughes}, {Kreckel}, {Kruijssen}, {Liu}, {Murphy}, {Pan}, {Pety}, {Razza},
  {Rosolowsky}, {Saito}, {Schruba}, {Usero}, {Watkins}, \&
  {Williams}}]{Querejeta2021}
{Querejeta}, M., {Schinnerer}, E., {Meidt}, S., {et~al.} 2021, \aap, 656, A133

\bibitem[{{Reid} {et~al.}(2014){Reid}, {Menten}, {Brunthaler}, {Zheng}, {Dame},
  {Xu}, {Wu}, {Zhang}, {Sanna}, {Sato}, {Hachisuka}, {Choi}, {Immer},
  {Moscadelli}, {Rygl}, \& {Bartkiewicz}}]{Reid2014}
{Reid}, M.~J., {Menten}, K.~M., {Brunthaler}, A., {et~al.} 2014, \apj, 783, 130

\bibitem[{{Renaud} {et~al.}(2015){Renaud}, {Bournaud}, {Emsellem}, {Agertz},
  {Athanassoula}, {Combes}, {Elmegreen}, {Kraljic}, {Motte}, \&
  {Teyssier}}]{Renaud2015}
{Renaud}, F., {Bournaud}, F., {Emsellem}, E., {et~al.} 2015, \mnras, 454, 3299

\bibitem[{{Rosolowsky} {et~al.}(2021){Rosolowsky}, {Hughes}, {Leroy}, {Sun},
  {Querejeta}, {Schruba}, {Usero}, {Herrera}, {Liu}, {Pety}, {Saito},
  {Be{\v{s}}li{\'c}}, {Bigiel}, {Blanc}, {Chevance}, {Dale}, {Deger}, {Faesi},
  {Glover}, {Henshaw}, {Klessen}, {Kruijssen}, {Larson}, {Lee}, {Meidt}, {Mok},
  {Schinnerer}, {Thilker}, \& {Williams}}]{Rosolowsky2021}
{Rosolowsky}, E., {Hughes}, A., {Leroy}, A.~K., {et~al.} 2021, \mnras, 502,
  1218

\bibitem[{{Saintonge} {et~al.}(2016){Saintonge}, {Catinella}, {Cortese},
  {Genzel}, {Giovanelli}, {Haynes}, {Janowiecki}, {Kramer}, {Lutz},
  {Schiminovich}, {Tacconi}, {Wuyts}, \& {Accurso}}]{Saintonge2016}
{Saintonge}, A., {Catinella}, B., {Cortese}, L., {et~al.} 2016, \mnras, 462,
  1749

\bibitem[{{Salim} {et~al.}(2007){Salim}, {Rich}, {Charlot}, {Brinchmann},
  {Johnson}, {Schiminovich}, {Seibert}, {Mallery}, {Heckman}, {Forster},
  {Friedman}, {Martin}, {Morrissey}, {Neff}, {Small}, {Wyder}, {Bianchi},
  {Donas}, {Lee}, {Madore}, {Milliard}, {Szalay}, {Welsh}, \& {Yi}}]{Salim2007}
{Salim}, S., {Rich}, R.~M., {Charlot}, S., {et~al.} 2007, \apjs, 173, 267

\bibitem[{{Salo} {et~al.}(2015){Salo}, {Laurikainen}, {Laine}, {Comer{\'o}n},
  {Gadotti}, {Buta}, {Sheth}, {Zaritsky}, {Ho}, {Knapen}, {Athanassoula},
  {Bosma}, {Laine}, {Cisternas}, {Kim}, {Mu{\~n}oz-Mateos}, {Regan}, {Hinz},
  {Gil de Paz}, {Menendez-Delmestre}, {Mizusawa}, {Erroz-Ferrer}, {Meidt}, \&
  {Querejeta}}]{Salo2015}
{Salo}, H., {Laurikainen}, E., {Laine}, J., {et~al.} 2015, \apjs, 219, 4

\bibitem[{Salvatier {et~al.}(2016)Salvatier, Wiecki, \& Fonnesbeck}]{pymc3}
Salvatier, J., Wiecki, T.~V., \& Fonnesbeck, C. 2016, PeerJ Computer Science,
  2, e55

\bibitem[{{Sammut} \& {Webb}(2010)}]{LOO}
{Sammut}, C. \& {Webb}, G.~I. 2010, {Encyclopedia of Machine Learning}

\bibitem[{{S{\'a}nchez} {et~al.}(2021){S{\'a}nchez}, {Barrera-Ballesteros},
  {Colombo}, {Wong}, {Bolatto}, {Rosolowsky}, {Vogel}, {Levy}, {Kalinova},
  {Alvarez-Hurtado}, {Luo}, \& {Cao}}]{Sanchez2021}
{S{\'a}nchez}, S.~F., {Barrera-Ballesteros}, J.~K., {Colombo}, D., {et~al.}
  2021, \mnras, 503, 1615

\bibitem[{{S{\'a}nchez} {et~al.}(2012){S{\'a}nchez}, {Kennicutt}, {Gil de Paz},
  {van de Ven}, {V{\'\i}lchez}, {Wisotzki}, {Walcher}, {Mast}, {Aguerri},
  {Albiol-P{\'e}rez}, {Alonso-Herrero}, {Alves}, {Bakos}, {Bart{\'a}kov{\'a}},
  {Bland-Hawthorn}, {Boselli}, {Bomans}, {Castillo-Morales}, {Cortijo-Ferrero},
  {de Lorenzo-C{\'a}ceres}, {Del Olmo}, {Dettmar}, {D{\'\i}az}, {Ellis},
  {Falc{\'o}n-Barroso}, {Flores}, {Gallazzi}, {Garc{\'\i}a-Lorenzo},
  {Gonz{\'a}lez Delgado}, {Gruel}, {Haines}, {Hao}, {Husemann},
  {Igl{\'e}sias-P{\'a}ramo}, {Jahnke}, {Johnson}, {Jungwiert}, {Kalinova},
  {Kehrig}, {Kupko}, {L{\'o}pez-S{\'a}nchez}, {Lyubenova}, {Marino},
  {M{\'a}rmol-Queralt{\'o}}, {M{\'a}rquez}, {Masegosa}, {Meidt},
  {Mendez-Abreu}, {Monreal-Ibero}, {Montijo}, {Mour{\~a}o}, {Palacios-Navarro},
  {Papaderos}, {Pasquali}, {Peletier}, {P{\'e}rez}, {P{\'e}rez}, {Quirrenbach},
  {Rela{\~n}o}, {Rosales-Ortega}, {Roth}, {Ruiz-Lara},
  {S{\'a}nchez-Bl{\'a}zquez}, {Sengupta}, {Singh}, {Stanishev}, {Trager},
  {Vazdekis}, {Viironen}, {Wild}, {Zibetti}, \& {Ziegler}}]{Sanchez2012}
{S{\'a}nchez}, S.~F., {Kennicutt}, R.~C., {Gil de Paz}, A., {et~al.} 2012,
  \aap, 538, A8

\bibitem[{{Santoro} {et~al.}(2022){Santoro}, {Kreckel}, {Belfiore}, {Groves},
  {Congiu}, {Thilker}, {Blanc}, {Schinnerer}, {Ho}, {Diederik Kruijssen},
  {Meidt}, {Klessen}, {Schruba}, {Querejeta}, {Pessa}, {Chevance}, {Kim},
  {Emsellem}, {McElroy}, {Barnes}, {Bigiel}, {Boquien}, {Dale}, {Glover},
  {Grasha}, {Lee}, {Leroy}, {Pan}, {Rosolowsky}, {Saito}, {Sanchez-Blazquez},
  {Watkins}, \& {Williams}}]{Santoro2021}
{Santoro}, F., {Kreckel}, K., {Belfiore}, F., {et~al.} 2022, \aap, 658, A188

\bibitem[{{Schinnerer} {et~al.}(2019){Schinnerer}, {Hughes}, {Leroy}, {Groves},
  {Blanc}, {Kreckel}, {Bigiel}, {Chevance}, {Dale}, {Emsellem}, {Faesi},
  {Glover}, {Grasha}, {Henshaw}, {Hygate}, {Kruijssen}, {Meidt}, {Pety},
  {Querejeta}, {Rosolowsky}, {Saito}, {Schruba}, {Sun}, \&
  {Utomo}}]{Schinnerer2019}
{Schinnerer}, E., {Hughes}, A., {Leroy}, A., {et~al.} 2019, \apj, 887, 49

\bibitem[{{Schinnerer} {et~al.}(2013){Schinnerer}, {Meidt}, {Pety}, {Hughes},
  {Colombo}, {Garc{\'\i}a-Burillo}, {Schuster}, {Dumas}, {Dobbs}, {Leroy},
  {Kramer}, {Thompson}, \& {Regan}}]{Schinnerer2013}
{Schinnerer}, E., {Meidt}, S.~E., {Pety}, J., {et~al.} 2013, \apj, 779, 42

\bibitem[{{Schlafly} \& {Finkbeiner}(2011)}]{Schlafly2011}
{Schlafly}, E.~F. \& {Finkbeiner}, D.~P. 2011, \apj, 737, 103

\bibitem[{{Schmidt}(1959)}]{Schmidt1959}
{Schmidt}, M. 1959, \apj, 129, 243

\bibitem[{{Schruba} {et~al.}(2019){Schruba}, {Kruijssen}, \&
  {Leroy}}]{Schruba2019}
{Schruba}, A., {Kruijssen}, J.~M.~D., \& {Leroy}, A.~K. 2019, \apj, 883, 2

\bibitem[{{Schruba} {et~al.}(2011){Schruba}, {Leroy}, {Walter}, {Bigiel},
  {Brinks}, {de Blok}, {Dumas}, {Kramer}, {Rosolowsky}, {Sandstrom},
  {Schuster}, {Usero}, {Weiss}, \& {Wiesemeyer}}]{Schruba2011}
{Schruba}, A., {Leroy}, A.~K., {Walter}, F., {et~al.} 2011, \aj, 142, 37

\bibitem[{{Schruba} {et~al.}(2010){Schruba}, {Leroy}, {Walter}, {Sandstrom}, \&
  {Rosolowsky}}]{Schruba2010}
{Schruba}, A., {Leroy}, A.~K., {Walter}, F., {Sandstrom}, K., \& {Rosolowsky},
  E. 2010, \apj, 722, 1699

\bibitem[{{Semenov} {et~al.}(2017){Semenov}, {Kravtsov}, \&
  {Gnedin}}]{Semenov2017}
{Semenov}, V.~A., {Kravtsov}, A.~V., \& {Gnedin}, N.~Y. 2017, \apj, 845, 133

\bibitem[{{Sheth} {et~al.}(2010){Sheth}, {Regan}, {Hinz}, {Gil de Paz},
  {Men{\'e}ndez-Delmestre}, {Mu{\~n}oz-Mateos}, {Seibert}, {Kim},
  {Laurikainen}, {Salo}, {Gadotti}, {Laine}, {Mizusawa}, {Armus},
  {Athanassoula}, {Bosma}, {Buta}, {Capak}, {Jarrett}, {Elmegreen},
  {Elmegreen}, {Knapen}, {Koda}, {Helou}, {Ho}, {Madore}, {Masters},
  {Mobasher}, {Ogle}, {Peng}, {Schinnerer}, {Surace}, {Zaritsky},
  {Comer{\'o}n}, {de Swardt}, {Meidt}, {Kasliwal}, \& {Aravena}}]{Sheth2010}
{Sheth}, K., {Regan}, M., {Hinz}, J.~L., {et~al.} 2010, \pasp, 122, 1397

\bibitem[{{Shetty} {et~al.}(2014){Shetty}, {Clark}, \& {Klessen}}]{Shetty2014}
{Shetty}, R., {Clark}, P.~C., \& {Klessen}, R.~S. 2014, \mnras, 442, 2208

\bibitem[{{Shi} {et~al.}(2011){Shi}, {Helou}, {Yan}, {Armus}, {Wu}, {Papovich},
  \& {Stierwalt}}]{Shi2011}
{Shi}, Y., {Helou}, G., {Yan}, L., {et~al.} 2011, \apj, 733, 87

\bibitem[{{Shi} {et~al.}(2018){Shi}, {Yan}, {Armus}, {Gu}, {Helou}, {Qiu},
  {Gwyn}, {Stierwalt}, {Fang}, {Chen}, {Zhou}, {Wu}, {Zheng}, {Zhang}, {Gao},
  \& {Wang}}]{Shi2018}
{Shi}, Y., {Yan}, L., {Armus}, L., {et~al.} 2018, \apj, 853, 149

\bibitem[{{Shu} {et~al.}(1987){Shu}, {Adams}, \& {Lizano}}]{Shu1987}
{Shu}, F.~H., {Adams}, F.~C., \& {Lizano}, S. 1987, \araa, 25, 23

\bibitem[{{Sorai} {et~al.}(2019){Sorai}, {Kuno}, {Muraoka}, {Miyamoto},
  {Kaneko}, {Nakanishi}, {Nakai}, {Yanagitani}, {Tanaka}, {Sato}, {Salak},
  {Umei}, {Morokuma-Matsui}, {Matsumoto}, {Ueno}, {Pan}, {Noma}, {Takeuchi},
  {Yoda}, {Kuroda}, {Yasuda}, {Yajima}, {Oi}, {Shibata}, {Seta}, {Watanabe},
  {Kita}, {Komatsuzaki}, {Kajikawa}, {Yashima}, {Cooray}, {Baji}, {Segawa},
  {Tashiro}, {Takeda}, {Kishida}, {Hatakeyama}, {Tomiyasu}, \&
  {Saita}}]{Sorai2019}
{Sorai}, K., {Kuno}, N., {Muraoka}, K., {et~al.} 2019, \pasj, 71, S14

\bibitem[{{Speagle} {et~al.}(2014){Speagle}, {Steinhardt}, {Capak}, \&
  {Silverman}}]{Speagle2014}
{Speagle}, J.~S., {Steinhardt}, C.~L., {Capak}, P.~L., \& {Silverman}, J.~D.
  2014, \apjs, 214, 15

\bibitem[{{Sun} {et~al.}(2020{\natexlab{a}}){Sun}, {Leroy}, {Ostriker},
  {Hughes}, {Rosolowsky}, {Schruba}, {Schinnerer}, {Blanc}, {Faesi},
  {Kruijssen}, {Meidt}, {Utomo}, {Bigiel}, {Bolatto}, {Chevance}, {Chiang},
  {Dale}, {Emsellem}, {Glover}, {Grasha}, {Henshaw}, {Herrera},
  {Jimenez-Donaire}, {Lee}, {Pety}, {Querejeta}, {Saito}, {Sandstrom}, \&
  {Usero}}]{Sun2020b}
{Sun}, J., {Leroy}, A.~K., {Ostriker}, E.~C., {et~al.} 2020{\natexlab{a}},
  \apj, 892, 148

\bibitem[{{Sun} {et~al.}(2020{\natexlab{b}}){Sun}, {Leroy}, {Schinnerer},
  {Hughes}, {Rosolowsky}, {Querejeta}, {Schruba}, {Liu}, {Saito}, {Herrera},
  {Faesi}, {Usero}, {Pety}, {Kruijssen}, {Ostriker}, {Bigiel}, {Blanc},
  {Bolatto}, {Boquien}, {Chevance}, {Dale}, {Deger}, {Emsellem}, {Glover},
  {Grasha}, {Groves}, {Henshaw}, {Jimenez-Donaire}, {Kim}, {Klessen},
  {Kreckel}, {Lee}, {Meidt}, {Sandstrom}, {Sardone}, {Utomo}, \&
  {Williams}}]{Sun2020}
{Sun}, J., {Leroy}, A.~K., {Schinnerer}, E., {et~al.} 2020{\natexlab{b}},
  \apjl, 901, L8

\bibitem[{{Sun} {et~al.}(2018){Sun}, {Leroy}, {Schruba}, {Rosolowsky},
  {Hughes}, {Kruijssen}, {Meidt}, {Schinnerer}, {Blanc}, {Bigiel}, {Bolatto},
  {Chevance}, {Groves}, {Herrera}, {Hygate}, {Pety}, {Querejeta}, {Usero}, \&
  {Utomo}}]{Sun2018}
{Sun}, J., {Leroy}, A.~K., {Schruba}, A., {et~al.} 2018, \apj, 860, 172

\bibitem[{{Tacconi} {et~al.}(2010){Tacconi}, {Genzel}, {Neri}, {Cox}, {Cooper},
  {Shapiro}, {Bolatto}, {Bouch{\'e}}, {Bournaud}, {Burkert}, {Combes},
  {Comerford}, {Davis}, {F{\"o}rster Schreiber}, {Garcia-Burillo},
  {Gracia-Carpio}, {Lutz}, {Naab}, {Omont}, {Shapley}, {Sternberg}, \&
  {Weiner}}]{Tacconi2010}
{Tacconi}, L.~J., {Genzel}, R., {Neri}, R., {et~al.} 2010, \nat, 463, 781

\bibitem[{{Utomo} {et~al.}(2018){Utomo}, {Sun}, {Leroy}, {Kruijssen},
  {Schinnerer}, {Schruba}, {Bigiel}, {Blanc}, {Chevance}, {Emsellem},
  {Herrera}, {Hygate}, {Kreckel}, {Ostriker}, {Pety}, {Querejeta},
  {Rosolowsky}, {Sandstrom}, \& {Usero}}]{Utomo2018}
{Utomo}, D., {Sun}, J., {Leroy}, A.~K., {et~al.} 2018, \apjl, 861, L18

\bibitem[{{Vazdekis} {et~al.}(2012){Vazdekis}, {Ricciardelli}, {Cenarro},
  {Rivero-Gonz{\'a}lez}, {D{\'\i}az-Garc{\'\i}a}, \&
  {Falc{\'o}n-Barroso}}]{Vazdekis2012}
{Vazdekis}, A., {Ricciardelli}, E., {Cenarro}, A.~J., {et~al.} 2012, \mnras,
  424, 157

\bibitem[{{Vazdekis} {et~al.}(2010){Vazdekis}, {S{\'a}nchez-Bl{\'a}zquez},
  {Falc{\'o}n-Barroso}, {Cenarro}, {Beasley}, {Cardiel}, {Gorgas}, \&
  {Peletier}}]{Vazdekis2010}
{Vazdekis}, A., {S{\'a}nchez-Bl{\'a}zquez}, P., {Falc{\'o}n-Barroso}, J.,
  {et~al.} 2010, \mnras, 404, 1639

\bibitem[{Vehtari {et~al.}(2017)Vehtari, Gelman, \& Gabry}]{Vehtari2017}
Vehtari, A., Gelman, A., \& Gabry, J. 2017, Statistics and Computing, 27, 1413

\bibitem[{{Walter} {et~al.}(2008){Walter}, {Brinks}, {de Blok}, {Bigiel},
  {Kennicutt}, {Thornley}, \& {Leroy}}]{Walter2008}
{Walter}, F., {Brinks}, E., {de Blok}, W.~J.~G., {et~al.} 2008, \aj, 136, 2563

\bibitem[{{Weilbacher} {et~al.}(2020){Weilbacher}, {Palsa}, {Streicher},
  {Bacon}, {Urrutia}, {Wisotzki}, {Conseil}, {Husemann}, {Jarno}, {Kelz},
  {P{\'e}contal-Rousset}, {Richard}, {Roth}, {Selman}, \&
  {Vernet}}]{Weilbacher2020}
{Weilbacher}, P.~M., {Palsa}, R., {Streicher}, O., {et~al.} 2020, \aap, 641,
  A28

\bibitem[{{Whitaker} {et~al.}(2012){Whitaker}, {van Dokkum}, {Brammer}, \&
  {Franx}}]{Whitaker2012}
{Whitaker}, K.~E., {van Dokkum}, P.~G., {Brammer}, G., \& {Franx}, M. 2012,
  \apjl, 754, L29

\bibitem[{{Williams} {et~al.}(2018){Williams}, {Gear}, \&
  {Smith}}]{Williams2018}
{Williams}, T.~G., {Gear}, W.~K., \& {Smith}, M. W.~L. 2018, \mnras, 479, 297

\bibitem[{{Williams} {et~al.}(2022{\natexlab{a}}){Williams}, {Kreckel},
  {Belfiore}, {Groves}, {Sandstrom}, {Santoro}, {Blanc}, {Bigiel}, {Boquien},
  {Chevance}, {Congiu}, {Emsellem}, {Glover}, {Grasha}, {Klessen}, {Koch},
  {Kruijssen}, {Leroy}, {Liu}, {Meidt}, {Pan}, {Querejeta}, {Rosolowsky},
  {Saito}, {S{\'a}nchez-Bl{\'a}zquez}, {Schinnerer}, {Schruba}, \&
  {Watkins}}]{Williams2022}
{Williams}, T.~G., {Kreckel}, K., {Belfiore}, F., {et~al.} 2022{\natexlab{a}},
  \mnras, 509, 1303

\bibitem[{{Williams} {et~al.}(2022{\natexlab{b}}){Williams}, {Kreckel},
  {Belfiore}, {Groves}, {Sandstrom}, {Santoro}, {Blanc}, {Bigiel}, {Boquien},
  {Chevance}, {Congiu}, {Emsellem}, {Glover}, {Grasha}, {Klessen}, {Koch},
  {Kruijssen}, {Leroy}, {Liu}, {Meidt}, {Pan}, {Querejeta}, {Rosolowsky},
  {Saito}, {S{\'a}nchez-Bl{\'a}zquez}, {Schinnerer}, {Schruba}, \&
  {Watkins}}]{Williams2021}
{Williams}, T.~G., {Kreckel}, K., {Belfiore}, F., {et~al.} 2022{\natexlab{b}},
  \mnras, 509, 1303

\bibitem[{{Wyder} {et~al.}(2009){Wyder}, {Martin}, {Barlow}, {Foster},
  {Friedman}, {Morrissey}, {Neff}, {Neill}, {Schiminovich}, {Seibert},
  {Bianchi}, {Donas}, {Heckman}, {Lee}, {Madore}, {Milliard}, {Rich}, {Szalay},
  \& {Yi}}]{Wyder2009}
{Wyder}, T.~K., {Martin}, D.~C., {Barlow}, T.~A., {et~al.} 2009, \apj, 696,
  1834

\bibitem[{{Zhang} {et~al.}(2017){Zhang}, {Yan}, {Bundy}, {Bershady}, {Haffner},
  {Walterbos}, {Maiolino}, {Tremonti}, {Thomas}, {Drory}, {Jones}, {Belfiore},
  {S{\'a}nchez}, {Diamond-Stanic}, {Bizyaev}, {Nitschelm}, {Andrews},
  {Brinkmann}, {Brownstein}, {Cheung}, {Li}, {Law}, {Roman Lopes}, {Oravetz},
  {Pan}, {Storchi Bergmann}, \& {Simmons}}]{Zhang2017}
{Zhang}, K., {Yan}, R., {Bundy}, K., {et~al.} 2017, \mnras, 466, 3217

\end{thebibliography}

\appendix
\section{Toy model to test the hierarchical fitting}
\label{sec:ToyModel}

In this section, we use a toy model to test the performance of the hierarchical fitting algorithm and also to explore the origin of the covariance between the parameters $C_{\star}$ and $C_{\rm mol}$. To this end, we define $20$ data subsets, each one having $3000$ randomly selected data points from our full data set ($\Sigma_{\star}$ and $\Sigma_{\rm mol}$). For each subset, a pair of  $C_{\star}$ and $C_{\rm mol}$ is randomly generated from normal distributions centered at $0.5$ and~$1$, respectively, with standard deviation of~$1$. $C_{\rm norm}$ is fixed to the value  measured in our data at $150$~pc for all subsets, to replicate the procedure we used for our data. The three coefficients are then used to generate simulated $\Sigma_{\rm SFR}$ values for each subset separately. A scatter of ${\sim}0.6$~dex (comparable to that found in our data) is added to the simulated $\Sigma_{\rm SFR}$ values.

Once the data subsets are created, we use the hierarchical fitting routine to find the best set of $C_{\star}$ and $C_{\rm mol}$ to describe each subset, whereas $C_{\rm norm}$ is fitted for all subsets simultaneously. In this test, the hyperprior distributions used are $C_{\star,\mu}\sim \mathcal{N}(0.5,1^{2})$, $C_{\star,\sigma}\sim \mathcal{H}(1^{2})$, $C_{\mathrm{mol},\mu}\sim \mathcal{N}(1,1^{2})$, $C_{\mathrm{mol},\sigma}\sim \mathcal{H}(1^{2})$, $C_{\mathrm{norm},\mu}\sim \mathcal{N}(0,2^{2})$ and $C_{\mathrm{norm},\sigma}\sim \mathcal{H}(2^{2})$.

\begin{figure*}[h!]
    \includegraphics[width = \textwidth]{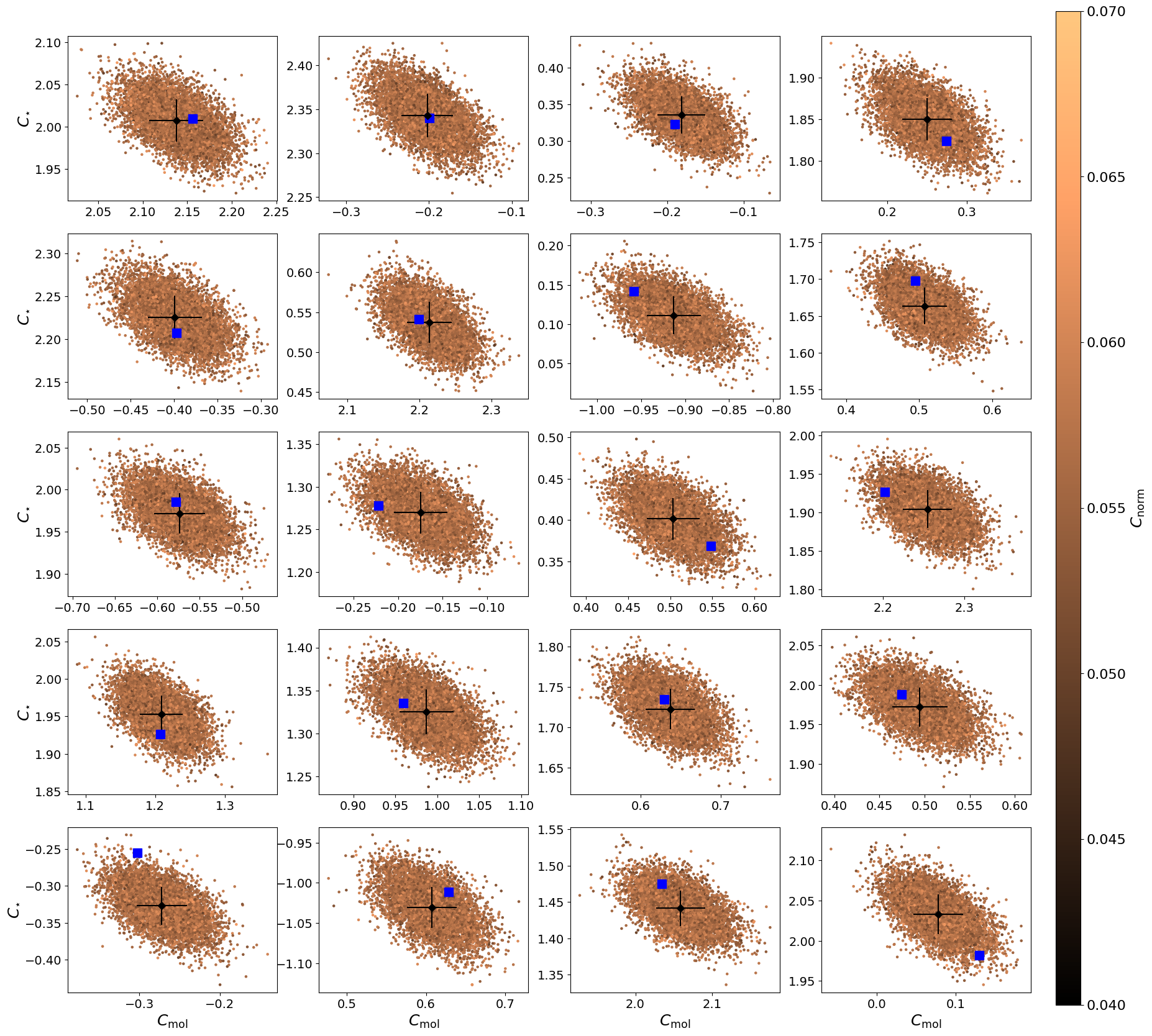}
    
    \caption{Results of the toy model fitting to test accuracy of the hierarchical fitting as well as origin of the $C_{\star} {-} C_{\rm mol}$ covariance. Each panel shows the posterior distribution obtained for a simulated data set. The black diamond and its error bar marks the median  and 1-sigma of the posterior distribution. The blue square marks the real pair of $C_{\star} {-} C_{\rm mol}$ used to generate each simulated data set. The color code  scales with the posterior value of $C_{\rm norm}$.}
    \label{fig:toy_model}
\end{figure*}

Figure~\ref{fig:toy_model} shows the plane conformed by the posterior distributions obtained for $C_{\star}$ and $C_{\rm mol}$ for each data subset. The color code represents the posterior of $C_{\rm norm}$. The black diamond marks the median of the posterior, whereas the blue square shows the real pair of  $C_{\star} {-} C_{\rm mol}$ used to generate each simulated data set. This test primarily demonstrates the accuracy of the hierarchical fitting, as in all cases the real values of the coefficients fall somewhere within the posterior distribution, typically within $1\sigma$ of the recovered one. Secondly, this test shows that the covariance between $C_{\star}$ and $C_{\rm mol}$ discussed in Sec.~\ref{sec:results} is partially an artifact resulting from the fitting, as we recover it even though the real coefficients were chosen independently. However, we can not rule out that the covariance arises as a result of an intrinsic correlation between $\Sigma_{\star}$ and $\Sigma_{\rm mol}$.

Nevertheless, this brings up the question of how many data points are required to break this covariance? And thus, how many galaxies would we require to break (or at least reduce) the covariance in the probed galactic environments?

Figure~\ref{fig:cov} provides some clues to these questions. It shows a larger covariance in the environments with fewer data points (centers) than in those environments probed with more pixels (disks and spiral arms), while the lowest covariance is seen in the full sample.  Then, if we would aim to reduce the covariance of the measurement in rings or centers (${\sim}500$ pixels) to the levels of spiral arms (${\sim}15{,}000$ pixels), we would need a sample on the order of $30$ times larger. However, if we restrict the observations exclusively to galaxies that show all different environments (such as NGC~1672 or NGC~4321), the number of additional galaxies would need to be about ${\sim}12$ times larger. However, this is only an estimation based on the area identified as ring or center in these two galaxies. The exact number will depend on the area occupied by these features in each galaxy.

\section{Choice of detection fraction threshold}
\label{sec:threshold_test}

Through this paper we adopt a detection fraction threshold of $60\%$ to our data, which means that we confine our analysis to those $\Sigma_{\star}$ ranges in which we have a detection fraction of $\Sigma_{\rm SFR}$ and $\Sigma_{\rm mol}$ higher than this threshold (see Sec.~\ref{sec:NDs} for a detailed description). Here we show how our results are influenced by the choice of detection fraction threshold.

\begin{figure}[h!]
    \includegraphics[width=\columnwidth]{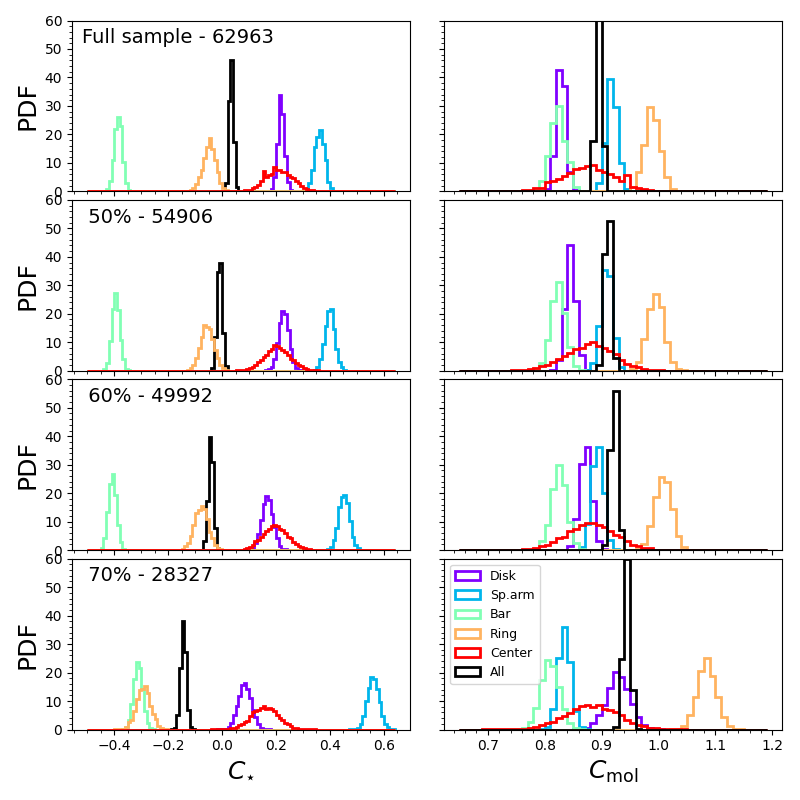}
    \caption{Posterior distributions obtained for the coefficients $C_{\star}$ and $C_{\rm mol}$, following the same methodology described in Sec.~\ref{sec:NDs}, but applying different levels of detection fraction threshold. The top two panels show the base case scenario, using the full sample without applying any detection fraction threshold. The following panels show our fiducial adopted value ($60\%$) and thresholds at $50\%$ and $70\%$. The threshold level and the number of pixels used for each measurement are indicated in the top left corner for each row.}
    \label{fig:threshold_test}
\end{figure}

Figure~\ref{fig:threshold_test} shows the posterior distributions obtained for the coefficients $C_{\star}$ and $C_{\rm mol}$ under different assumptions of detection fraction threshold. The top panel shows the distributions obtained without applying any detection fraction threshold (i.e., using the full sample). The bottom panels show our fiducial adopted value ($60\%$), $50\%$, and $70\%$.

Firstly, we note that the posteriors of nondisk environments only change slightly from using the full sample to applying a threshold of $60\%$ in the detection fraction (first three rows). This is because at the $60\%$ level, the fraction of pixels dropped from these environments is very small (${<}5\%$), as explained in Sec.~\ref{sec:NDs}. On the other hand, disks show a clear steepening of the $C_{\rm mol}$ slope in the same range of detection fraction. This behavior is consistent with the expected role of N/Ds in the measurement on the slope (Fig.~\ref{fig:Det_frac}).

We also note that only $5000$ data points are lost when going from $50\%$ to $60\%$, while going from $60\%$ to $70\%$ leads to drop ${\sim}22{,}000$ additional data points. Furthermore, at the $70\%$ threshold level, the fraction of data points removed from spiral arms and rings rises to ${\sim}11\%$ and ${\sim}35\%$, respectively. As a result of this higher number of dropped data points, and the significant truncation of the $\Sigma_{\star}$ range in which these environments are probed, their posteriors change drastically at this detection fraction level, becoming unstable with respect to less strict threshold levels. Thus, we conclude here that a $60\%$ threshold (fiducial value) provides a good balance between minimizing the impact of N/Ds on our analysis, and keeping the integrity of our data.

\section{Global versus per-environment detection fraction threshold}
\label{sec:threshold_test_global_vs_single}

In Sec.~\ref{sec:NDs} we present our methodology to minimize the impact of N/Ds in the slope measurement, by restricting our analysis to the $\Sigma_{\star}$ in which each environment has a detection fraction higher than $60\%$. Here, we present our results under a slightly different approach. Instead of defining the used $\Sigma_{\star}$ range for each environment separately, we use the same $\Sigma_{\star}$ range for all environments, defined on the detection fraction of the full sample (black line in Fig.~\ref{fig:Det_frac}.)

As our full sample is statistically dominated by the disk environment, this implies that the $\Sigma_{\star}$ range at which the rest of the environments are probed is truncated to roughly match the $\Sigma_{\star}$ range in which disks satisfy the detection fraction threshold. In practice, this has an impact mainly in the sampling of spiral arms and rings, which due to their inherently higher detection fractions (Fig.~\ref{fig:Det_frac}), can be probed in a larger range of  $\Sigma_{\star}$ values following our fiducial approach.

\begin{figure}[h!]
    \includegraphics[width=\columnwidth]{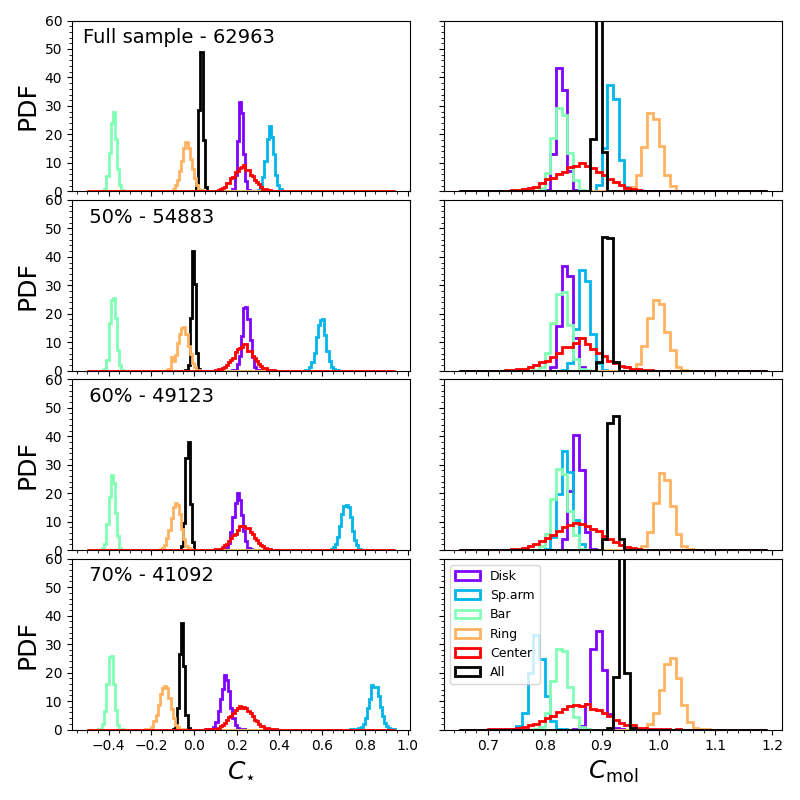}
    \caption{Posterior distributions obtained for the coefficients $C_{\star}$ and $C_{\rm mol}$, restricting the $\Sigma_{\star}$ range of different galactic environments according to the detection fraction of the full sample, for different levels of detection fraction threshold. The top two panels shows the base case scenario, using the full sample without applying any detection fraction threshold. The following panels show our fiducial adopted value ($60\%$) and thresholds at $50\%$ and $70\%$. The threshold level and the number of pixels used for each measurement are indicated in the top left corner for each row.}
    \label{fig:threshold_test_global_vs_single}
\end{figure}

Figure~\ref{fig:threshold_test_global_vs_single} shows the obtained posterior distributions of the $C_{\star}$ and $C_{\rm mol}$ coefficients in each environment, for four different levels of detection fraction threshold (similar to Fig.~\ref{fig:threshold_test}). Due to the truncation of the  $\Sigma_{\star}$ range, the posterior of the spiral arms becomes unstable, and highly dependant on the threshold level applied, reaching almost a value of $C_{\rm mol}\approx1$ for a detection fraction threshold of $70\%$. Thus, we opt for a ``dynamic'' $\Sigma_{\star}$ range, chosen for each environment separately, as our fiducial approach.

We note that the total number of data points in the sample at this threshold is drastically increased with respect to the fiducial scenario. This is because additional data points from the disk environment are included in the sample (as the detection fraction of the full sample is slightly higher than that for the disk only), at the cost of excluding data points from other environments. 

Nevertheless, we find that environmental differences, in terms of $C_{\star}$ and $C_{\rm mol}$, persist under this different methodology, and thus, qualitatively our conclusions are robust against this different treatment of our data.

\section{Correlation of $\langle \sigma_{\mathrm{H}\alpha} \rangle$ and  $\langle f_{\mathrm{mol}} \rangle$ with coefficients calculated in individual environments of single galaxies}
\label{sec:corr_Ha_GasFrac}

\begin{figure*}[h!]
    \includegraphics[width=\textwidth]{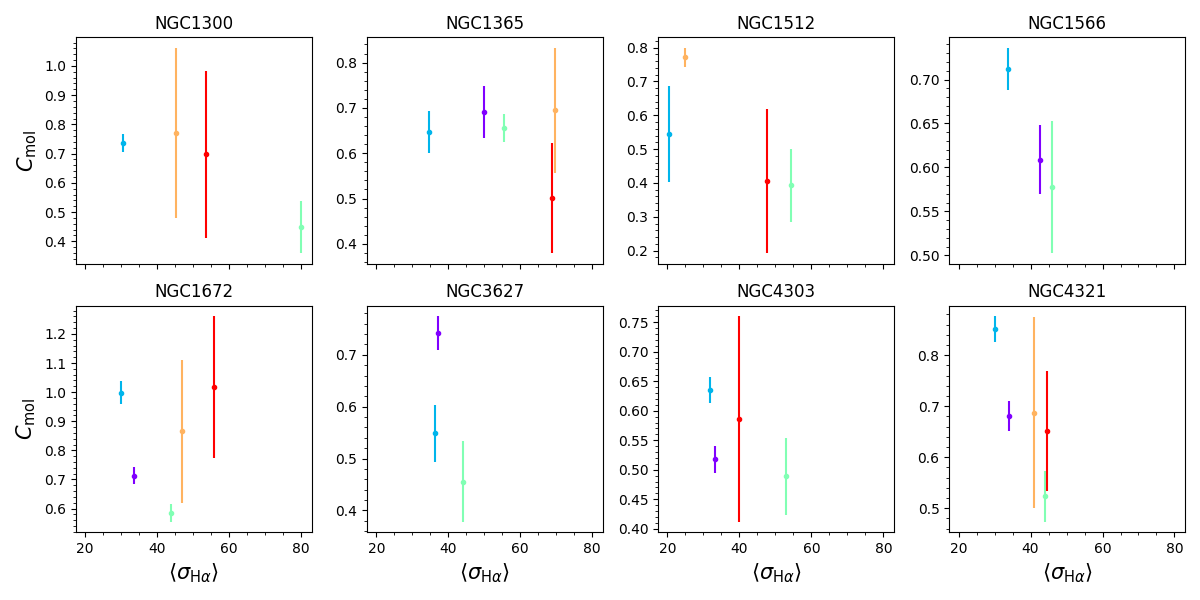}
    \caption{$C_{\rm mol} {-} \langle \sigma_{\mathrm{H}\alpha} \rangle$ correlation across different galactic environments, for the galaxies that satisfy our single-galaxy selection criteria. The color code for each environment is the same as used in Fig.~\ref{fig:Det_frac}.}
    \label{fig:C1_Ha}
\end{figure*}

Here we show the correlation between the $C_{\star}$ and $C_{\mathrm{mol}}$ coefficients derived for individual environments in single galaxies (see Sec.~\ref{sec:third_param}) and the second-highest Pearson coefficient parameters in Fig.~\ref{fig:OverallP}. Figure~\ref{fig:C1_Ha} shows the $C_{\mathrm{mol}}$ calculated for each environment, as a function of the average $\sigma_{\mathrm{H}\alpha}$  within that environment, for each galaxy in the subsample defined in Sec.~\ref{sec:third_param}. This correlation could be an imprint of how the dynamical state of the gas might play role in regulating the level of local SFR.

\begin{figure*}[h!]
    \includegraphics[width=\textwidth]{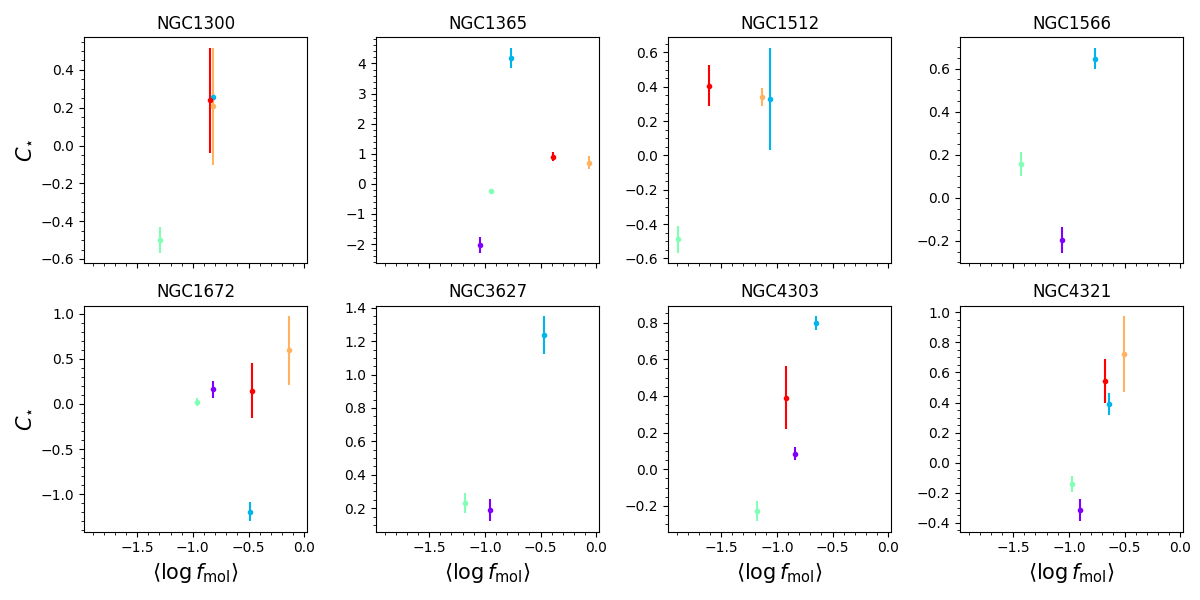}
    \caption{$C_{\star} {-} \langle f_{\mathrm{mol}} \rangle$ correlation across different galactic environments, for the galaxies that satisfy our single-galaxy selection criteria. The color code for each environment is the same as used in Fig.~\ref{fig:Det_frac}.}
    \label{fig:C0_GasFrac}
\end{figure*}

For comparison, Fig.~\ref{fig:C0_GasFrac} shows the correlation between $C_{\star}$ and $\langle f_{\mathrm{mol}} \rangle$ for each environment, for the same galaxies. A positive correlation implies that environments with higher gas fractions also show typically lower depletion times.

\end{document}